%% file: main.tex
\pdfoutput=1
\documentclass[acmtog]{acmart}
\acmSubmissionID{951}
\usepackage{diagbox}
\usepackage{verbatim}
\usepackage{amssymb}
\usepackage{latexsym}
\usepackage{lineno}
\usepackage{xcolor}
\usepackage{tabularx}
\usepackage{xspace}
\usepackage{color}
\graphicspath{{figures/}}
\usepackage{amsmath,bm}
\usepackage{psfrag}
\usepackage{mathtools}
\usepackage{hyperref}

\usepackage{multirow, tabularx}
\newcolumntype{Y}{>{\centering\arraybackslash}X}
\usepackage{capt-of}%
\usepackage{array, boldline, makecell, booktabs}

\newcommand{\Mod}[1]{\ (\mathrm{mod}\ #1)}
\usepackage{colortbl}
\usepackage{array}

\usepackage{pifont}% http://ctan.org/pkg/pifont
\newcommand{\cmark}{\ding{51}}%
\newcommand{\xmark}{\ding{55}}%

\usepackage{psfrag}
\usepackage{forest}
\usepackage{makecell}
\usepackage{physics}
\usepackage{algorithm}
\usepackage{threeparttable}
\usepackage{graphicx}
\usepackage{subcaption}
\usepackage{flushend}
\usepackage{wrapfig}
\usepackage{algpseudocode}
\usepackage{fancyhdr}
\usepackage{algpseudocode}
\input{bo_macro}

\usepackage{url}
\usepackage{xcolor}
\definecolor{newcolor}{rgb}{.8,.349,.1}

\usepackage{enumitem}
\usepackage{svg}
\svgpath{{./img/}} % <- using \svgpath to avoid warning

\usepackage[normalem]{ulem}

\newcommand{\gradF}[1]{\nabla \mathcal{F}}
\newcommand{\surfmask}[1]{\chi_S^\text{surf}}
\newcommand{\inmask}[1]{\chi_S^\text{in}}
\newcommand{\solidmask}[1]{\chi_S}
\newcommand{\xp}[1]{\bm x^p}
\newcommand{\midvortp}[1]{\bm \omega_b^p}
\newcommand{\initvortp}[1]{\bm \omega_a^p}
\newcommand{\currvortp}[1]{\bm \omega_c^p}
\newcommand{\currvortg}[1]{\bm \omega_c^g}
\newcommand{\currvelg}[1]{\bm u_c^g}
\newcommand{\Tab}[1]{\mathcal{T}^p_{[a, b]}}
\newcommand{\Tbc}[1]{\mathcal{T}^p_{[b, c]}}
\newcommand{\Fab}[1]{\mathcal{F}^p_{[a, b]}}
\newcommand{\Fbc}[1]{\mathcal{F}^p_{[b, c]}}
\newcommand{\Fac}[1]{\mathcal{F}^p_{[a, c]}}
\newcommand{\Tac}[1]{\mathcal{T}^p_{[a, c]}}
\newcommand{\T}[1]{\mathcal{T}}
\newcommand{\F}[1]{\mathcal{F}}
\newcommand{\ce}[1]{\centering}
\newcommand{\gradFbc}[1]{\gradF{}_{[b, c]}^p}
\newcommand{\gradFac}[1]{\gradF{}_{[a, c]}^p}
\newcommand{\omegavisc}[1]{\nu \Delta \bm \omega}
\newcommand{\normv}[1]{\bm u_{Sn}^g}
\newcommand{\tanv}[1]{\bm u_{St}^g}
\newcommand{\currdeltagammag}[1]{\delta \Gamma^g}
\newcommand{\gammapac}[1]{\gamma^p_{a \rightarrow c}}
\newcommand{\gammapbc}[1]{\gamma^p_{b \rightarrow c}}
\newcommand{\figref}[1]{Figure~\ref{#1}}

\newcommand{\rev}[1]{{#1}}
\newcommand{\revv}[1]{{#1}}

\setcopyright{cc}
\setcctype{by}
\acmJournal{TOG}
\acmYear{2025} \acmVolume{44} \acmNumber{4} \acmArticle{91} \acmMonth{8} \acmPrice{}\acmDOI{10.1145/3731198}

\begin{document}
% \begin{frontmatter}
% \footnote{}

\author{Sinan Wang}
\email{swang3081@gatech.edu}
\affiliation{
\institution{Georgia Institute of Technology}
\country{USA}
}

\author{Junwei Zhou}
\email{zjw330501@gmail.com}
\affiliation{
\institution{University of Michigan Ann Arbor}
\country{USA}
}

\author{Fan Feng}
\email{fan.feng.gr@dartmouth.edu}
\affiliation{
\institution{Dartmouth College}
\country{USA}
}

\author{Zhiqi Li}
\email{zli3167@gatech.edu}
\affiliation{
\institution{Georgia Institute of Technology}
\country{USA}
}

\author{Yuchen Sun}
\email{yuchen.sun.eecs@gmail.com}
\affiliation{
\institution{Georgia Institute of Technology}
\country{USA}
}

\author{Duowen Chen}
\email{dchen322@gatech.edu}
\affiliation{
\institution{Georgia Institute of Technology}
\country{USA}
}

\author{Greg Turk}
\email{turk@cc.gatech.edu}
\affiliation{
\institution{Georgia Institute of Technology}
\country{USA}
}

\author{Bo Zhu}
\email{bo.zhu@gatech.edu}
\affiliation{
\institution{Georgia Institute of Technology}
\country{USA}
}

\title{Fluid Simulation on Vortex Particle Flow Maps}
\begin{abstract}
We propose the \textbf{V}ortex \textbf{P}article \textbf{F}low \textbf{M}ap (VPFM) method to simulate incompressible flow with complex vortical evolution in the presence of dynamic solid boundaries. The core insight of our approach is that vorticity is an ideal quantity for evolution on particle flow maps, enabling significantly longer flow map distances compared to other fluid quantities like velocity or impulse. To achieve this goal, we developed a hybrid Eulerian-Lagrangian representation that evolves vorticity and flow map quantities on vortex particles, while reconstructing velocity on a background grid. The method integrates three key components: (1) a vorticity-based particle flow map framework, (2) an accurate Hessian evolution scheme on particles, and (3) a solid boundary treatment for no-through and no-slip conditions in VPFM. These components collectively allow a substantially longer flow map length (\textbf{3}-\textbf{12} times longer) than the state-of-the-art, enhancing vorticity preservation over extended spatiotemporal domains. We validated the performance of VPFM through diverse simulations, demonstrating its effectiveness in capturing complex vortex dynamics and turbulence phenomena.

\end{abstract}




\begin{teaserfigure}
\centering
\includegraphics[width=1.02\textwidth]{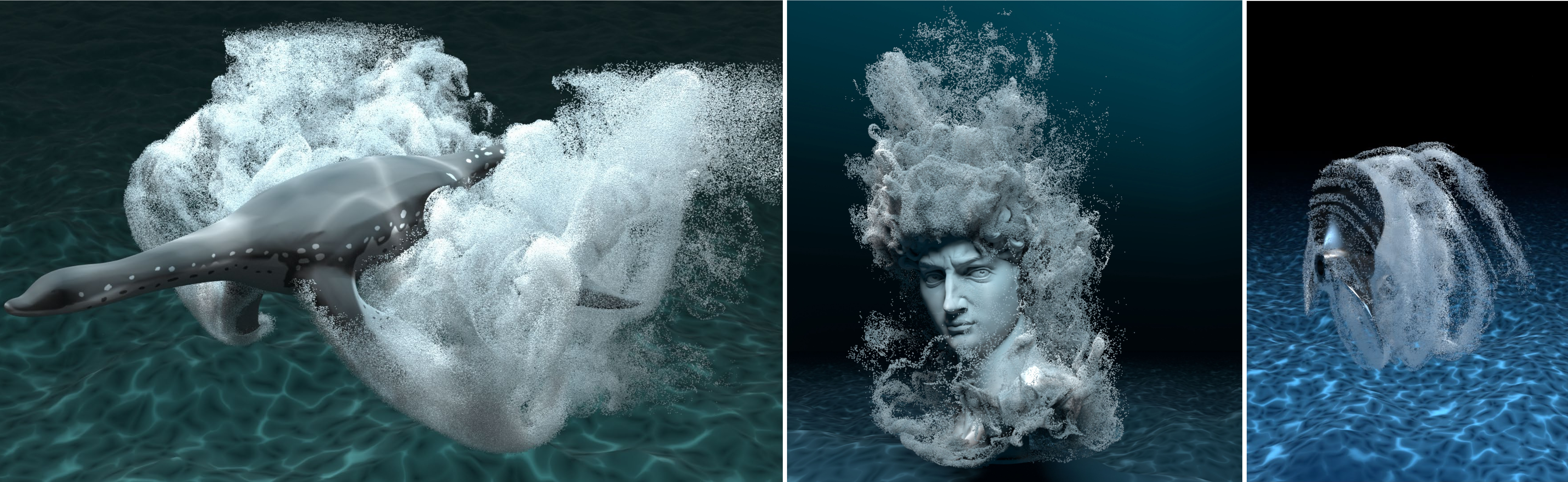}
% \figvspace
\caption{Bubbles visualize the flow around a plesiosaur swimming (left), the David statue (middle) and a propeller turning (right).}
\label{fig:teaser}
\end{teaserfigure}

\maketitle

% \begin{figure}
%     \centering
%     \includegraphics[width=0.5\textwidth]{teaser/0599.jpg}
%     \caption{Head of Michelangelo's David}
%     \label{fig:teaser}
% \end{figure}

\section{INTRODUCTION}
Vortex particles have been recognized as an essential numerical tool for simulating incompressible fluids in both computational physics and computer graphics over decades, accommodating a broad variety of applications ranging from investigating the wake-vortex street behind a fishtail in fluid mechanics to reproducing the vortical evolution of a smoke plume in movie visual effects. Among these simulation scenarios, \textit{vortex particles} have been established as a bridge connecting the flow physics and numerical stencils. As stated  by \citet{cottet2000vortex}: "\textit{The close link of numerics and physics is the essence of vortex methods}." On the physical side, each vortex particle can be seen as a vorticity-carrying fluid element that transports vorticity and induces velocity; on the numerical side, each vortex particle is treated as an element that discretizes the space and performs time integration. 

An interesting difference of vortex methods in computational physics versus computer graphics lies in how the term \textit{\rev{vortex particle methods}} is interpreted in each domain. In computational physics, \textit{\rev{vortex particle methods}}  broadly encompass a family of methods derived from the vorticity-form Navier–Stokes equations. One of the most widely used methods based on vortex particles is the "Vortex-In-Cell" (VIC) method \cite{christiansen1973numerical}, where particles serve as traditional Lagrangian samplers to transport vorticity, while a background grid handles local differentiation such as vorticity stretching and viscosity (resembling a vorticity-form PIC/FLIP approach for Navier-Stokes). However, in computer graphics, "\rev{vortex particle methods}" signify something distinct and limited: they either refer to \rev{seeding vortex particles} into a fluid domain for vorticity confinement (\citet{bridson2015fluid} refers to these as "spin particles") \cite{selle2005vortex, pfaff2009synthetic}, and evolved passively with the background flow that is solved using the traditional velocity-form Navier-Stokes equations; Or, they are limited to pure Lagrangian vortex methods \cite{yaeger1986combining, gamito1995two, angelidis2005simulation, park2005vortex}. Unlike \rev{vortex particle methods} in computational physics, where the VIC methods directly extend the PIC/FLIP framework from velocity to vorticity, in computer graphics, they do not constitute such a direct extension. This disparity raises an intriguing question: why has the graphics community primarily limited its use of vortex particles to applications like vorticity confinement or pure particle-based vortex methods? In particular, why hasn’t the graphics community adopted the VIC method as widely as they have embraced PIC/FLIP?

The answer to this question is quite simple: the VIC method does not demonstrate a significant advantage in preserving vorticity, which is often a critical focus in graphics algorithms.
%(as demonstrated in a comparison with a traditional VIC method \cite{rossinelli2008vortex} in Section \ref{subsubsec:compare_optimal_flowmap}, where its ability to preserve vorticity is shown to be limited)
Although it might seem that VIC could gain an advantage over purely Eulerian methods by using particles to advect vorticity, its vorticity preservation still falls short of pure particle-based vortex methods and it lacks vortical details compared to pure Lagrangian vortex methods \cite{bridson2015fluid}. This limitation persists, despite the significant computational conveniences offered by the grid, such as enabling cheap finite difference calculations, using fast Poisson solvers, and avoiding the need for Biot–Savart summations, making VIC methods unappealing to the graphics community.

%Evolving vorticity on particles does not inherently guarantee vorticity conservation. On the contrary, there is no evidence that the vortical structures evolved using vortex particles and their background grid outperform alternatives such as velocity-based PIC/FLIP. This renders the additional implementation complexities --- such as converting vorticity to velocity and handling vorticity-form boundary conditions --- unjustifiable and inefficient, especially considering that vorticity-form simulations do not naturally preserve vorticity and vorticity can always be reconstructed from velocity by taking the curl.

This paper focuses on reviving the VIC method in graphics by refurbishing traditional VIC numerics from the modern particle flow map perspective. The cornerstone of our method is established on the following facts and insights: (1) Vorticity, as a type of line element \cite{wang2024eulerian} or 2-form \cite{nabizadeh2022covector, 10.1145/3592402}, can be naturally transported over long-term flow maps \cite{ yin2021characteristic, yin2023characteristic}. (2) Particle trajectories are an ideal representation of such flow maps due to their unique bidirectional nature in spacetime (see \cite{zhou2024eulerian, li2024particle, chen2024solid}). (3) Extending traditional "vortex particles" to "vortex particle flow maps" by evolving not only their vorticity but also higher-order quantities, such as vorticity gradient, flow-map Jacobians and Hessians, has the potential to improve vorticity preservation compared to prior methods. (4) Most importantly, vorticity can serve as a better gauge variable than others, especially considering its potential in facilitating a long-term flow map (discussed in Section \ref{disc:vorticity_less_singular}).

%owing to its geometric interpretation (i.e., vorticity evolution can be expressed as a mapping between distinct frames), its physical stability (i.e., vorticity does), and therefore, its potential in extending the flow map length in turbulent simulations (which has been infeasible for gauge variables such as velocity and impulse).

Motivated by these observations, we propose the \textbf{V}ortex \textbf{P}article \textbf{F}low \textbf{M}ap (VPFM) method to unlock the potential of vortex particles in supporting long-term flow maps. Our system consists of three key components: (1) a vortex particle flow map framework for transporting vorticity with flow maps carried by moving particles, (2) a novel flow map Hessian solution evolved on particle flow maps,  and (3) a boundary treatment solution for VPFM that achieves accurate no-through conditions and approximated no-slip conditions. These components collectively enables a robust, long-term flow map that is \textbf{3}-\textbf{12} times longer than the state-of-the-art. Under challenging long-term flow maps, our method remains \textbf{indefinitely stable} in 2D benchmarks as other methods quickly fail, while in 3D benchmarks, our method preserves vortical structures up to \textbf{30} times longer than the state-of-the-art. Through a wide range of validation tests and simulation examples across different vortex dynamics scenarios, we demonstrate the efficacy of our vortex particle flow map method in producing and preserving spatiotemporally coherent vortical structures, effectively revitalizing the traditional VIC method -- previously on the periphery of graphics applications -- into a state-of-the-art approach for producing physically accurate and visually appealing vortical flow motions.

\section{RELATED WORK}
\subsection{Flow Map Methods}
\label{subsec:relate_flowmap}
%Flow map, or characteristic mapping method is a family of advection methods, which was first proposed by \citet{wiggert1976numerical}, and introduced to the graphics community by \citet{tessendorf2011characteristic}. Flow maps were first computed by virtual particles \cite{hachisuka2005combined, tessendorf2015advection}, and used to advect velocity \cite{sato2017long, sato2018spatially}. An Eulerian scheme for efficiently computing flow maps were proposed by \cite{qu2019efficient}. Recently, gauge variables, such as impulse \cite{nabizadeh2022covector, deng2023neural}, and vorticity \cite{yin2023characteristic, yin2021characteristic, wang2024eulerian, 10.1145/3592402}, were advected on Eulerian flow maps. Lagrangian or particle flow map methods \cite{li2024lagrangian, zhou2024eulerian} evolve flow map Jacobians to compute the stretching term for impulse, compared to traditional approaches that are based on finite difference or interpolation via neighboring particles \cite{cortez1998accuracy, sancho2024impulse, park2005vortex, christiansen1973numerical}. In order to achieve a long-term particle flow map, we further extend this to a vortex particle flow map method, which evolves flow map Jacobians and Hessian to compute vorticity stretching and evolve vorticity gradient on particles.
Flow map techniques, or characteristic mapping methods, first proposed by \citet{wiggert1976numerical} and introduced to computer graphics by \citet{tessendorf2011characteristic}, are accurate advection techniques that effectively maintain vortex structures.
%They were first used to advect velocity \cite{sato2017long, sato2018spatially, qu2019efficient}, and then impulse \cite{nabizadeh2022covector, deng2023neural, zhou2024eulerian}, and vorticity \cite{wang2024eulerian, yin2021characteristic, yin2023characteristic, 10.1145/3592402}.
%They become unstable, or show artifacts when flow map length is further extended, as presented in Section \ref{subsubsec:compare_same_long_flow_map}. In contrast, our method is able to use flow map length of \textbf{240} for a 2D experiment, and \textbf{60} for a 3D experiment, representing a \textbf{12$\times$} and \textbf{3$\times$} improvement, and achieved the state-of-the-art performance, as shown in Section \ref{subsubsec:compare_optimal_flowmap}.
%Specifically, \textbf{5} is used in \cite{qu2019efficient, nabizadeh2022covector, 10.1145/3592402}, and a maximum of \textbf{20} is used in \cite{deng2023neural, wang2024eulerian, zhou2024eulerian}, compared to a maximum of \textbf{240} for ours. 
However, none of the existing flow map methods can achieve a robust and long-term flow map. The reasons behind are listed as follows: (1) They either use Eulerian flow maps \cite{qu2019efficient, wang2024eulerian, deng2023neural, nabizadeh2022covector, yin2021characteristic, yin2023characteristic, 10.1145/3592402}, incurring interpolation errors, and distortion in flow map quantities; (2) Or, they choose a less suitable variable to couple with flow maps. Specifically, velocity $\bm{u}$ \cite{qu2019efficient, sato2018spatially, sato2017long}, or impulse \(\bm m\) \cite{deng2023neural, nabizadeh2022covector, zhou2024eulerian} \rev{(originally introduced to graphics in~\cite{feng2022impulse})} will introduce larger singularities and instability (see Section~\ref{disc:vorticity_less_singular}) and cannot support long-term flow maps. (3) When using particle flow maps, prior methods have inadequately handled the flow map Hessian term. Specifically, \citet{zhou2024eulerian} directly omits this term. \citet{sancho2024impulse} approximates this term via temporarily sampled points. \rev{By} contrast, we accurately evolve this term on particle flow maps, \rev{resulting} in a perfect flow map Hessian. \rev{Recently, CO-FLIP~\cite{nabizadeh2024coflip} introduced a structure-preserving method based on a modified Hamiltonian system, albeit at the expense of increased computational cost due to the pseudoinverse solve in the P2G transfer.}

% attempted to handle the flow map Hessian term by interpolating it through the aggregation of neighboring particles. However, this approach led to poor performance and introduced asymmetry, ultimately resulting in the omission of

\subsection{Vortex Methods}
\rev{Vortex methods} reformulate the Navier-Stokes equations by using vorticity as the primary variable. These methods offer more direct control over fluid circulation, quantified by vorticity, which is defined as: \(\bm \omega = \nabla \times \bm u\). Initially, they emerged as purely Lagrangian approaches, leveraging their circulation-preserving nature. In these methods, vorticity is carried by particles \cite{selle2005vortex,cottet2000vortex,park2005vortex,angelidis2017multi}, filaments \cite{angelidis2005simulation,weissmann2010filament,ishida2022hidden,padilla2019bubble}, segments \cite{xiong2021incompressible}, sheets \cite{brochu2012linear,pfaff2012lagrangian,da2015double}, and Clebsch level sets \cite{chern2016schrodinger,chern2017inside,yang2021clebsch,xiong2022clebsch}. Eulerian vortex methods, while less common due to their tendency to introduce numerical dissipation, offer advantages such as straightforward finite difference computations on \rev{the grid} and efficient Poisson solvers. To mitigate the dissipation inherent in Eulerian methods, several circulation-preserving approaches \rev{\cite{elcott2007stable,wang2024eulerian, yin2021characteristic, yin2023characteristic, azencot2014functional}} have been developed. \rev{\citet{10.1145/3592402} employs \textit{Functional Fluids} \cite{azencot2014functional} in 2D and \textit{Covector Fluids} \cite{nabizadeh2022covector} in 3D to help conserve vorticity, and derives the missing dynamics for harmonic (cohomology) components of the flow on non-simply-connected domains.} Hybrid vortex methods were introduced to combine the benefits of both frameworks, where a Lagrangian vortex scheme solves advection on particles to reduce numerical dissipation, while the underlying Eulerian grid facilitates vortex stretching, viscosity handling, and Poisson solving. A notable hybrid scheme is the Vortex-In-Cell (VIC) method \cite{christiansen1973numerical}. Further advancements, such as remeshed VIC methods \cite{ould2001blending,gallizio2009analytical,van2011comparison,mimeau2015vortex}, relocate vortex particles periodically to regularize their distribution. In the graphics community, hybrid methods such as \cite{zhang2014pppm} leverage an underlying grid to accelerate the Biot-Savart summation for velocity reconstruction. \rev{By} comparison, our hybrid vortex method solves both advection and vortex stretching on particles via long-term particle flow maps, which yields a more accurate advection system for vortex methods. \rev{We refer the reader to the reviews \cite{mimeau2021review, koumoutsakos2005multiscale} for comprehensive overviews of particle systems and vortex methods.}

\subsection{Solid Boundary Condition for Vortex Methods}
For velocity, the no-through solid boundary condition (BC) (\(\bm u \cdot \bm n = \bm u_\text{solid} \cdot \bm n\)) can be enforced by a voxelized pressure projection process \cite{foster1996realistic, foster2001practical, houston2003unified, rasmussen2004directable}, which results in stair-stepped artifacts, especially on \rev{a coarser grid}. Later, some Symmetric Positive Semi-Definite (SPSD) cut cell methods are proposed for velocity, such as those based on the finite volume method \cite{ng2009efficient}, or a variational framework \cite{batty2007fast}. The treatment of solid boundaries is a fundamental aspect of vortex methods, as solid objects and viscosity are the primary sources of vorticity generation.
\paragraph{Solid BC for Lagrangian vortex method} The very first successful boundary treatment for pure Lagrangian vortex method is the vortex sheet/vortex blob methods \cite{chorin1973numerical, chorin1978vortex} (2D) and \cite{chorin1980vortex, ploumhans2002vortex} (3D), where they construct vortex sheets at boundaries to vanish the slip velocities. In graphics, \citet{zhang2014pppm} adopted this method to shed vortices. Later on, BEM or panel methods \cite{hess1964calculation, koumoutsakos1994boundary, koumoutsakos1995high, willis2006unsteady} were proposed based on the vorticity flux boundary conditions, which was adopted by \citet{park2005vortex} in graphics.  Pure Lagrangian vortex methods are naturally suited for curved geometries but face three challenges: (1) solving dense, ill-conditioned systems to determine panel strengths, (2) computing vortex stretching and viscosity using unstructured particles, and (3) the computational cost of Biot-Savart summation for velocity reconstruction. While acceleration techniques like the Fast Multipole Method (FMM) \cite{greengard1987fast}, PPPM \cite{zhang2014pppm}, and precomputed panel solvers \cite{xie2018precomputed} can help, they are complex to implement and typically less efficient than fast Poisson solvers.
%For Lagrangian vortex methods, there were vortex sheet/blob methods \cite{chorin1973numerical, chorin1978vortex} (2D) and \cite{chorin1980vortex, ploumhans2002vortex} (3D), that creates a vortex sheet at boundaries to cancel the slip velocity. In computer graphics, this was used by \citet{zhang2014pppm} to enforce the no-slip conditions (\(\bm u = \bm u_\text{solid}\)) and shed vortices. Instead of vortex sheets/blobs, BEM or Panel methods \cite{koumoutsakos1994boundary, koumoutsakos1995high, hess1964calculation}  enforce no-slip conditions by vorticity flux boundary conditions. which was introduced to graphics community by \citet{park2005vortex}. While Lagrangian vortex methods inherently excel at accurately handling curved geometries, they require solving a dense and computationally expensive linear system to determine panel strengths. \citet{xie2018precomputed} accelerates the process using a precomputed panel solver but is limited to rigid bodies.
%For hybrid/Eulerian vortex methods,
%\paragraph{Solid BC for Hybrid/Eulerian Vortex Methods}
\begin{table}[t]
\hspace*{-1em}
\caption{Comparison of hybrid/Eulerian vortex methods in the literature. These methods solve the velocity reconstruction on \rev{the grid}. Spa.\&SPSD stands for Sparse and Symmetric Positive Semi-Definite system for velocity reconstruction. Moving BC refers to a moving solid boundary. \rev{SVR stands for Single Velocity Reconstruction. While iterative Brinkmann penalization methods require multiple velocity reconstructions, our approach requires only one by relaxing the no-slip constraints.}}
\centering\small
\begin{tabular}{l@{\hspace{0mm}}c@{\hspace{2mm}}c@{\hspace{2mm}}c@{\hspace{2mm}}c@{\hspace{2mm}}c@{\hspace{2mm}}c}
\hlineB{3}
Method & Cut Cell & No-slip & Spa.\&SPSD & 3D & Moving BC & \rev{SVR} \\
\hlineB{2}
%\cite{rossinelli2008vortex} & \xmark & \xmark & \cmark & \xmark & \xmark\\

\cite{rasmussen2011multiresolution} & \xmark & \cmark & \cmark & \xmark & \xmark & \cmark\\
\cite{hejlesen2015iterative} & \xmark & \cmark & \cmark  & \xmark & \xmark & \xmark\\
\cite{spietz2017iterative} & \xmark & \cmark & \cmark & \cmark & \xmark & \xmark\\
\cite{marichal2016immersed} & \cmark & \cmark & \xmark  & \xmark & \xmark & \cmark\\
\cite{gillis2018fast} & \cmark & \cmark & \xmark  & \cmark & \xmark & \cmark\\
\cite{wang2024eulerian} & \xmark & \xmark & \cmark  & \cmark & \cmark & \cmark\\
Ours & \cmark & \cmark & \cmark & \cmark & \cmark & \cmark\\
\hlineB{3}
\end{tabular}
%\captionsetup{aboveskip=40pt}
%\captionsetup{aboveskip=40pt}
\label{tab:vortex_related_work}
\end{table}

\paragraph{Solid BC for Hybrid/Eulerian vortex method} The immersed boundary methods (IBM) \cite{peskin1972flow, peskin1977numerical} were first used for velocity, and later on introduced to hybrid vortex methods, as vortex IBM in VIC methods \cite{cottet2004advances, poncet2009analysis}. However, IBM cannot exactly enforce the no-through or the no-slip (\(\bm u  = \bm u_\text{solid} \)) conditions (e.g., fluid may penetrates solid). As part of IBM, the Brinkmann penalization was proposed and developed by \citet{caltagirone1994interaction, angot1999penalization}, and first used in vortex methods by \citet{kevlahan2001computation}. A traditional way of using vortex Brinkmann penalization \cite{rasmussen2011multiresolution, mimeau2015vortex} is by penalizing the voxelized solid domain, and the no-through and no-slip boundary conditions will be enforced simultaneously. However, neither can be exactly enforced (especially for moving objects), without the iterative vortex Brinkmann penalization \cite{hejlesen2015iterative, spietz2017iterative}, which requires multiple Poisson solves in one time step and is considered expensive. Moreover, these methods rely on voxelized solvers, resulting in stair-stepped artifacts. The immersed interface method (IIM) \rev{\cite{calhoun2002cartesian, leveque1994immersed, li2001immersed, leveque1997immersed}}, which uses jump conditions across the solid interface and achieves second order accuracy, was introduced to vortex methods as vortex IIM in \cite{marichal2016immersed, gillis2018fast}. However, despite its accuracy and ability to handle cut cell geometry, IIM is not SPSD and powerful tools like multi-grid pre-conditioned conjugate gradient (MGPCG) solvers cannot be applied. To suit the use of graphics, we extend the SPSD cut cell method for velocity-pressure based methods \cite{ng2009efficient, bridson2015fluid} to vortex methods, and propose a cheap and simplified Brinkmann penalization scheme for approximating the no-slip conditions. \rev{Table~\ref{tab:vortex_related_work} outlines several hybrid and Eulerian vortex methods.}

%To suite the purposes of computer graphics, we want to enforce exactly the no-through condition, and provide users control on the extent of vortex shedding (no-slip conditions), while being sparse and SPSD so that it can be solved

%Moreover, these methods cannot handle cut cell geometry. In contrast, we want to exactly enforce the no-through boundary condition for cut cell geometry, while being able to mimic the behavior of vortex shedding induced by the no-slip boundary conditions. Given that motivation, we propose a modified Brinkmann penalization scheme for enforcing the no-slip boundary conditions. 
% \begin{table}[htbp]
% \centering\small
% \caption{Flow map based methods.}
% \begin{tabular}{>{\centering\arraybackslash}m{1.28cm}>{\centering\arraybackslash}m{3.6cm}>{\centering\arraybackslash}m{3cm}}
% \hlineB{3}
% Variable & Eulerian &  Lagrangian or E-L \\
% \hlineB{2}
% Impulse \(\bm m\) &  CF \cite{nabizadeh2022covector}, NFM \cite{deng2023neural} &  LCF \cite{li2024lagrangian}, PFM \cite{zhou2024eulerian} \\
% \hlineB{2}
% Vorticity \(\bm \omega\) & EVM \cite{wang2024eulerian}  &  \textcolor{red}{N/A $\rightarrow$ \textbf{(Ours)}} \\
% \hlineB{2}
% Velocity \(\bm u\) & BiMocq \cite{qu2019efficient}, LTSL \cite{sato2017long} & \sinan{check}\\
% \hlineB{3}
% \end{tabular}
% \vspace{5pt}
% \label{tab: relate_work_tab}
% \end{table}

\newcolumntype{z}{X}
\newcolumntype{s}{>{\hsize=.25\hsize}X}
\begin{table}
\caption{Summary of the main symbols and notations.}%\bo{This table should be put at the beginning of Sec 3. Do we need a paragraph explaining naming convention?}}
\centering
\small
\begin{tabularx}{0.47\textwidth}{scz}
\hlineB{2.5}
Notation & Type & Definition\\
\hlineB{2.5}
\hspace{12pt}$\bm{X}$ & vector & material point position at initial state\\
\hline
\hspace{12pt}$\bm{x}$ & vector & material point position at time t\\
\hline
\hspace{12pt}$t$ & scalar & time\\
\hline
\hspace{12pt}$\tau$ & scalar & dummy variable\\
\hline
\hspace{12pt}$\bm{\phi}$ & vector & forward map\\
\hline
\hspace{12pt}$\bm{\psi}$ & vector & backward map\\
% \hline
% $\bm{\phi}(X,t)$ & vector & position induced from $X$ by $\bm{\phi}$ at time $t$ \\
% \hline
% $\bm{\psi}(x,t)$ & vector & position backtraced from $x$ by $\bm{\psi}$ at time $t$\\
\hline
\hspace{12pt}$\mathcal{F}$ & matrix & forward map Jacobian\\
\hline
\hspace{12pt}$\mathcal{T}$ & matrix & backward map Jacobian\\
\hline
\hspace{12pt}$\nabla \mathcal{F}$ & matrix & forward map Hessian\\
\hline
\hspace{12pt}$\nabla \mathcal{T}$ & matrix & backward map Hessian\\
\hline
\hspace{12pt}$\bm{u}$ & vector & velocity\\
\hline
\hspace{12pt}$\bm{m}$ & vector & impulse\\
\hline
\hspace{12pt}$\bm{\omega}$ & vector & vorticity\\
\hline
\hspace{12pt}$\bm{\Psi}$ & vector & vector potential (streamfunction in 2D)\\
\hline
\hspace{12pt}$\Phi$ & scalar & harmonic function\\
\hline
\hspace{12pt}$\Gamma$ & scalar & path integral\\
\hline
\hspace{12pt}$\alpha$ & scalar & fluid fraction\\
\hline
\hspace{12pt}$\chi$ & scalar & mask\\
\hline
\hspace{12pt}$p$ & scalar & pressure\\
\hline
\hspace{12pt}$s$ & scalar & interpolation weight\\
% \hline
% $\nabla\varphi$ & vector & an arbitrary gradient between $\bm{u}$ and $\bm{m}$\\
% \hline
% $\mathcal{N}$ & SNF & neural buffer\\
% \hline
% $S$ & scalar & sizing field\\
\hline
\hspace{12pt}$n$ & scalar & number of steps between reinitializations\\
\hlineB{2.5}
\end{tabularx}
\captionsetup{aboveskip=20pt}
\label{tab:notation_table}
\end{table}

\begin{figure*}[h]
    \centering
    \includegraphics[width=1.0\textwidth]{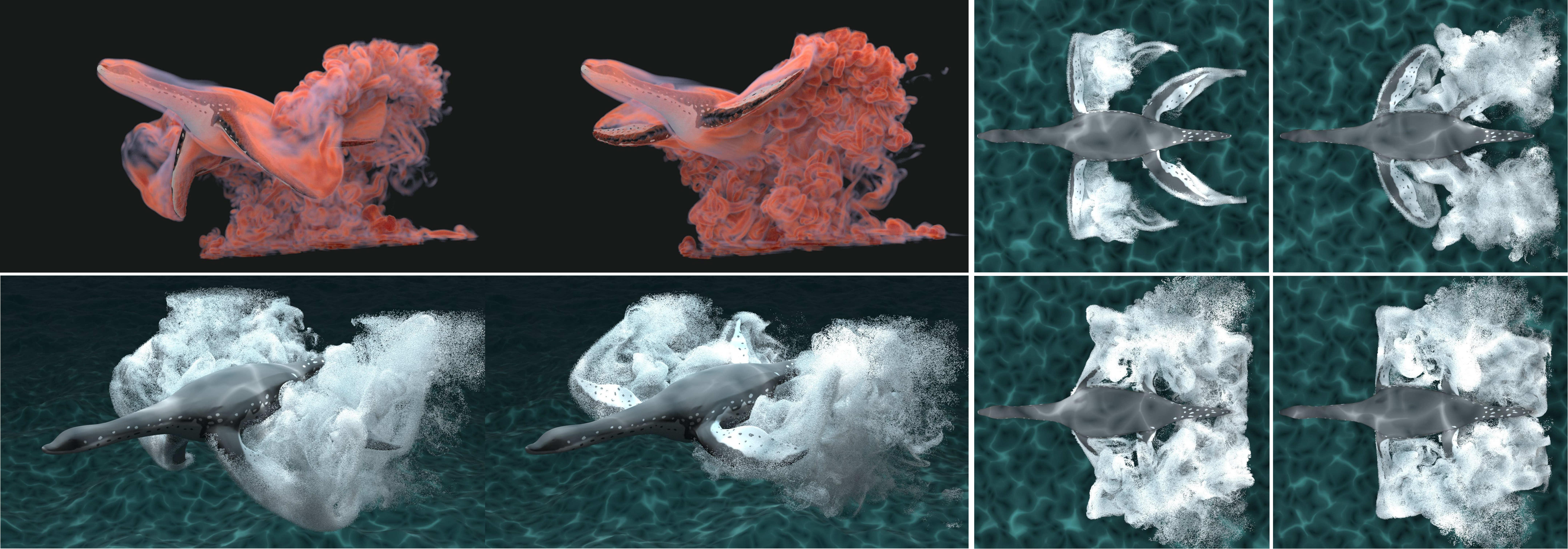}
    \caption{A plesiosaur propels through water by flapping its flippers. The top left images display the vorticity of the fluid during the plesiosaur's movement. Bubbles are generated around each of the four flippers of the plesiosaur. The bottom left images show side views of the bubble flow during the plesiosaur's movement, while the four images on the right provide a top-down view.}
    \label{fig:plesiosaur_224}
\end{figure*}
% \vspace{-0.2in}
\section{PHYSICAL MODEL}
\input{math}
\section{VORTEX PARTICLE FLOW MAP WITH HESSIAN}
%\bo{Should we introduce PFM here? I haven't seen the introduction of particle anywhere.}
In this section, we present the framework of the vortex particle flow map augmented by an evolved Hessian term. % detailing the general framework for a Eulerian-Lagrangian vortex method, as well as the implementation of the flow map Jacobians \(\mathcal{F}\) and \(\mathcal{T}\), as well as the Hessian \(\nabla \mathcal{F}\). We also demonstrate how these components are applied to vorticity advection and Particle-to-Grid transfer.

\subsection{Particle Flow Map}
Leveraging the observation that a particle trajectory inherently represents a perfect flow map, the particle flow map method was introduced in \cite{zhou2024eulerian}, enabling accurate bidirectional flow maps on particles.

Given a particle trajectory spanning from time $a$ to $c$, any intermediate time $t$ (where $t \in [a, c]$) can serve as the starting point of the particle flow map, with $c$ being the endpoint regardless of the choice of $t$. Particles carry fluid quantities such as $\bm\omega$, $\bm m$, $\nabla\bm\omega$, and $\nabla\bm m$, and the flow map quantities $\mathcal{F}$, $\mathcal{T}$, $\nabla\mathcal{F}$, and $\nabla\mathcal{T}$ at the same time. These flow map quantities evolve with the particles, mapping the fluid quantities from time \(t\) to \(c\). Unlike prior methods, where the stretching term is computed using finite differences, the particle flow map method computes it directly via flow map Jacobians, mapping it from the initial state.

\subsection{Vorticity on Flow Maps}
\rev{Vorticity} can be evolved with flow map as \cite{cortez1995impulse}:
\begin{equation}
    \label{eq:evolve_omega}
    \bm \omega(\bm x,t)=\mathcal{F}_t(\bm x)\,\bm{\omega}(\bm \psi(\bm x),0).
\end{equation}
% \begin{equation}
%     \label{eq:evolve_imp}
%     \bm \omega(\bm X,0)=\mathcal{T}^{T}_t(\bm x)\,\bm{\omega}(\bm \phi(\bm X),t),
% \end{equation}
As will be needed in the particle-to-grid (P2G) transfer (Section \ref{subsec:p2g_vort}), taking \rev{gradients} on both sides, we describe the evolution of the vorticity gradient, $\bm \nabla \bm \omega$, using the flow map as follows: 
\begin{equation}
    \label{eq:evolve_grad_omega}
    \bm \nabla \bm \omega(\bm x,t) = \mathcal{F}_t\,\bm \nabla_{\bm \psi} \bm{\omega}(\bm \psi(\bm x),0)\,\mathcal{T}_t + \bm \nabla \mathcal{F}_t\,\bm{\omega}(\bm \psi(\bm x),0).
\end{equation}
For Eq. \eqref{eq:evolve_grad_omega}, the second term on the right-hand side involves the product of a \(3 \times 3 \times 3\) tensor and a \(3 \times 1\) vector. To eliminate any ambiguity, we expand it explicitly using component (index) notation. For all subsequent calculations, we use \(k\) as the summation index. Thus, we have \(
    \label{eq:expand_hessian}
    \bigl((\bm \nabla \mathcal{F}_t)\,\bm{\omega}(\bm \psi(\bm x),0)\bigl)_{il} \,=\, \sum_k (\bm \nabla \mathcal{F}_t)_{ikl}\,\bm{\omega}_{k}(\bm \psi(\bm x),0).\)

\begin{figure*}[h]
    \centering
    \includegraphics[width=1.0\textwidth]{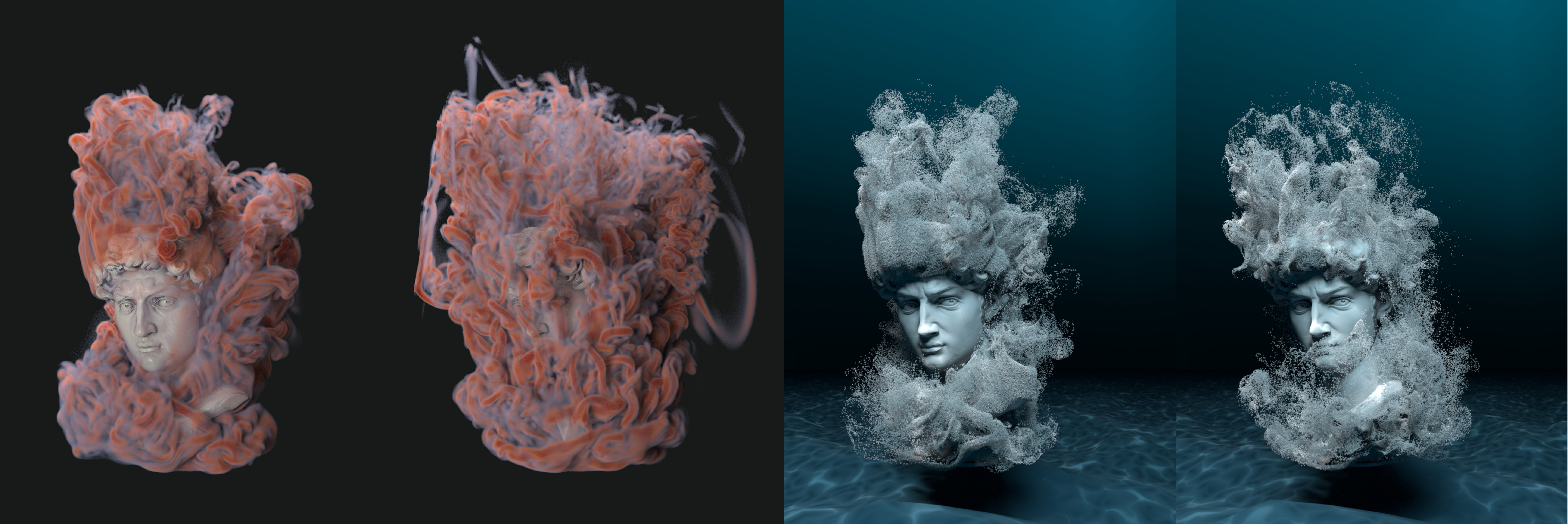}
    \caption{Fluid flows upward from directly beneath the head of Michelangelo's David. The two images on the left illustrate the vorticity as the fluid moves past the sculpture. Bubbles are generated around the base of the sculpture and at the base of the hair, and the two images on the right depict the flow of these bubbles past the sculpture.}
    \label{fig:david}
\end{figure*}

\subsection{Hessian Evolution}
% In traditional impulse-based particle flow map methods \cite{zhou2024eulerian, sancho2024impulse}, this variable was computed via direct interpolation among neighboring particles, an approach rooted in Smoothed Particle Hydrodynamics (SPH) \cite{ihmsen2014sph, morris1997modeling, monaghan2005smoothed}. Specifically,
% \begin{equation}
%     A_i = \sum_j A_j W_{ij},
% \end{equation}
% where \(A_i\) is any quantity carried by particle \(i\), and \(W_{ij}\) is a kernel function. Taking the gradient, we obtain the original formulation of the spatial derivative \(\nabla A_i = \sum_j A_j \nabla W_{ij}\), which is improved and modified by \cite{morris1997modeling, monaghan2005smoothed}:
% \begin{equation}
%     \nabla A_i = \sum_j (A_i + A_j) \nabla W_{ij},
% \end{equation}
% More details can be found in \cite{ihmsen2014sph, morris1997modeling, monaghan2005smoothed}.

% However, previous particle flow map approaches \cite{zhou2024eulerian, sancho2024impulse} overlooked the fact that \textit{the Hessian, \(\nabla \mathcal{F}\) and \(\nabla \mathcal{T}\), can also be evolved explicitly by the velocity field and the Jacobians of the flow map}, just like how \(\mathcal{T}\) and \(\mathcal{F}\) are evolved. Although the corresponding evolution equation is a bit more complex, it ultimately reduces to efficient matrix multiplications, avoiding neighbor queries, or extra temporary particles required by prior methods.
Now we present our novel method for solving \(\nabla \mathcal{F}_t\). Instead of trying to compute/interpolate the Hessian term using the evolved Jacobian via neighboring particles or temporarily sampled particles \cite{sancho2024impulse, zhou2024eulerian}, we directly evolve it \rev{(derivation provided in the supplementary material)}, similar to how \(\mathcal{T}\) and \(\mathcal{F}\) are evolved by particle flow maps in Eq.~\eqref{eq:evole_FT}:
\begin{multline}
\label{eq:gradF_evolution}
\frac{D \left(\frac{\partial \mathcal{F}_{ij}(\bm{x}, t)}{\partial x_l}\right)}{D t}
= \Bigg[\frac{D (\nabla \mathcal{F})}{D t}\Bigg]_{ijl} \\
% = -\frac{\partial u_k(\bm{x}, t)}{\partial x_l} 
%     \frac{\partial \mathcal{F}_{ij}(\bm{x}, t)}{\partial x_k} 
%   + \frac{\partial u_i(\bm{x}, t)}{\partial x_k} 
%     \frac{\partial \mathcal{F}_{kj}(\bm{x}, t)}{\partial x_l} \\
%   + \frac{\partial^2 u_i(\bm{x}, t)}{\partial x_l \partial x_k} 
%     \mathcal{F}_{kj}(\bm{x}, t) \\
= -(\nabla \mathcal{F})_{ijk} (\nabla \bm{u})_{kl} 
  + (\nabla \bm{u})_{ik} (\nabla \mathcal{F})_{kjl} 
  + (\nabla \nabla \bm{u})_{ilk} \mathcal{F}_{kj}.
\end{multline}

\rev{By "compute/interpolate," we refer to the process of interpolating a quantity using spatially sampled particles and an interpolation kernel, with its gradient computed by differentiating the interpolation kernel, as seen in methods like APIC \cite{jiang2015affine} and SPH \cite{ihmsen2014sph}. Our method naturally circumvents issues arising from “unstructured particles.” In practice, finite differences should be computed on structured fields (e.g., uniform grids), not on unstructured, randomly distributed particles, which can lead to instability and asymmetry.} We demonstrate the effectiveness of our approach through a 3D trefoil knot comparison, detailed in Section \ref{para:trefoil} and \figref{fig:dTdx}, where our evolved Hessian \rev{exhibits} smoother vortex tubes and rings, with stabler simulation, compared to the Hessian obtained through interpolation. Additionally, we show that our evolved Hessian almost \textbf{doubles} the stable simulation time under challenging long flow maps in Section \ref{subsubsec:compare_same_long_flow_map}.

% \begin{figure}
%     \centering
%     \includegraphics[width=0.3\textwidth]{illustration/adaptiveFlowMap_grid_1.pdf}
%     \caption{VPFM illustration. Blue circles represent vortex particles, yellow circles represent vorticity, green circles represent velocity. The black curve depicts the trajectory of a vortex particle over time. Every \(n\) steps, vorticity is reinitialized through a G2P process (illustrated at the bottom left), while a P2G process transfers information from particles to \rev{the grid} at the current time step (shown at the top right).}
%     \label{fig:flowmap_illustration}
% \end{figure}

\subsection{Discrete Model for Vorticity Transport}
We now present our discrete model for solving the evolution equations for the vorticity, its gradient and the flow map quantities. The ultimate goal of this subsection is to obtain the current advected vorticity on grid. In this framework, vortex particles carry the vorticity \(\bm \omega\), its gradient \(\nabla \bm \omega\), the flow map Jacobians \(\mathcal{T, F}\), and the flow map Hessian \(\nabla \mathcal{F}\). The key steps are summarized as follows:
%The vorticity advection and stretching are solved on particles, while the velocity reconstruction and viscosity are solved on grids. 
\begin{enumerate}[leftmargin=*, labelindent=0pt]
    \item \textit{Reinitialization.} Vortex particles are redistributed uniformly every \(n\) steps, and quantities are transferred from the grid onto the particles (G2P). % over the computational domain to maintain a well-resolved distribution

    \item \textit{Advection.} Vortex particles are advected, with the carried quantities updated by the evolution equations.

    \item \textit{Particle-to-Grid (P2G) Transfer.} Vorticity \(\bm \omega\) carried by particles is interpolated back onto the grid, aided by its gradient \(\nabla \bm \omega\).

    % \item \textit{Viscosity and External Forces} \\
    % The vorticity is updated to account for viscosity and external forces (detailed in Section \ref{sec:visc_ext_force}).

    % \item \textit{Velocity Reconstruction} \\
    % The velocity is reconstructed from the vorticity on the grid, enforcing the no-through solid boundary conditions (detailed in Section \ref{subsec:cutcell_nothrough}).

    % \item \textit{No-slip Boundary Condition} \\
    % The no-slip conditions are approximated by a simplified Brinkmann penalization model (detailed in Section \ref{subsec:noslip}).
\end{enumerate}

\begin{figure}[t]
    \centering
    \begin{subfigure}[t]{0.49\linewidth}
        \centering
        \includegraphics[width=\linewidth]{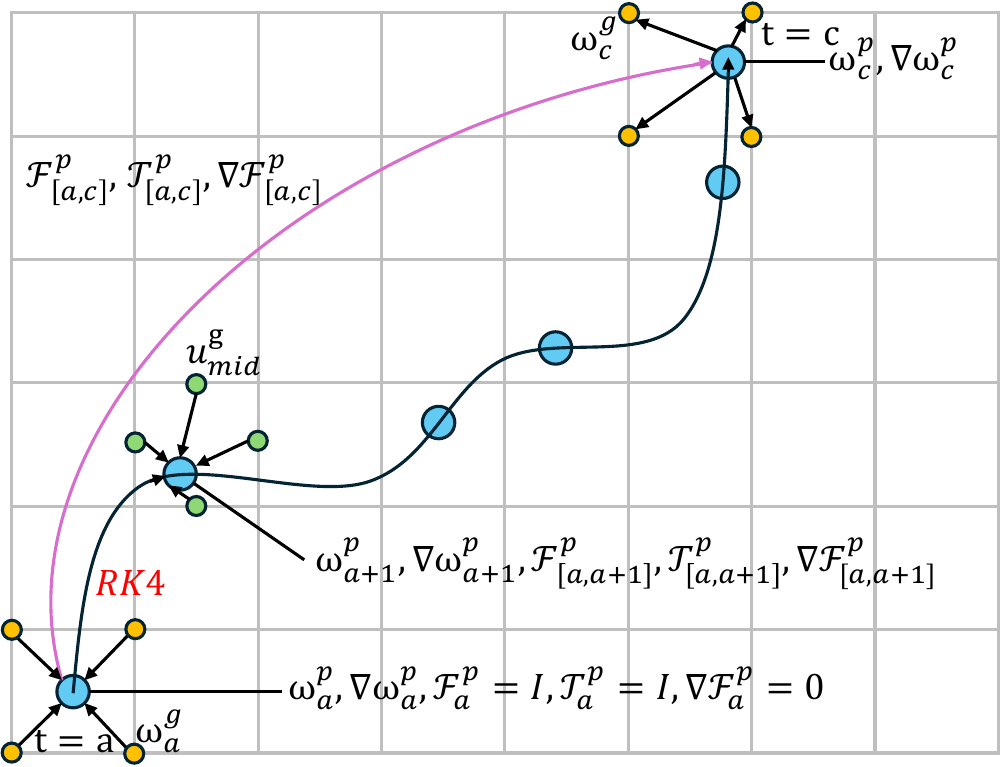}
        \caption{\rev{VPFM illustration. Blue: vortex particles; Yellow: vorticity; Green: velocity. Black curve: trajectory of a vortex particle over time. Every \(n\) steps, vorticity is reinitialized via G2P (bottom left); P2G is applied at the current time step (top right).}}
        \label{fig:flowmap_illustration}
    \end{subfigure}
    \hfill
    \begin{subfigure}[t]{0.49\linewidth}
        \centering
        \includegraphics[width=\linewidth]{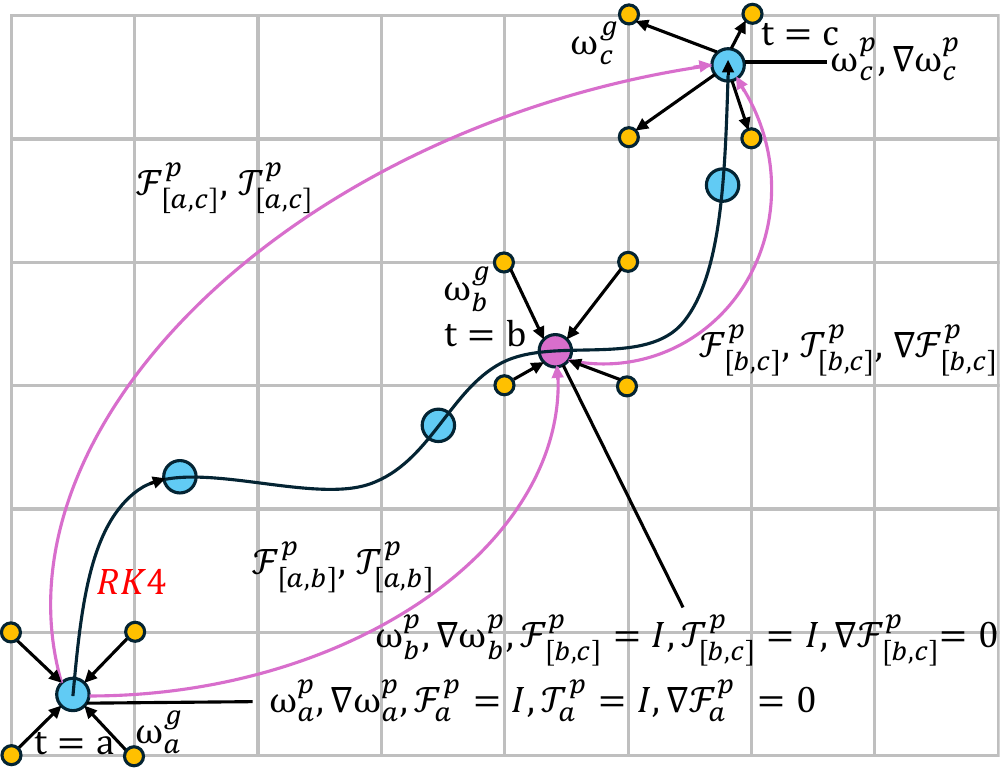}
        \caption{\rev{Adaptive flow map illustration. The full trajectory (left purple curve), with a total length denoted as \(n^L\), is divided into two connected segments, separated by the purple circle. The shorter segment (upper right) is denoted \(n^S\).}}
        \label{fig:adaptive_flowmap}
    \end{subfigure}
    \caption{\rev{Illustration of VPFM (left) and the adaptive flow map (right).}}
    \label{fig:flowmap_combined}
\end{figure}

We will first explain step (1), followed by step (2) and (3).
\paragraph{Reinitialization with G2P Transfer}
\rev{The set of all simulation steps is divided into consecutive segments} of length \(n\), where \(n\) denotes the flow map length. At the start of each simulation segment, we perform reinitialization by: \textbf{1}. uniformly redistributing vortex particles across the entire computational domain,
\textbf{2}. transferring quantities from the grid to the particles through a Grid-to-Particle (G2P) step, and  
%We adopt the approach of \cite{jiang2015affine} and \cite{zhou2024eulerian} for a fair comparison.
%The same weighting kernels are used as well, with details provided in Appendix \ref{subsec:interp_kernel}.
\textbf{3}. reinitializing flow map quantities.
The G2P step transfers vorticity and its gradient from grid nodes to vortex particles, denoted by the yellow circles (grid nodes) and blue circles (vortex particles) in \figref{fig:flowmap_illustration}. Specifically, a particle's vorticity is given by \cite{zhou2024eulerian, jiang2015affine}:
\begin{equation}
\label{eq:g2p_vort}
    \bm{\omega}^p \; \gets \; \sum_{i} s^{ip}\,\bm{\omega}^{g,i},
\end{equation}
where \( \bm{\omega}^p \), \( \bm{\omega}^{g,i} \) are the vorticity at particle \(p\) and grid node \(i\). \( s^{ip} \) is the interpolation weight between grid node \(i\) and particle \(p\). \rev{We use a quadratic B-spline kernel \cite{steffen2008analysis} (see supplementary).}
%(Detailed in Supplementary  \ref{subsec:interp_kernel}).

The gradient of vorticity at the particle is computed as:
\begin{equation}
\label{eq:g2p_vort_grad}
    \nabla \bm{\omega}^p \;\gets\; \sum_{i} \nabla s^{ip}\,\bm{\omega}^{g,i}.
\end{equation}
Here, \( \nabla s^{ip} \) represents the gradient of the interpolation weight.

When reinitializing flow map quantities, the flow map Jacobians are reinitialized as the identity matrix for every particle \(p\), i.e., \(\mathcal{F}^p = \bm{I}\) and \(\mathcal{T}^p = \bm{I}\). The flow map Hessian for each particle \(p\) is reinitialized to zero, i.e., \(\nabla \mathcal{F}^p = 0\).

\begin{figure*}[h]
    \centering
    \includegraphics[width=1.0\textwidth]{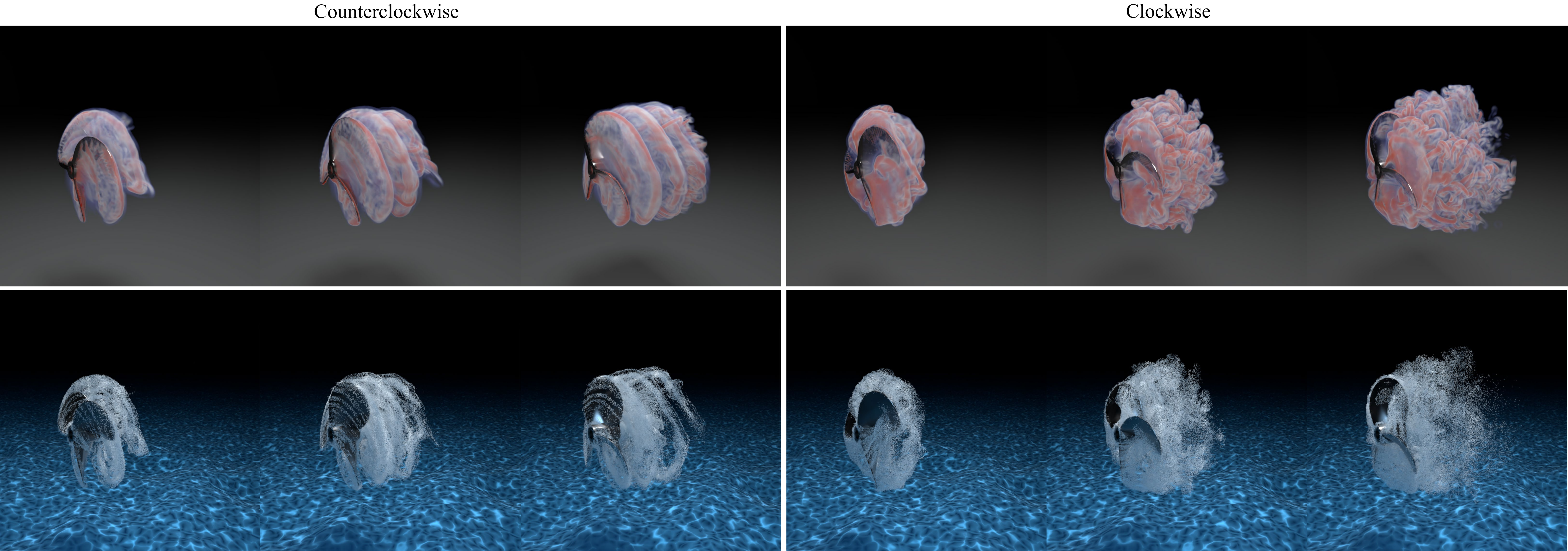}
    \caption{The propeller rotates, with the inflow passing from left to right. The images on the left depict the propeller rotating counterclockwise, while those on the right show clockwise rotation. The upper images on both sides illustrate the fluid vorticity during the propeller's motion, and the lower images display the bubbles generated by the propeller's rotation. Notably, a spiral vortex is formed during counterclockwise rotation, whereas the clockwise rotation generates turbulence phenomenon.}
    \label{fig:propeller_26}
\end{figure*}

\paragraph{Advection}
% \begin{wrapfigure}{r}{0.25\textwidth}
% \centering
% \includegraphics[width=0.25\textwidth]{illustration/adaptiveFlowMap_grid_1.pdf}
% \captionsetup{aboveskip=4pt}
% \caption{VPFM illustration.}
% \label{fig:flowmap_illustration}
% %\vspace{-0.1in}
% \end{wrapfigure}

The vortex particle's trajectory \(\gamma^p\) itself serves as a flow map. Following the notation from \cite{zhou2024eulerian}, let \(a\) be the initial time step, \(c\) the current time step, and \(b\) an intermediate time step. We define a vortex particle's trajectory from \(a\) to \(c\) as
\begin{equation}
    \gamma^p_{a \rightarrow c} 
    \;=\; \bigl[\bm x^p (a), \,\dots, \,\bm x^p (b), \,\dots, \,\bm x^p (c)\bigr],
\end{equation}
where \(\bm x^p (t)\) denotes the position of particle \(p\) at time step \(t\). As shown in \figref{fig:flowmap_illustration}, during a time step, for instance from \(a\) to \(a+1\), an RK4 scheme is used for the numerical integration of the particle positions \(\xp{}\), the flow map Jacobians \(\T{}^p\), \(\F{}^p\) using Eq.\eqref{eq:evole_FT} , and the Hessian \(\gradF{}^p\) using Eq.\eqref{eq:gradF_evolution}. The velocities used in RK4 are interpolated from the grid, denoted by the green circles in \figref{fig:flowmap_illustration}. By numerically integrating from \(a\) to \(c\), we can get the current vorticity on particles:
\begin{equation}
    \currvortp{} = \Fac{} \,\initvortp{},
    \label{eq:omega_evolve_adaptive}
\end{equation}
where \(\Fac{}\) is the forward flow map Jacobian corresponding to the trajectory \(\gammapac{}\), \(\currvortp{}\) and \(\initvortp{}\) are the current and initial vorticity on particles.
To evolve the vorticity gradient, we have:
\begin{equation}
    \nabla \bm \omega_c^p = \Fac{} \,\nabla\bm\omega_a^p \,\Tac{} \;+\; \gradFac{}\,\bm{\omega}_a^p,
    \label{eq:gradomega_evolve_noadaptive}
\end{equation}
where \(\nabla \currvortp{}\) denotes the current vorticity gradient on the particles, and \(\Fac{}\), \(\Tac{}\), \(\gradFac{}\) are the forward Jacobian, backward Jacobian, and forward Hessian corresponding to the trajectory \(\gammapac{}\).

Having obtained the current vortivity \(\currvortp{}\) and its gradient \(\nabla \currvortp{}\) on particles, we are now ready for the Particle-to-Grid (P2G) transfer.

\paragraph{P2G Transfer}
\label{subsec:p2g_vort}
The P2G step computes the current vorticity on \rev{grid cells} by interpolating the particle values back onto the Eulerian grid \cite{jiang2015affine}:
\begin{equation}
\label{eq:vort_P2G}
    \bm{\omega}^{g,i} = \displaystyle \sum_{p} s^{ip} \Bigl(\bm{\omega}^p \;+\; \nabla\bm{\omega}^p \cdot \bigl(\bm{x}^{g,i} - \bm{x}^{p}\bigr)\Bigr)\,/\,\displaystyle \sum_{p} s^{ip} \,,
\end{equation}
where \( \bm{\omega}^p,\nabla \bm{\omega}^p \) are the vorticity and its gradient at particle \(p\).
    %\item \( \nabla \bm{\omega}^p \) is the gradient of vorticity at particle \(p\).
\( \bm{x}^{g, i} \) and \(\bm{x}^{p}\) are the positions of grid node \(i\) and particle \(p\).
    %\item \( s^{ip} \) is again the interpolation weight from particle \(p\) to grid node \(i\).
%The numerical benefits of this scheme are demonstrated by APIC \cite{jiang2015affine}.

\paragraph{\rev{Interpolation Kernel Near Boundaries}}
\rev{Near solid boundaries, the kernel remains unchanged, and the BC for the Jacobian or Hessian translates to the BC for velocities during the RK4. We choose the velocity inside the solid to be the solid velocity, following NFM and PFM. 
One may extrapolate those velocities for inviscid fluids.}

\begin{figure*}[t]
    \centering
    \includegraphics[width=1.0\textwidth]{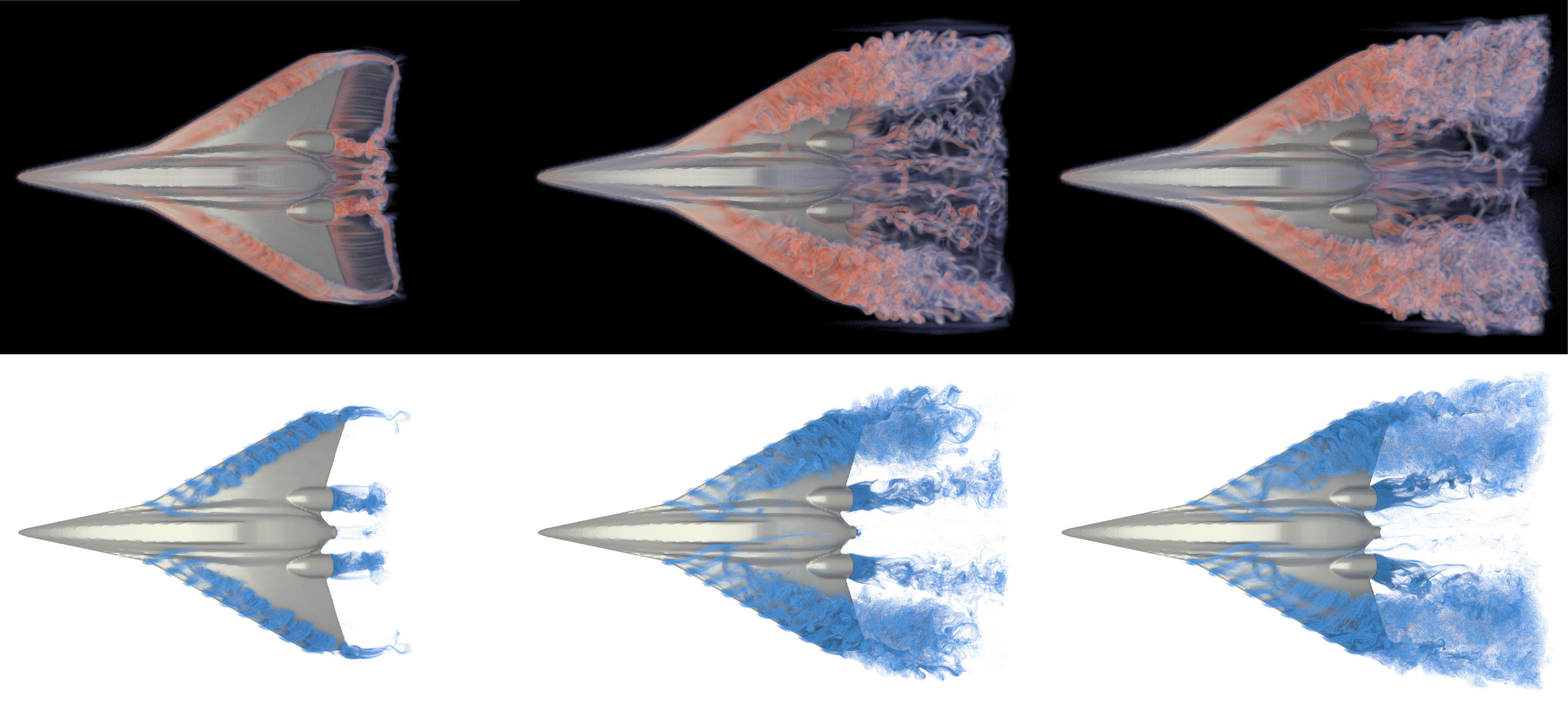}
    \caption{Aircraft and vortex lift. The upper images illustrate the fluid vorticity around the aircraft during flight, from where we observe "vortex lift" \cite{anderson2010aircraft}, and our result is similar to the experimental result given in \cite{delery2001robert}. The lower images show the movement of smoke \rev{particles} generated at the aircraft’s wings, tail nozzle, and tail fins throughout its flight path.}
    \label{fig:airplane}
\end{figure*}

\subsection{Adaptive Flow Map for Different Quantities}
% \begin{wrapfigure}{r}{0.25\textwidth}
% \centering
% \includegraphics[width=0.25\textwidth]{illustration/adaptiveFlowMap_grid_2.pdf}
% \captionsetup{aboveskip=4pt}
% \caption{Adaptive flow map.}
% \label{fig:adaptive_flowmap}
% %\vspace{-0.1in}
% \end{wrapfigure}

% \begin{wrapfigure}{r}{0.25\textwidth}
% \centering
% \includegraphics[width=0.25\textwidth]{illustration/adaptive_flow_map.pdf}
% % \captionsetup{aboveskip=8pt}
% \caption{Adaptive flow map length illustration.}
% \label{fig:adaptive_flowmap}
% %\vspace{-0.1in}
% \end{wrapfigure}
% As discussed in \cite{zhou2024eulerian},
As shown in an ablation study in Section \ref{para:longshort_ablation}, and also discussed in \cite{zhou2024eulerian}, higher-order quantities such as \(\nabla \bm{\omega}\) require a shorter flow map length for advection compared to lower-order quantities like \(\bm{\omega}\). \rev{This is because higher-order terms involve additional differentiation, which amplifies numerical errors and may lead to instability when advected over long flow maps.} In this section, we detail how this requirement is met within the VPFM framework.

Specifically, we maintain two distinct segments of the flow map Jacobians: one corresponding to the full trajectory \(\gamma^p_{a \rightarrow c}\) and another to the shorter segment \(\gamma^p_{b \rightarrow c}\). An illustration is shown in Figure~\ref{fig:adaptive_flowmap}. When advecting the vorticity gradient \(\nabla \bm \omega\), we use the flow map Jacobian and Hessian associated with the shorter trajectory \(\gamma^p_{b \rightarrow c}\):
\begin{equation}
    \nabla \bm \omega_c^p = \Fbc{} \,\nabla\bm\omega_b^p \,\Tbc{} \;+\; \gradFbc{}\,\bm{\omega}_b^p,
    \label{eq:gradomega_evolve_adaptive}
\end{equation}
where \(\nabla \currvortp{}\) denotes the current vorticity gradient on the particles, and \(\Fbc{}\), \(\Tbc{}\), \(\gradFbc{}\) are the forward Jacobian, backward Jacobian, and forward Hessian from time \(b\) to time \(c\). 

For the vorticity \(\bm \omega\), however, we use the Jacobian corresponding to the longer trajectory \(\gamma^p_{a \rightarrow c}\), which preserves the vorticity with less numerical dissipation, as given in Eq.~\eqref{eq:omega_evolve_adaptive}.

A natural implementation for maintaining both \(\Fac{}\) and \(\Fbc{}\) relies on Jacobian connection (derivation provided in the supplementary material):
%(derivation given in Appendix \ref{subsec:Jacobian_connect}):
\begin{equation}
    \label{eq:connect_Jacobian}
    \begin{dcases}
    \mathcal{F}_{[a, c]}^p = \mathcal{F}_{[b, c]}^p \mathcal{F}_{[a, b]}^p, \\
    \mathcal{T}_{[a, c]}^p = \mathcal{T}_{[a, b]}^p \Tbc{}.
    \end{dcases}
\end{equation}

% We also note that the Hessian does not require “connection,” since only the Hessian for the shorter trajectory \(\gamma^p_{b \rightarrow c}\) is needed. 
\input{sec5_solid_bc}
\input{sec6_visc_extforce}
\input{sec7_time_int}

\section{VALIDATION}
\label{sec:validation}
In this section, we first validate that VPFM achieves a robust, long-term flow map using the evolved Hessian. This is followed by the validation of our cut cell system for enforcing no-through boundary conditions and the simplified Brinkmann penalization for approximating no-slip boundary conditions. Experimental details are given in the supplementary material.
\input{flowmap_compare_failurepoint}

% \begin{figure}
% \includegraphics[width=0.48\textwidth]{cutcell/cutcell_14.pdf}
% \caption{Comparison of vortex rings induced by the ball between cut-cell and voxelized results at different resolutions. The stair-stepping artifacts, which are prominent in voxelized results, are largely mitigated by our cut-cell method.}
% \label{fig:cutcell}
% % \vspace{-0.2in}
% \end{figure}

% \begin{figure}
% \includegraphics[width=0.48\textwidth]{cutcell/13_cutcell_gt_compare.pdf}
% \caption{\sinan{Pending} Comparison with a grid-aligned ground truth.}
% \label{fig:cutcell_gt_compare}
% % \vspace{-0.2in}
% \end{figure}

\subsection{No-Through Boundary Condition}
\label{subsec:cutcell_validate}
\paragraph{Vortex Ring Passes by a Static Ball (3D)}
With a resolution of \(128 \times 128 \times 128\), a static ball is encountering an inflow \rev{with \(velocity = 0.1\)}  from the left in Figure~\ref{fig:vortex_ring_ball}. We verify the effectiveness of our cut cell method by comparing the vortex rings passing by the ball. With our cut cell method, the vortex rings become more circular, while the voxelized results appear blocky and angular.

\begin{figure}
    \centering
    \includegraphics[width=0.48\textwidth]{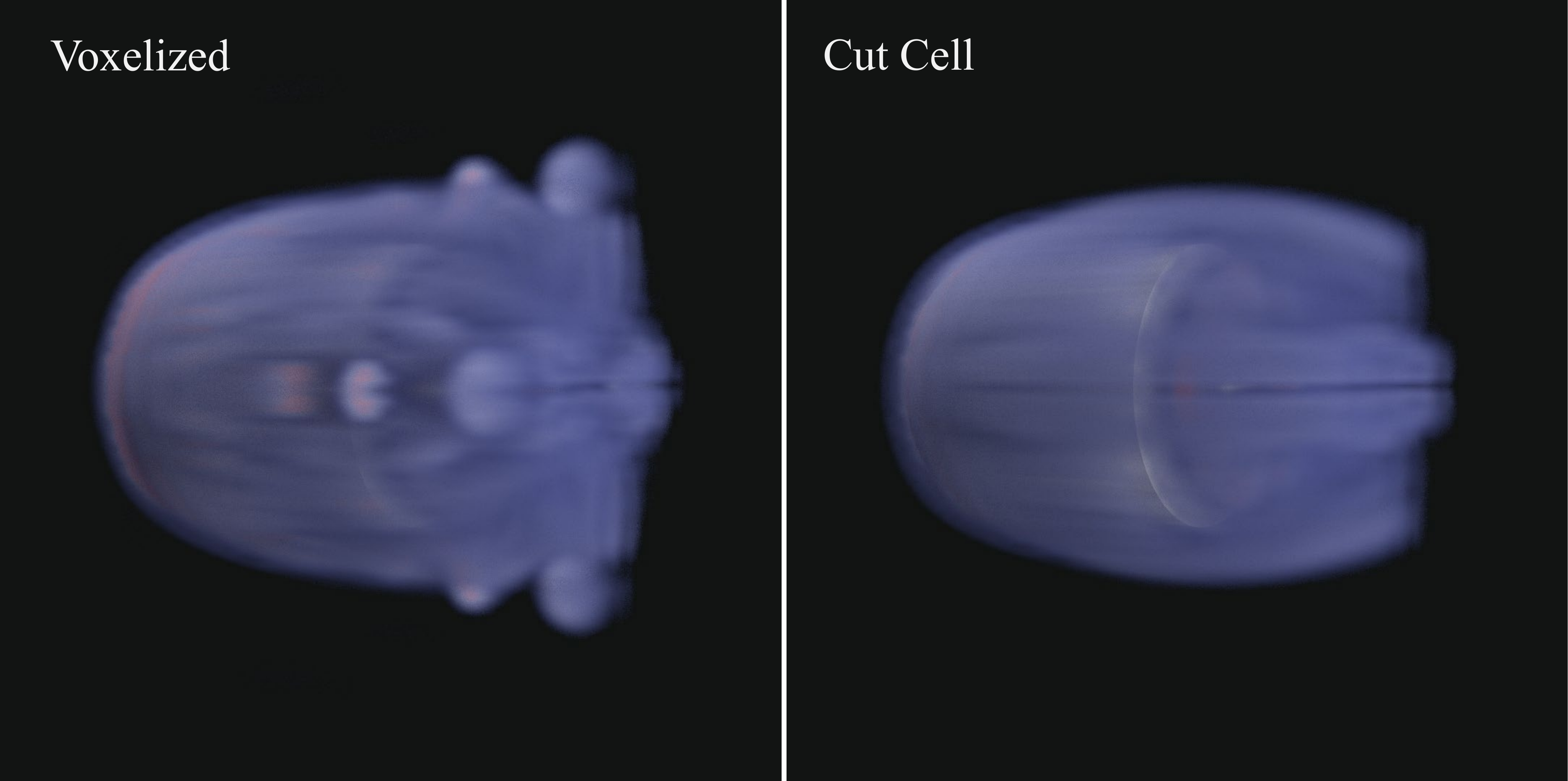}
    \caption{Flow around a cylinder. A stair-stepped vortex shedding pattern is observed with the voxelized solver, while the flow passes around the cylinder smoothly when our cut cell method is used.}
    \label{fig:cylinder_cutcell}
\end{figure}

\paragraph{Flow around a Static Cylinder (3D)}
With a resolution of \(64 \times 64 \times 64\), a cylinder encounters an inflow \rev{with \(velocity = 0.1\)} from the left, as shown in Figure~\ref{fig:cylinder_cutcell}. With our cut cell method, the flow passes smoothly around the cylinder, whereas the voxelized results exhibit stair-stepped shedding vortices.

\subsection{No-Slip Boundary Condition}
\label{subsec:validate_brinkmann}

\begin{figure}
\includegraphics[width=0.48\textwidth]{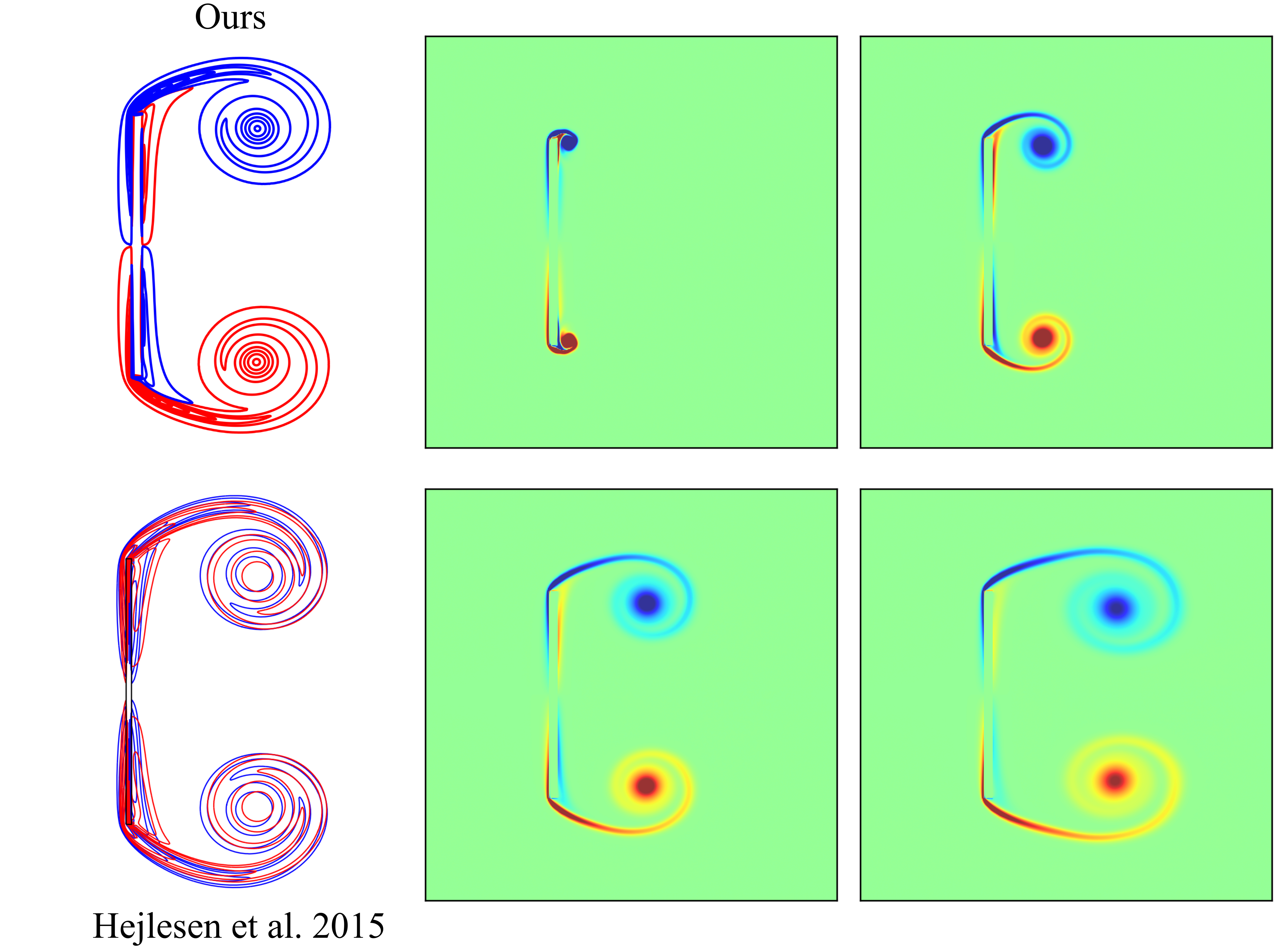}
\caption{At a relatively low Reynolds number Re = 1000, a flow passes around a thin plate. The vorticity contour on the left compares our results with the iterative Brinkmann penalization method \cite{hejlesen2015iterative}, showing high similarity between the two. On the right, four additional \rev{time frames of} vorticity visualizations illustrate the entire process.}
\label{fig:thinplate}
% \vspace{-0.2in}
\end{figure}

\paragraph{Flow around a Thin Plate (2D)}
As depicted in \figref{fig:thinplate}, at \(Re = 1000\), a thin plate with a thickness of \(1/50\) (within a domain of size \(1 \times 1\)) experiences inflow with \(velocity = 0.1\) from the left, leading to the generation of vortices at both ends of the plate. The left column of \figref{fig:thinplate} compares the vorticity contours produced by our method with that from the iterative Brinkmann penalization approach \cite{hejlesen2015iterative}, while the right two columns display four frames of the vorticity visualization. This experiment demonstrates that our simplified Brinkmann penalization scheme achieves results comparable to those of the iterative Brinkmann penalization approach, validating the accuracy of our simplified no-slip boundary conditions at low Reynolds numbers.

\begin{figure}[h]
\includegraphics[width=0.48\textwidth]{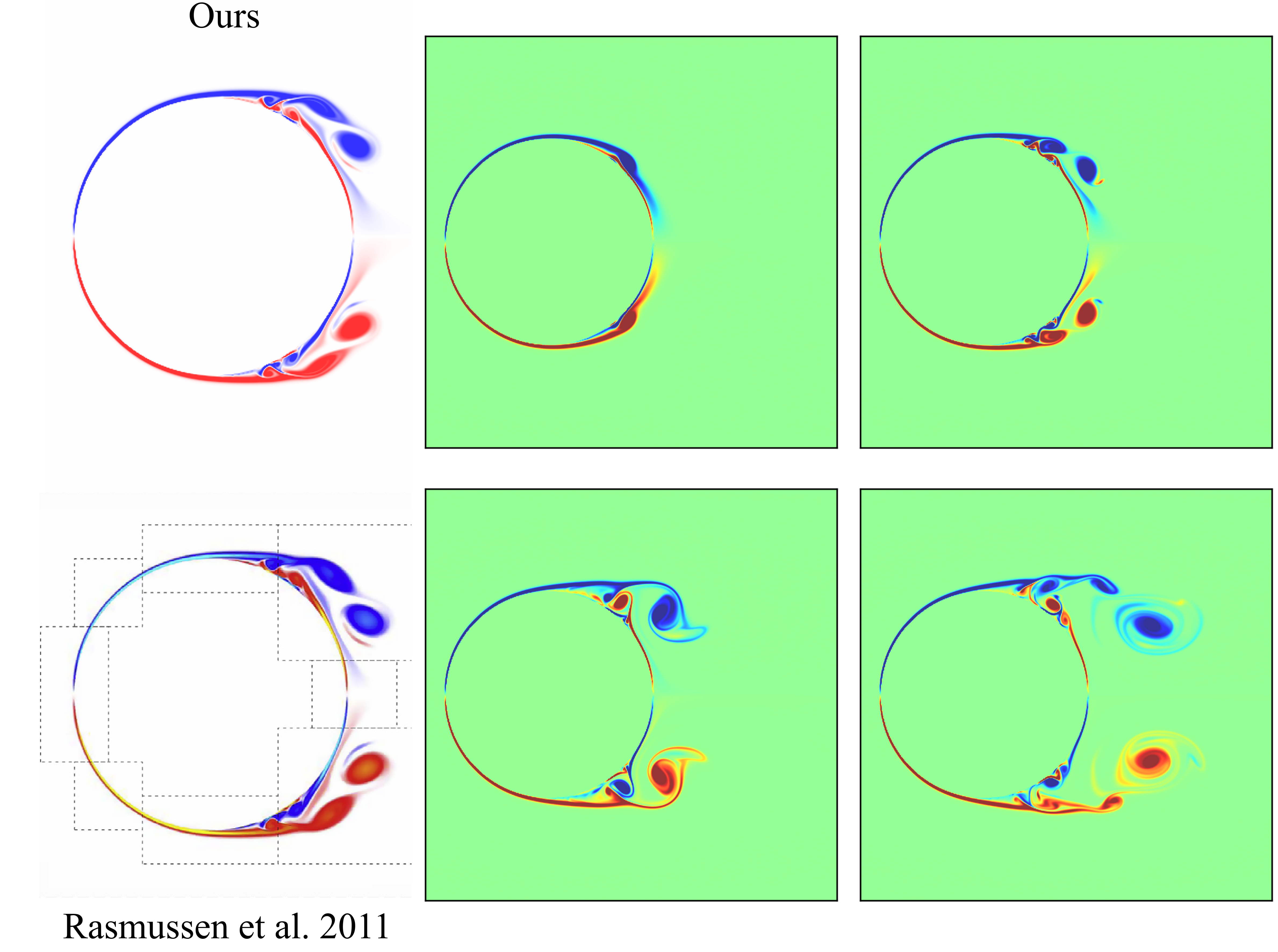}
\caption{At a high Reynolds number Re = 9500, a flow passes around a disk. We compare our results with the multi-resolution Brinkmann penalization Vortex-In-Cell method \cite{rasmussen2011multiresolution} on the left, observing a high level of consistency between the two. \rev{On the right, four additional time frames of vorticity visualizations illustrate the entire process.}}
\label{fig:disk}
% \vspace{-0.2in}
\end{figure}

\paragraph{Flow around a Disk (2D)} 
As illustrated in \figref{fig:disk}, at \(Re = 9500\), a disk \rev{with radius 0.25} subjected to an inflow from the left generates a vortex-shedding pattern. The vorticity visualization (left column of \figref{fig:disk}) aligns closely with the results from a multi-resolution Brinkmann penalization VIC method \cite{rasmussen2011multiresolution}, validating the effectiveness of our simplified no-slip boundary conditions at high Reynolds numbers. \rev{Four additional vorticity frames are shown in the right two columns.}

\begin{figure}[h]
    \centering
    \includegraphics[width=0.5\textwidth]{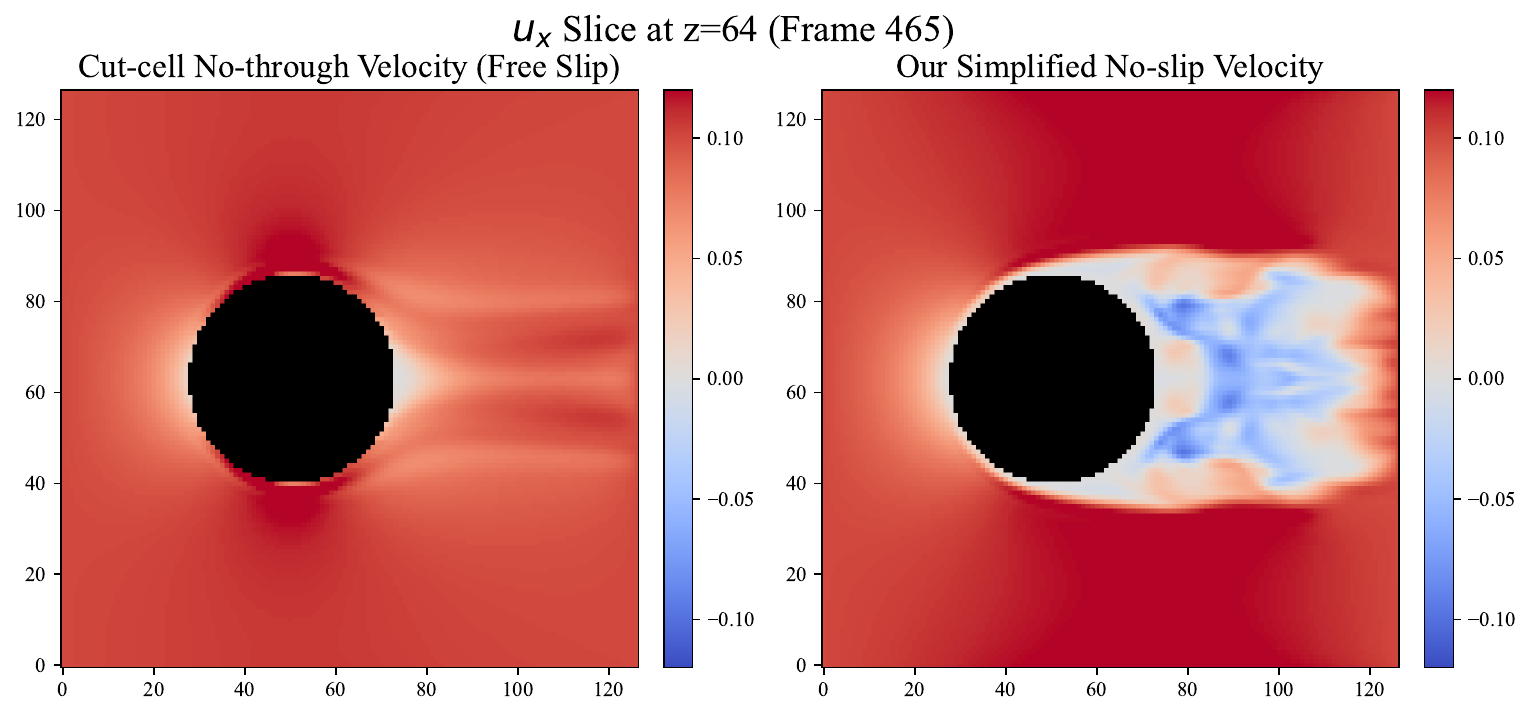}
    \caption{A static ball is subjected to an inflow from the left with velocity being 0.1. The x-component velocity slice at \(z=64\) is compared with and without our simplified Brinkmann penalization for enforcing no-slip conditions. With our simplified Brinkmann penalization enabled, the velocity around the static ball becomes zero, whereas under free-slip conditions, the velocity near the ball remains nonzero.}
    \vspace{-0.1in}
    \label{fig:no_slip_velocity}
\end{figure}

\paragraph{Flow around a Static Ball (3D)}
A static ball is encountering an inflow from left with \(velocity = 0.1\). We verify the effectiveness of our simplified Brinkmann penalization for no-slip conditions by comparing the a slice of the velocity at the same frame in \figref{fig:no_slip_velocity}. With our no-slip condition, the velocity around the ball is fully canceled by the shedded vortices and becomes zero as expected, by setting \(\lambda = \frac{1}{\delta t}\).

\subsection{Viscosity}
\label{subsec:validate_visc}
\paragraph{Lid-driven Cavity Flow (2D)}
\begin{figure*}
    \centering
    \includegraphics[width=1.04\textwidth]{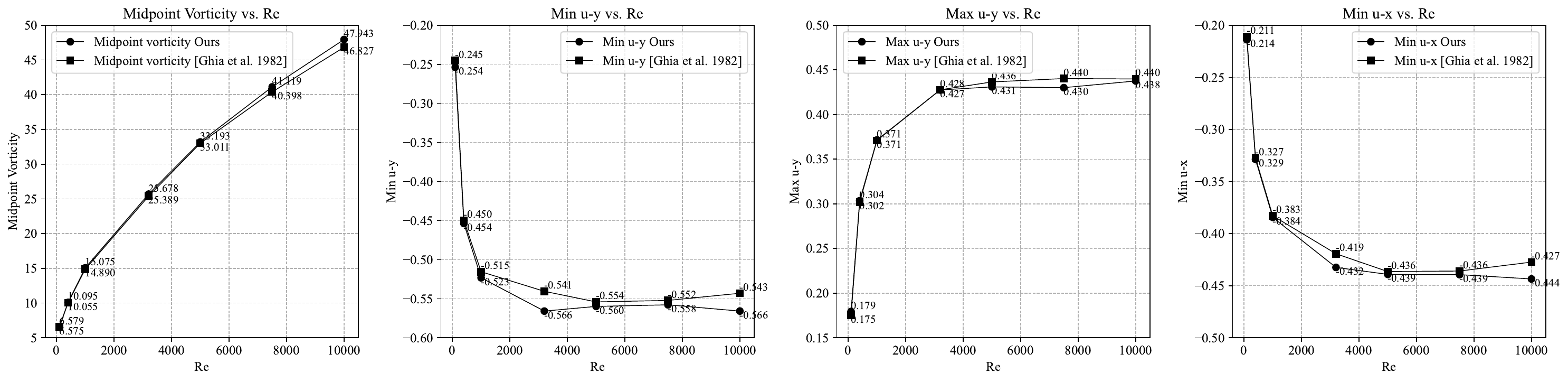}
    \caption{Quantitative comparison for lid-driven cavity flow between ours and \citet{ghia1982high}. The four figures (from left to right) shows the comparison of the midpoint vorticity on the top boundary, the minimum and maximum y-velocity along the horizontal centerline of the cavity, and the minimum x-velocity along the vertical centerline. High consistency is observed between our results and \citet{ghia1982high}.}
    \vspace{-0.1in}
    \label{fig:cavity_vort_vel_compare}
\end{figure*}

% \begin{figure}
% % \hspace{-0.28\textwidth}
% \includegraphics[width=0.48\textwidth]{velocity_field_plot/middle_vorticity_final.pdf}
% \caption{Comparison of the midpoint vorticity on the moving boundary with \cite{ghia1982high} under different Re.}
% \label{fig:midomega}
% % \vspace{-0.2in}
% \end{figure}

% \begin{figure}
% % \hspace{-0.28\textwidth}
% \includegraphics[width=0.48\textwidth]{velocity_field_plot/min_v_final.pdf}
% \caption{Comparison of the minimum y-component of velocity along the horizontal line through the center of the cavity with \cite{ghia1982high} under different Re.}
% \label{fig:minv}
% % \vspace{-0.2in}
% \end{figure}

% \begin{figure}
% % \hspace{-0.28\textwidth}
% \includegraphics[width=0.48\textwidth]{velocity_field_plot/max_v_final.pdf}
% \caption{Comparison of the maximum y-component of velocity along the horizontal line through the center of the cavity with \cite{ghia1982high} under different Re.}
% \label{fig:maxv}
% % \vspace{-0.2in}
% \end{figure}

% \begin{figure}
% % \hspace{-0.28\textwidth}
% \includegraphics[width=0.48\textwidth]{velocity_field_plot/min_u_final.pdf}
% \caption{Comparison of the minimum x-component of velocity along the vertical line through the center of the cavity with \cite{ghia1982high} under different Re.}
% \label{fig:minu}
% % \vspace{-0.2in}
% \end{figure}
As shown in Figure \ref{fig:lid_driven}, we present our results of the converged 2D lid-driven cavity flow with various Reynolds numbers (Re) = 100, 400, 1000, 3200, 5000, 7500 and 10000, compared with \cite{ghia1982high}. In this well-known benchmark, the top wall is moving with \(velocity=1\), which drives the generation of \rev{vorticity} in the cavity. The vorticity at the midpoint of the moving (top) boundary is compared with \cite{ghia1982high} in the leftmost subfigure of Figure~\ref{fig:cavity_vort_vel_compare}. The second and third subfigures compare the minimum and maximum y-velocity along the horizontal centerline of the cavity, while the rightmost subfigure compares the minimum x-velocity along the vertical centerline. It can be seen from the comparisons that our results are highly consistent with those of \cite{ghia1982high}, both quantitatively and qualitatively, demonstrating the accuracy of the VPFM framework as a whole and its capability to effectively handle viscosity. The boundary condition used for this experiment \rev{follows that of \cite{ghia1982high} and} is given in the supplementary material.

\subsection{\rev{Order of Accuracy}}
\label{subsec:conv_order}
\begin{figure*}
    \centering
    \includegraphics[width=1.0\textwidth]{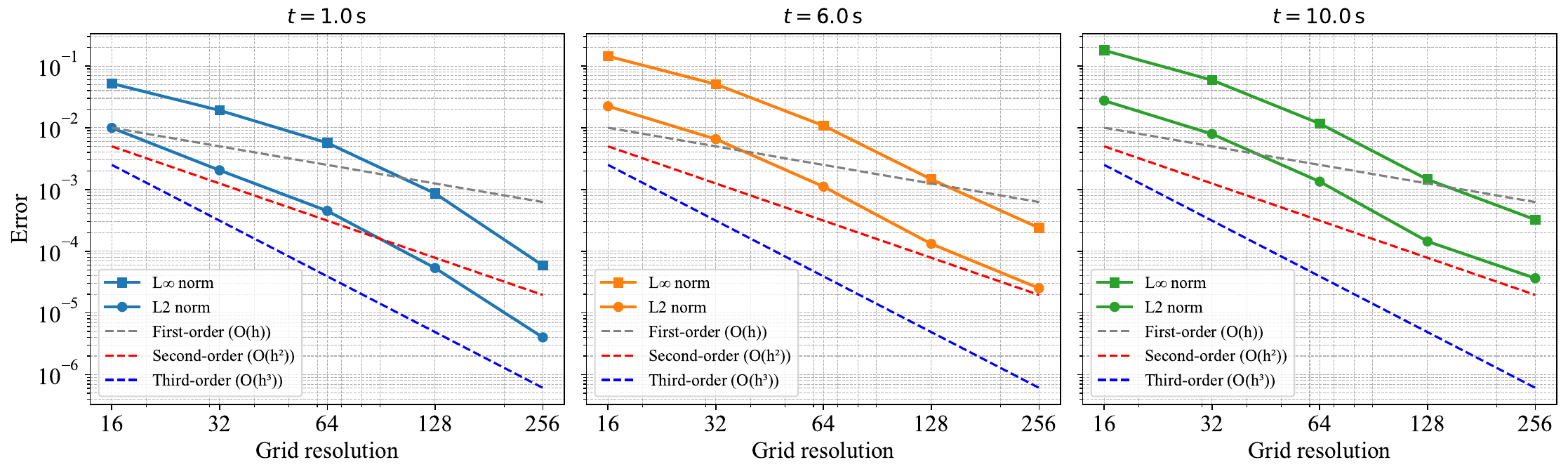}
    \caption{\rev{Convergence rate for the 2D Taylor-Green vortex experiment in $L_2$ and $L_\infty$ Norms.}}
    \label{fig:conv_plot}
\end{figure*}
\rev{As shown in Figure~\ref{fig:conv_plot}, we illustrate the order of accuracy of our method using a 2D Taylor-Green vortex experiment. In numerical analysis, the \emph{order of accuracy} refers to the convergence rate of a numerical solution to the exact solution. A method is said to be of \(n\)-th order accuracy in space if the error is proportional to the \(n\)-th power of the grid spacing \(\Delta x\) \cite{leveque1998finite, strikwerda2004finite}.
In this experiment, we compare our solution against the analytical solution of the Navier--Stokes equations for the Taylor-Green vortex with \(\nu = 0.005\), measuring errors under both the \(L_2\) and \(L_\infty\) norms. We perform simulations with resolutions of 16, 32, 64, 128, and 256, and compare the solutions at \(t = 1\), \(t = 6\), and \(t = 10\). All runs use the same flow map length \(n^L = 20\), a CFL number of 0.4 and a quadratic B-spline interpolation kernel \cite{steffen2008analysis}.
For \(t = 1\), our method shows third-order convergence from 64 to 256 and second-order convergence from 16 to 64. At \(t = 6\), it shows third-order convergence from 32 to 128, second-order convergence from 16 to 32, and approximately 2.5th-order convergence from 128 to 256. For \(t = 10\), the method again shows third-order convergence from 32 to 128 and second-order convergence from both 16 to 32 and 128 to 256. Overall, our method achieves roughly 2.5th-order accuracy across these tests. By comparison, \citet{nabizadeh2024coflip} reports third-order convergence at \(t = 1\) under the same settings. We emphasize that our estimate of order of accuracy is based on the empirical convergence rate, which is commonly used in practice to characterize the order of a method. A formal theoretical analysis is beyond the scope of this work.
}

\subsection{Memory and Time Cost}

\rev{We present a detailed breakdown of wall-clock time, GPU memory usage, and average convergence iterations for various 3D experiments of our method in Table~\ref{tab: time_mem}, comparing results with and without the use of the Hessian.} The performance test is conducted on a machine with an Intel Core i9-14900KF processor, 64 GB of RAM, and an NVIDIA GeForce RTX 4090 GPU with 24 GB of memory. Our code is implemented through Taichi \cite{hu2019taichi}. \rev{We observe that incorporating the Hessian increases GPU memory consumption, while introducing minimal runtime overhead. Notably, the Hessian mainly affects the runtime performance through the advection step. The current performance bottleneck lies in the P2G transfer, which accounts for nearly 50\% of the total computation time. This is primarily due to the atomic operations involved in P2G, which can potentially be accelerated using techniques proposed in GPU-MPM \cite{gao2018gpu}.
Notably, although our method solves three Poisson equations—compared to only one in PFM \cite{zhou2024eulerian}—it still runs faster, albeit with increased GPU memory usage.
The speedup stems solely from employing a more efficient Poisson solver implemented in CUDA and C++ and all other steps are implemented via Taichi. The increased memory usage is primarily attributed to two factors: first, the inclusion of the Hessian; and second, limitations in Taichi's interoperability with PyTorch—specifically, Taichi does not currently support the conversion of Taichi arrays to PyTorch tensors without copying the underlying data. As a result, an intermediate copy must be made to bridge the Taichi code with the CUDA implementation, increasing memory consumption.} Since the Poisson solver is not a contribution of this work, we will not discuss its details further.

\subsection{Ablation Study}
% \paragraph{Compare Hessian Evolution and Computation}
% \label{para:hessian_compare}
% We demonstrate the effectiveness of our approach through a 3D experiments (trefoil knot) in Figure~\ref{fig:dTdx}, where the vorticity of the same frame is compared. Notably, evolving the Hessian—as opposed to computing it from neighbors—yields smoother, more symmetric vortex tubes and rings.

% \paragraph{Vortex ring pass a static Cube}
% A static cube is encountering an inflow. The vortex ring induced by the cube show stair-stepping artifacts without the cut-cell enabled, as shown in Figure~\ref{fig:cutcell_gt_compare}.

% \paragraph{Varying flow map length for Hessian term}
% \begin{figure}
% % \hspace{-0.28\textwidth}
% \includegraphics[width=0.48\textwidth]{ablation/ablation_hessianmap_hessian.pdf}
% \caption{Varying flow map length for Hessian term.}
% \label{fig:ablation_hessianflowmap_hessian}
% % \vspace{-0.2in}
% \end{figure}

\paragraph{Impact of Short Flow Map Length \(n^S\) with Fixed \(n^L\)}
\label{para:longshort_ablation}
\begin{figure}
% \hspace{-0.28\textwidth}
\includegraphics[width=0.48\textwidth]{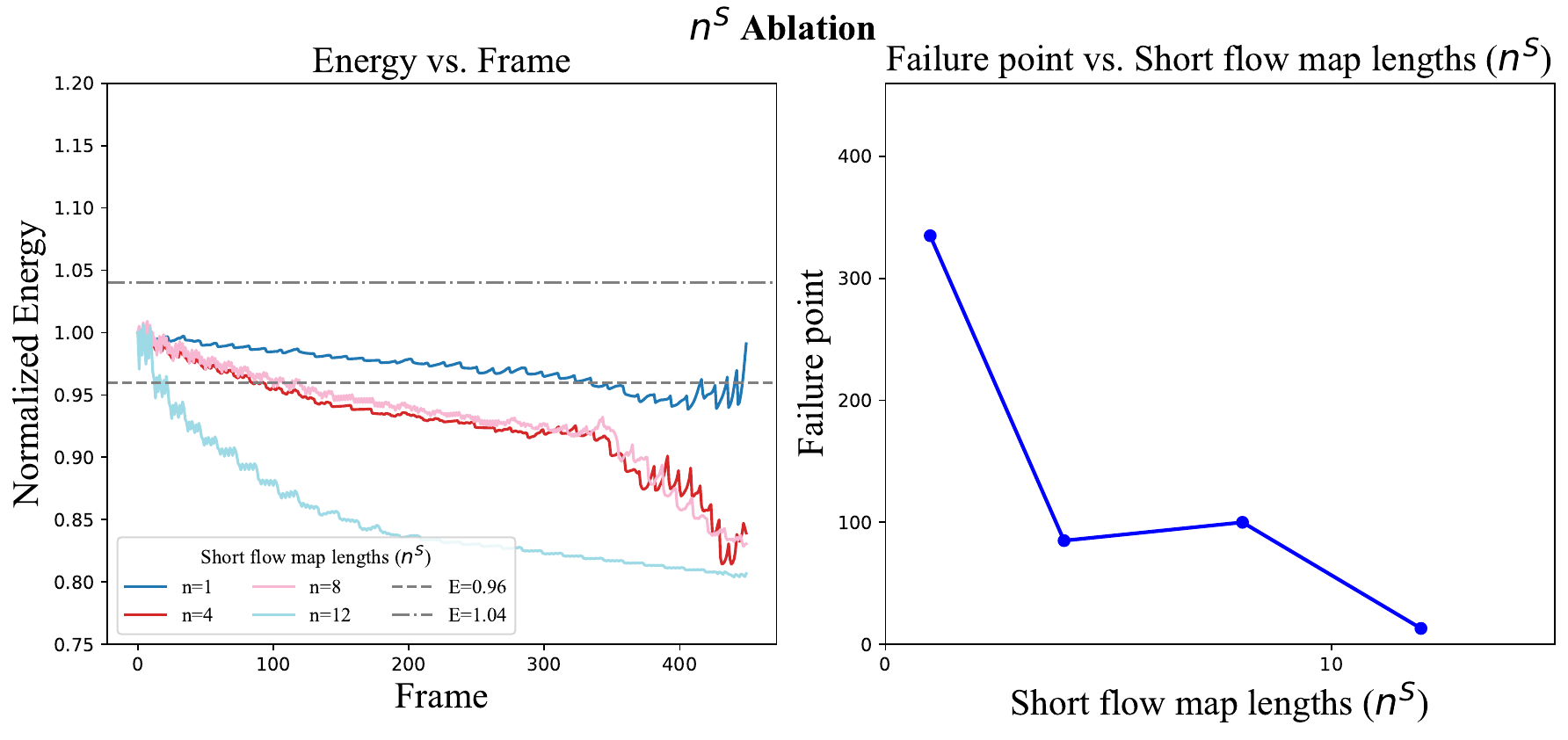}
\caption{Effect of varying short flow map length with \(n^L = 60\) in a 3D leapfrog setting. We find that using \(n^S=1\) gives the best result.}
\label{fig:ablation_shortflowmap_hessian}
% \vspace{-0.2in}
\end{figure}
In a 3D leapfrog setting, we fix the flow map length \(n^L\) to be 60, and vary the short flow map length \(n^S\) as 1, 4, 8 and 12 respectively, we found that using 1 gives the best result, as shown in Figure~\ref{fig:ablation_shortflowmap_hessian}.

\begin{figure}[h]
    \centering
    \includegraphics[width=0.48\textwidth]{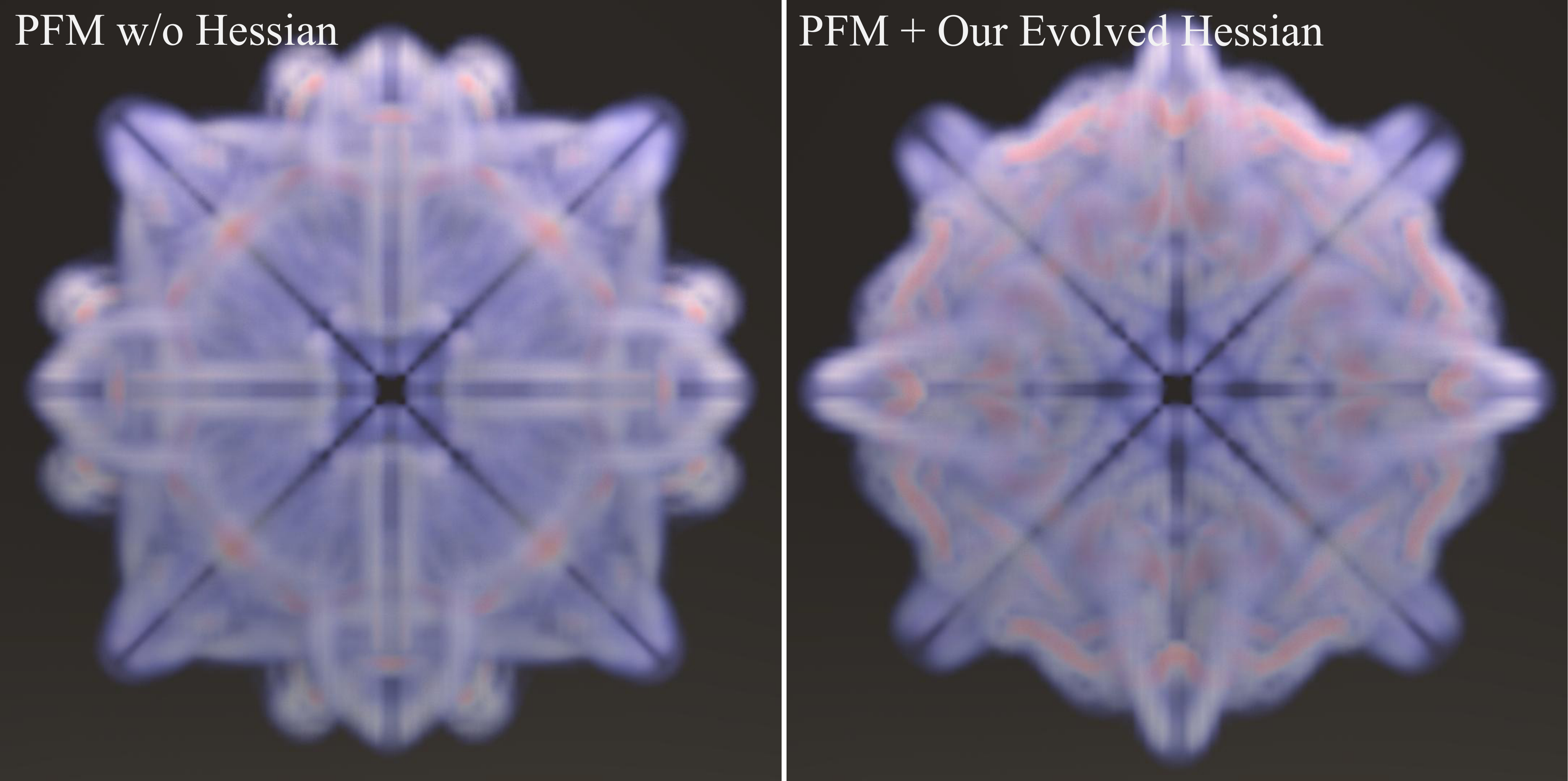}
    \caption{At the fifth leap, the pair of leapfrogging vortex rings has merged in the simulation using PFM. However, with the incorporation of our evolved Hessian term, PFM successfully keeps the rings separate at the same frame.}    \label{fig:ablation_pfm_hessian}
\end{figure}

\paragraph{Enhance PFM with Our Evolved Hessian Term}
\label{para:enhance_pfm_with_hessian}
In PFM \cite{zhou2024eulerian}, it was shown that incorporating the Hessian term by interpolating \(\mathcal{T}\) from neighboring particles does not improve its performance, and thus this term was omitted. \rev{By} contrast, our novel Hessian evolution scheme can be integrated into PFM to achieve enhanced performance. By utilizing the evolution equation of $\nabla \mathcal{T}$ and an RK4 marching scheme analogous to that used for $\nabla \mathcal{F}$ (see supplementary material for details), we evolve the Hessian term $\nabla \mathcal{T}^p_{[b, c]}$ on particles. This previously omitted term is then incorporated into the P2G process of PFM as $(\nabla \mathcal{T}^p_{[b, c]})^T\bm{m}_b$, effectively addressing the gap in prior methods. We demonstrate the effectiveness of the evolved Hessian in PFM through the 3D leapfrog benchmark, using the same setup as described in Section~\ref{subsubsec:compare_optimal_flowmap}. As illustrated in Figure~\ref{fig:ablation_pfm_hessian} and Figure~\ref{fig:3d_leapfrog_compare}, incorporating our evolved Hessian enables PFM to better preserve vortex structures and maintain the separation of two vortex rings at the fifth leap.

\begin{table*}
\caption{\rev{A detailed breakdown of wall-clock time, GPU memory usage and the average convergence iteration number. To compare results with and without Hessian, values are reported in the format “with $\bm{/}$ w/o” where applicable. The performance of 3D leapfrog is tested on the one with the shorter domain.}}
\centering\small
\begin{tabularx}{\textwidth}{Y | Y | Y | Y | Y | Y | Y | Y}
\hlineB{2}
3D Experiment & Time (s, with $\bm{/}$ w/o Hessian) &
GPU Mem. (GB, with $\bm{/}$ w/o Hessian) &
Converge Iteration &
P2G Time (s, with $\bm{/}$ w/o Hessian) &
Poisson Solve Time (s, with $\bm{/}$ w/o Hessian) &
Advection Time (s, with $\bm{/}$ w/o Hessian) &
Midpoint Time (s, with $\bm{/}$ w/o Hessian) \\
\hlineB{1.5}
\textbf{Hopf Link} & 0.25 $\bm{/}$ 0.24 & 7.37 $\bm{/}$ 5.26 & 11.33 & 0.10 $\bm{/}$ 0.10 & 0.048 $\bm{/}$ 0.049 & 0.031 $\bm{/}$ 0.017 & 0.053 $\bm{/}$ 0.052\\
\hlineB{1}
\textbf{Leapfrog} & 0.43 $\bm{/}$ 0.40 & 13.34 $\bm{/}$ 9.95 & 10.00 & 0.21 $\bm{/}$ 0.21 & 0.057 $\bm{/}$ 0.056 & 0.063 $\bm{/}$ 0.035 & 0.065 $\bm{/}$ 0.064\\
\hlineB{1}
\textbf{Trefoil} & 0.26 $\bm{/}$ 0.25 & 7.36 $\bm{/}$ 5.29 & 12.00 & 0.10 $\bm{/}$ 0.10 & 0.051 $\bm{/}$ 0.051 & 0.031 $\bm{/}$ 0.018 & 0.055 $\bm{/}$ 0.056\\
\hlineB{2}
\end{tabularx}
\label{tab: time_mem}
\end{table*}

\section{EXAMPLES}

% \begin{figure}
%     \centering
%     \includegraphics[width=0.5\textwidth]{3D_experiments/normalized_energy_curves.pdf}
%     \caption{Trefoil energy curve}
%     \label{fig:trefoil_energy}
% \end{figure}

\paragraph{Head-on Vortex Collision (3D)}
Two vortex rings are positioned face-to-face with opposite-signed vorticity, leading to a collision that generates secondary vortices, as illustrated in \figref{fig:headon}. \rev{The first ring is centered at $(0.1, 0.5, 0.5)$, and the second is centered at $(0.4, 0.5, 0.5)$ with domain size $(0.5, 1, 1)$. Both rings have a major radius of $0.06$, a mollification support (minor radius) of $0.016$, lie in the $y$–$z$ plane, and are initialized with vorticity strengths of $\pm 2 \times 10^{-2}$.} In this experiment, the flow map length is set to 20, demonstrating that even with a relatively short flow map, our method effectively captures and preserves the intricate details of the secondary vortices.

\paragraph{Head of Michelangelo's David (3D)}
We submerged a sculpture of the head of Michelangelo's David in water and generated bubbles at the base of the sculpture and at the base of David's hair, while introducing an upward inflow directly beneath the sculpture. Figure~\ref{fig:david} illustrates the generation of bubbles on the sculpture, which then follow the inflow as it moves past the sculpture. \rev{Particles are passively advected and rendered to visually resemble bubbles.}

\paragraph{Swimming Plesiosaur (3D)}
Figure~\ref{fig:plesiosaur_224} illustrates the extinct aquatic reptile, Meyerasaurus (a type of plesiosaur). Paleontologists have long debated how these creatures coordinated their four flippers in order to swim~\cite{liu2015computer}. In this figure, the plesiosaur generates \rev{turbulent} flow by flapping its flippers as the inflow moves from the creature's head towards its tail. Bubbles are generated on each of the Meyerasaurus's four flippers and move through the water following the turbulence flow generated by the Meyerasaurus's flapping motion. Note that previous methods can only use a flow map length \(n^L\) of 8 (NFM, EVM) or 12 (PFM) for moving solid objects, while we can extend this to 30 in this experiment.

\paragraph{Rotating Propeller (3D)}
Figure~\ref{fig:propeller_26} illustrates an underwater propeller rotating with an incoming flow from the left. During the propeller's rotation, bubbles are generated on the propeller's edges and transported by the combined effects of the propeller's motion and the inflow. When the propeller rotates counterclockwise, the bubbles follow a spiral trajectory, and helical tip vortices are observed, which is similar to the results given in a Large Eddy Simulation (LES) \cite{kumar2017large}. During clockwise rotation, the vorticity and bubbles show a turbulence pattern. Similar to the swimming plesiosaur, the flow map length \(n^L\) here is also 30, exceeding the state-of-the-art maximum of 12 for moving solid boundaries.

\paragraph{Aircraft (3D)}
Figure~\ref{fig:airplane} presents an aircraft subjected to an incoming flow from the left, with an angle-of-attack being \(20^\circ\). Smoke is generated at the aircraft’s wings, tail nozzle, and tail fins, moving with the flow as it passes over the aircraft. The phenomenon "vortex lift" \cite{anderson2010aircraft} on two leading edges of the aircraft is observed, which is consistent with the experiment in \cite{delery2001robert}.

\section{DISCUSSION}
\label{sec:discussion}
\paragraph{Harmonic component} \citet{10.1145/3592402} observed that the harmonic component \(\bm u_{h}\) possesses its own dynamics, necessitating a coupled treatment with the vorticity. \rev{Specifically, the harmonic component \(\bm u_h\) cannot be simply treated as the gradient of the harmonic functions on non-simply connected domains. We refer the reader to their paper for more details.} In our setting, however, the presence of moving obstacles, viscosity, and no-slip conditions prevents the direct application of their formulation, as these aspects remain unaddressed in their work. Since addressing this complex coupling is beyond the scope of our work, we do not attempt to solve it here. Nonetheless, exploring harmonic components' dynamics under these conditions is an exciting avenue for future research.

\paragraph{Vorticity is more suitable than impulse for long flow maps.} %\cite{wang2024eulerian, yin2021characteristic, yin2023characteristic, 10.1145/3592402}
\label{disc:vorticity_less_singular}
Experimental validation for this statement has been given in Section \ref{subsubsec:compare_same_long_flow_map}, where under the same challenging long flow map length, the impulse-based simulation exploded prematurely. Now we provide theoretical insights into this statement. As noted by \citet{cortez1995impulse}, the differences in velocity reconstruction processes result in greater singularities for impulse \(\bm m\) compared to vorticity. Specifically, for velocity \(\Tilde{\bm u}\) or impulse \(\bm m\), the solenoidal velocity \(\bm u\) is reconstructed through a pressure-projection process:
\begin{equation}
    \rev{
    \bm u = \bm m - \nabla (\Delta^{-1} \left(\nabla \cdot \bm m\right)).
    }
\end{equation}
\rev{By} contrast, the velocity reconstruction from vorticity \(\bm \omega\) yields:
\begin{equation}
    \bm u = -\nabla \times \Delta^{-1} \bm \omega.
\end{equation}
Notably, the additional differentiation in the \rev{impulse-based methods} implies that it is more singular, compared to \rev{vortex methods}, making it less suitable for long flow maps. \citet{wang2024eulerian} attempted to support this statement by demonstrating that impulse values grow unbounded over time without reinitialization, while vorticity remains stable. However, their work did not include a complete simulation to directly show the failure of impulse-based methods or the robustness of vorticity-based methods. \rev{By} contrast, we provide comprehensive experimental results in Section \ref{sec:validation}, showcasing various scenarios that validate this claim.

\paragraph{\rev{Stability of VPFM}} 
\rev{We discuss several numerical practices for VPFM's stability: (1) \textbf{Flow map length.} In 3D simulations, the flow map length \(n^L\) should generally not exceed 60. \rev{By} contrast, for 2D simulations, this constraint is largely lifted in the absence of viscosity. (2) \textbf{Time step.} Although vortex methods are not theoretically constrained by the time step in the same way as impulse or velocity-based methods, we adopt a consistent strategy for fair comparison with NFM, PFM, and EVM. Specifically, the time step is controlled using the CFL number and the maximum velocity in the domain. We typically use \(\text{CFL} = 0.5\) for 3D and \(\text{CFL} = 1.0\) for 2D, which we have found to yield stable results. (3) \textbf{Penalization coefficient \(\lambda\).} This coefficient controls the amount of vorticity injected by solid objects into the domain and thus directly affects stability. For static objects, we usually use \(\lambda = \frac{1}{\delta t}\). For moving objects, depending on their speed, we typically choose \(\lambda\) in the range of \(\frac{1}{100 \delta t}\) to \(\frac{1}{2 \delta t}\). (4) \textbf{Flow map Jacobian.} Both the projection method from CO-FLIP and the Jacobian-aware blending from IPIC aim to enforce the determinant of the flow map Jacobian to be 1, thereby improving stability. We have tested the projection method from CO-FLIP and observed nearly identical results. This is likely because our method already maintains the Jacobian determinant close to 1. In scenarios where the determinant deviates significantly from 1, such strategies may provide benefits. (5) \textbf{Viscosity.} We currently employ an explicit scheme for handling viscosity, which may lead to instability when viscosity is significant. An implicit scheme can be adopted, at the cost of treating the vorticity boundary condition globally \cite{we1996vorticity}.}

\paragraph{VPFM in 2D}
\label{para:vpfm_2D}
The main focus of this paper is on the 3D case. Here, we provide additional details for the 2D case. Specifically, in 2D, vorticity reduces to a scalar field (aligned with the \(z\)-direction), and vortex stretching is absent. Consequently, the vorticity carried by a vortex particle remains unchanged after its initialization. For the first term in the evolution of vorticity gradient (Eq.~\eqref{eq:evolve_grad_omega}), the forward Jacobian \(\mathcal{F}\) will be dropped and only the backward Jacobian \(\mathcal{T}\) is left. The second term in Eq.~\eqref{eq:evolve_grad_omega} is fully dropped. This result can be obtained by considering 2D as a special case of 3D, with vorticity \(\bm{\omega} = (0,0,\omega_z)\) and velocity \(\bm{u} = (u_x, u_y, 0)\). 

\section{CONCLUSION AND LIMITATIONS}
In this paper, we presented VPFM, a hybrid vortex method framework that revisits and extends the traditional VIC family of methods through a modern particle flow map perspective. The long-standing challenge of insufficient vortex preservation in VIC methods, that has limited the adoption of VIC in computer graphics, is addressed. By integrating VIC with particle flow maps, the strengths of both approaches are leveraged in a complementary manner: the particle flow map framework enhances VIC's ability to preserve vorticity, while the VIC foundation, in turn, provides a robust platform that refines and elevates the capabilities of flow map methods. This synergy enables our approach to achieve robust, long-term flow maps that outperforms state-of-the-art methods in maintaining vortex structures and overall stability. Additionally, we propose a cut cell approach derived from velocity-based methods and tailored for vortex simulations, ensuring accurate treatment of curved boundaries. To approximate the no-slip condition in a straightforward yet effective manner, we introduce a simplified Brinkmann penalization scheme, which is both cost-effective and easy to implement.
Overall, VPFM stands as a promising new direction for fluid simulations in computer graphics, offering a flexible, efficient, and powerful toolset for capturing complex vortex dynamics over extended time frames. 

However, our method is subject to several limitations. First, it does not currently support two-way solid-fluid coupling. Furthermore, it is not equipped to handle free surface scenarios, and the enforcement of the no-slip condition remains an approximation. Lastly, the exact harmonic component's dynamics are not accounted for. These limitations highlight promising directions for future research within the VPFM framework.

\begin{acks}
We sincerely thank the anonymous reviewers for their valuable feedback and Ruicheng Wang for his help with environment setup. We are also grateful to the authors of the IPIC paper for running simulations with our initial velocity fields and sharing the results. We thank Yuting Gu for providing the plesiosaur model. Georgia Tech authors acknowledge NSF IIS \#2433322, ECCS \#2318814, CAREER \#2420319, IIS \#2433307, OISE \#2433313, and CNS \#1919647 for funding support. We credit the Houdini education license for video animations.
\end{acks}
% \begin{figure*}
% \centering
% \includegraphics[width=0.995\textwidth]{velocity_field_plot/lid-driven_low.pdf}
% \\
% \caption{Comparison of our reseults with \cite{ghia1982high} under different Re.}
% \label{fig:liddrivelow}
% % \vspace{-0.1in}
% \end{figure*}

% \input{math}
% \input{limitation}

% \clearpage
% \newpage
{
% \small
% \nocite{*} 
\bibliography{refs_ML_sim.bib, refs_INR.bib, refs_flow_map.bib, refs_simulation.bib, refs_vortex_method}
\bibliographystyle{ACM-Reference-Format}
% \printbibliography
}

% \newpage
% \input{appendix}

\end{document}

%% file: bo_macro.tex
\usepackage{xcolor}
\usepackage{graphicx}

\newif\ifverbose
\verbosetrue

\ifverbose

\else

\newcommand{\vb}[1]{}
\fi

%% file: math.tex
% \section{Math}
\subsection{Naming Convention}
Bold symbols represent vector fields, while regular symbols denote scalar fields. Superscripts usually indicate whether a quantity is associated with particles or the grid, and subscripts typically describe the physical meaning of the quantity. Main symbols and notations are summarized in Table~\ref{tab:notation_table}.

\subsection{Vortex Method Preliminaries}
\paragraph{Fluid Motion}
%\bo{I hope we can start this section from the top of the column.}
\rev{Assuming constant density}, we start with the velocity-form incompressible Navier-Stokes equations:
%\begin{equation}
\begin{align}
    \frac{D \bm u }{D t} = - \frac{1}{\rho} \nabla p + \nu \Delta \bm u, \label{eq:ns1} \\
    \nabla \cdot \bm u = 0, \label{eq:ns2}
\end{align}
%\end{equation}
where $\bm u, \rho, p, \nu$ represent velocity, density, pressure, and kinematic viscosity respectively. And \(\frac{D}{Dt} = \frac{\partial}{\partial t} + \bm u \cdot \nabla\) is the material derivative. The first equation specifies momentum conservation, and the second equation is the incompressibility condition. 
We obtain its vorticity form by taking the curl of the momentum equation and substituting the incompressibility condition:
\begin{equation}
 \frac{D \bm \omega }{D t} =  \rev{(\bm{\omega}\cdot\nabla)\bm{u}} + \nu \Delta \bm \omega,
 \label{eq:vor}
\end{equation}
where $\rev{(\bm{\omega}\cdot\nabla)\bm{u}}$ denotes the vortex-stretching term and \(\nu \Delta \bm \omega\) denotes the viscosity term. %\bo{explain Re, better to use $\nu$ here}
\paragraph{Velocity Reconstruction}
According to the Helmholtz decomposition and the incompressible condition Eq.~\eqref{eq:ns2}, the velocity field \(\bm u\) can be decomposed into a solenoidal vortical component \(\bm u_{\omega}\) and a solenoidal irrotational (i.e., harmonic) component \(\bm u_{h}\) \cite{cottet2000vortex}:
\begin{equation}
    \bm u = \underbrace{-\nabla \Phi}_{\bm u_{h}} + \underbrace{\nabla \times \bm \Psi}_{\bm u_{\omega}}.
    \label{eq:helmholtz}
\end{equation}

The solenoidal vortical component \(\bm u_{\omega}\) can be reconstructed from a vector potential \(\bm \Psi\) (i.e., streamfunction in 2D) induced by the bulk vorticity \(\bm \omega\) by further assuming the vector potential \(\bm \Psi\) is divergence-free (\(\nabla \cdot \bm \Psi = 0\)). \rev{This reconstruction involves solving three Poisson equations, one for each spatial dimension}:
\begin{equation}
\label{eq:stream_poisson}
        \rev{\Delta} \bm \Psi_d = -\bm \omega_d \text{, for } d = \rev{x,y,z,}
        %\bm{\Tilde{u}} = \nabla \times \bm \Psi.
\end{equation}
\rev{subject to the BC} \(\rev{\bm \Psi = 0}\) \rev{and}  \(\rev{\nabla \cdot \bm \Psi = 0}\) \rev{assuming no solid objects are present}. Then, a harmonic function \(\Phi\), whose gradient is the harmonic component \(\bm u_h\), is solved to enforce the no-through \rev{BC}:
\begin{equation}
    \label{eq:solve_harmonic_add}
        \rev{\Delta} \,\Phi = 0, \\
\end{equation}
subject to the Neumann \rev{BC}:
\begin{equation}
\label{eq:harmonic_neumann}
        \frac{\partial \Phi}{\partial n} = (\bm u_{\omega} - \bm u_{\text{solid}}) \cdot \bm n,
\end{equation}
where \(\bm n\) is the normal direction of the solid boundaries pointing outwards, and \(\bm u_{\text{solid}}\) is the solid boundary velocities. A discussion regarding the harmonic component is provided in Section \ref{sec:discussion}.

\subsection{Flow Map Preliminaries}
We define a velocity field $\bm u(\bm x,t)$ in the fluid domain $\Omega$ which specifies the velocity at a given location $\bm x$ and time $t$. Consider a material point $\bm X \in \Omega$ at time $t=0$, We define the forward flow map $\bm \phi(:,t):\Omega\rightarrow\Omega$ as
\begin{equation}
    \label{eq:phi_def}
    \begin{dcases}
    \frac{\partial \bm \phi(\bm X,t)}{\partial t}=\bm u[\bm \phi(\bm X,t),t],\\
    \bm \phi(\bm X,0) = \bm X, \\
   \bm  \phi(\bm X,t) = \bm x,
    \end{dcases}
\end{equation}
which traces the trajectory of the point, moving from its initial position $\bm X$ at time $0$ to its location at time $t$, represented by $\bm x$. Its inverse mapping $\bm \psi(:,t):\Omega\rightarrow\Omega$ is defined as
\begin{equation}
    \label{eq:psi_def}
    \begin{dcases}
    \bm \psi(\bm x,0) = \bm x, \\
   \bm  \psi(\bm x,t) = \bm X,
    \end{dcases}
\end{equation}
which maps $\bm x$ at $t$ to $\bm X$ at $0$. To characterize infinitesimal changes in the flow map and its inverse mapping, we compute their Jacobian matrices as
\begin{equation}
    \begin{dcases}
        \mathcal{F}(\bm X, t)= \frac{\partial \bm \phi (\bm X, t)}{\partial \bm X}, \\
        \mathcal{T}(\bm x, t) = \frac{\partial \bm \psi (\bm x, t)}{\partial \bm x}.
    \end{dcases}
    \label{eq:FT}
\end{equation}
The evolution of $\mathcal{F}$ and $\mathcal{T}$, under the Lagrangian view, satisfies
\begin{equation}
\label{eq:evole_FT}
\begin{dcases}
\frac{D \mathcal{F}}{D t}=\bm \nabla\bm{u}\mathcal{F},\\
\frac{D \mathcal{T}}{D t}=-\mathcal{T}\bm \nabla\bm{u}.
\end{dcases}
\end{equation}
%\bo{Refer to some previous papers for details.}
We refer readers to \cite{cortez1995impulse, fung1977first} for more details.
%wang2024eulerian, zhou2024eulerian, deng2023neural, li2024lagrangian, li2024particle

%% file: sec5_solid_bc.tex
\begin{wrapfigure}{r}{0.1\textwidth}
\centering
\includegraphics[width=0.1\textwidth]{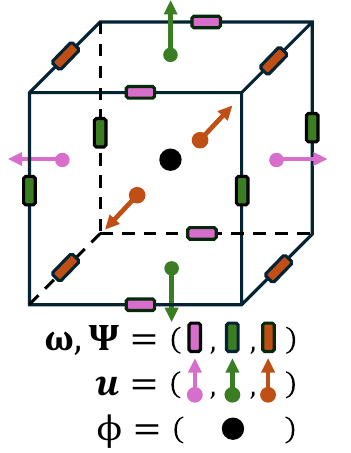}
% \captionsetup{aboveskip=8pt}
\caption{Discrete storage.}
\label{fig:discrete}
%\vspace{-0.1in}
\end{wrapfigure}
\section{NO-THROUGH BOUNDARY CONDITION}
\label{sec:solid_bc}
% With the integral evolution of \(\nabla \bm \omega\) on particles with the evolved flow map Hessian, we achieve superior vortex preservation and signifextend the flow map length by up to two times. 
% The effectiveness of vortex preservation naturally depends on the presence of vortices in the first place. This raises the question of how vortices can be captured or generated. A common approach is vortex shedding, which occurs due to solid objects and viscosity.
% \vspace{-0.1in}

In this section, we demonstrate a Symmetric Positive Semi-Definite (SPSD) cut cell system for velocity reconstruction in order to enforce more accurate no-through conditions on curved solid boundaries. While being a natural extension of the cut cell method developed for traditional velocity-pressure-based approaches \cite{bridson2015fluid, ng2009efficient}, to the best of our knowledge, this is the first SPSD cut cell method for velocity reconstruction on \rev{the grid} in vortex methods. \rev{We note that \citet{Ando:2015:streamfunc} also proposed an SPSD system for solving the vector potential with cut cells. However, their method does not involve vorticity and differs fundamentally from vortex formulations.} Two 3D experiments will be presented to validate the effectiveness of our method in Section \ref{subsec:cutcell_validate}.
% Additionally, we introduce a simplified Brinkmann penalization scheme to simulate visually convincing no-slip solid boundary conditions, which, while simplified, closely mimic the correct physical behavior.
%\vspace{-0.1in}

\subsection{Discrete Storage}

We illustrate our discrete storage in 3D by Figure~\ref{fig:discrete}. The vorticity \(\bm \omega\) and the vector potential \(\bm \Psi\) are stored on voxel edges, the velocity \(\bm u\) is stored on faces, and the harmonic function \(\Phi\) is stored on centers. The 2D discrete storage scheme is illustrated in the left of Figure~\ref{fig:fluid_frac}, where the scalar vorticity \(\omega\) is stored on cell nodes, velocity \(\bm u\) is stored on edges and the harmonic function \(\Phi\) is stored on centers.

\subsection{Solenoidal Vortical Component Reconstruction}
Let \(D\) denote the whole computational domain \rev{(a rectangular box here)}. In order to reconstruct the solenoidal vortical component of the velocity, \(\bm u_{\omega}\), we first solve the vector potential \(\bm \Psi\) assuming no solid obstacles are present:
\begin{equation}
\label{eq:solve_psi_numerical}
    \begin{dcases}
        \rev{\Delta} \bm \Psi_d = - \bm \omega_d, & \text{in } D, \text{for }d =\rev{x,y,z},\\
        \bm \Psi = 0, & \text{on } \partial D, \\
        \nabla \cdot \bm \Psi = 0, & \text{on } \partial D.
    \end{dcases}
\end{equation}
\(\bm u_{\omega}\) can then be computed by:
\(
    \bm u_{\omega} = \nabla \times \bm \Psi.
\)
\rev{When solving the system, we need to use \(\bm \Psi\) that lies on and outside \(\partial D\). For \(\bm \Psi\) on \(\partial D\), we impose \(\bm \Psi=0\). For \(\bm \Psi\) outside \(\partial D\), we need the condition \(\nabla\cdot \bm \Psi=0\) to pin it down. We refer to the same equation in Appendix B of \cite{10.1145/3592402} and vorticity BC \cite{we1996vorticity} for details. We note that simply setting \(\bm \Psi=0\) outside also works in practice.} The Laplacian operator in Eq.~\eqref{eq:solve_psi_numerical} is implemented via five-point stencil in 2D and seven-point stencil in 3D. In 2D cases, the first equation of Eq.~\eqref{eq:solve_psi_numerical} only involves a scalar Poisson (\(d = \rev{z}\)), and the third equation of Eq.~\eqref{eq:solve_psi_numerical} is dropped. An AMGPCG solver is used to solve Eq.~\eqref{eq:solve_psi_numerical}.

\begin{figure*}[t]
    \centering
    \includegraphics[width=1.0\textwidth]{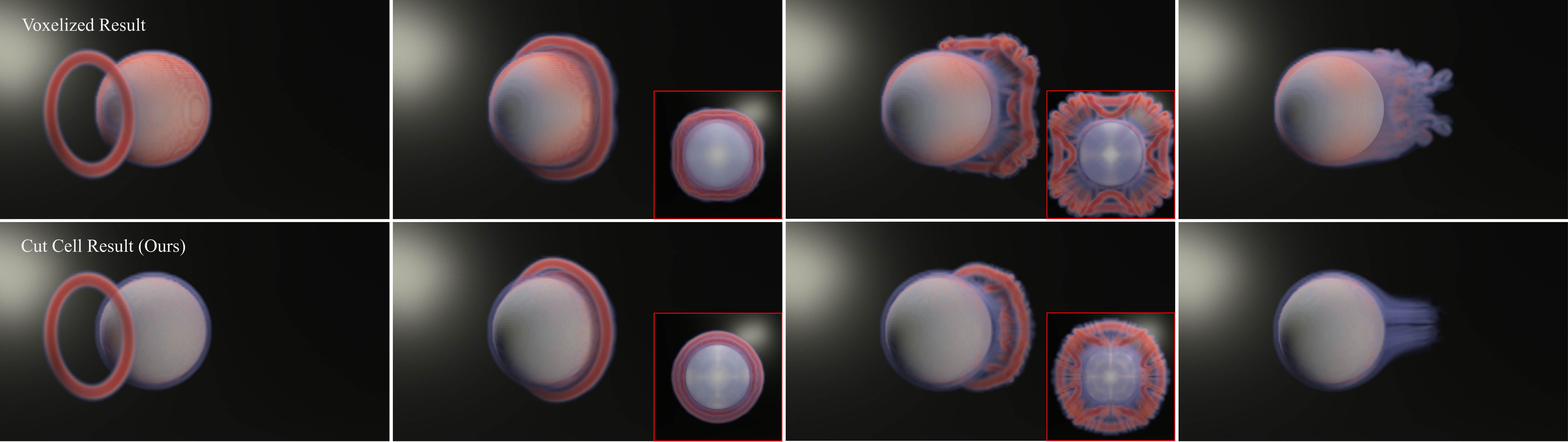}
    \caption{Vortex ring passing by a ball. Without our cut cell method, the vortex ring exhibits blocky and angular patterns, whereas our cut cell method produces a much smoother and more circular result. The subfigures present back views of the simulation.}
    \label{fig:vortex_ring_ball}
\end{figure*}

\subsection{Harmonic Component for Cut Cell Geometry}
\label{subsec:cutcell_nothrough}
% Alternative cut cell approaches for vortex methods include the immersed interface method (IIM) \cite{marichal2016immersed} and vortex panel methods \cite{park2005vortex}. However, both are computationally expensive and complex, making them unsuitable for computer graphics applications.
Given the solenoidal vortical component \(\bm u_{\omega}\), our cut cell no-through velocity can then be constructed \rev{from} the solution of the harmonic function \(\Phi\), \rev{obtained} by solving the Laplace equation (Eq.~\eqref{eq:solve_harmonic_add}) with the Neumann \rev{BC} (Eq.~\eqref{eq:harmonic_neumann}) \rev{treated using} the finite volume method \cite{ng2009efficient}. %For enforcing our cut cell no-through boundary conditions, we replace the Neumann boundary condition in Eq.~\eqref{eq:harmonic_neumann} by the finite volume method \cite{ng2009efficient}.
%Same as the velocity-pressure cut cell condition in \cite{bridson2015fluid}
Given a control volume \(C\) and its boundary \(\partial C\), we start with the incompressibility condition in its integral form:
\begin{equation}
    \label{eq:incompre_finite_vol}
\oiint_{\partial C} \bm{u} \cdot \bm{n} \, dS = 0.
\end{equation}
Replacing the Neumann boundary condition in Eq.~\eqref{eq:harmonic_neumann} by Eq.~\eqref{eq:incompre_finite_vol}, we obtain the continuous system:
\begin{equation}
    \label{eq:harmonic_solve_cutcell}
    \begin{dcases}
    \rev{\Delta} \, \Phi = 0, & \text{in } \Omega, \\
    \bm u = \bm u_{\omega} - \nabla \Phi,  & \text{in } \Omega,\\
    \oiint_{\partial C} \bm{u} \cdot \bm{n} \, dS = 0, & \text{on } \partial\Omega.
    \end{dcases}
\end{equation}

\begin{figure}
    \centering
    \includegraphics[width=0.45\textwidth]{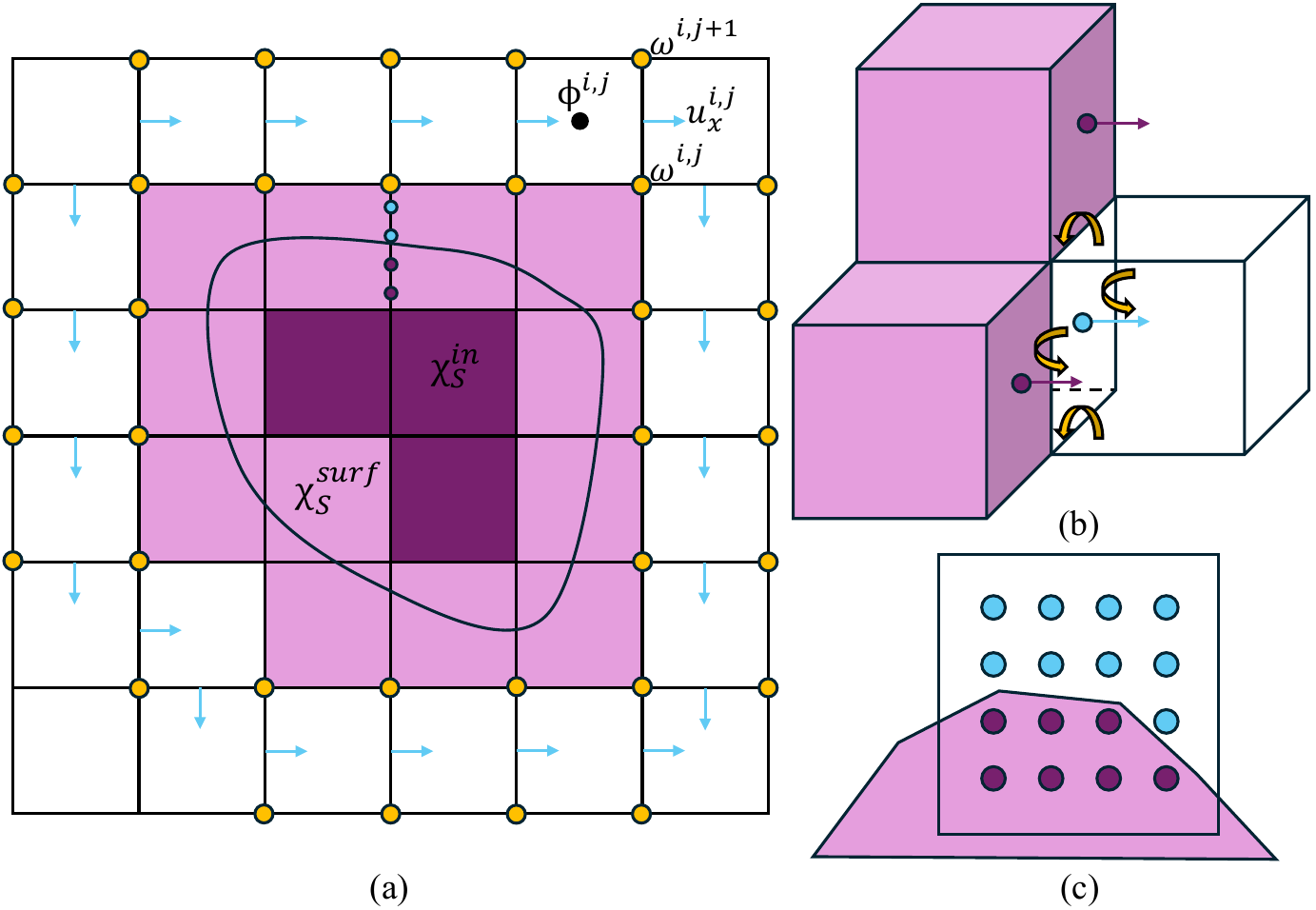}
    %\vspace{-0.7cm}
    \caption{The left figure shows an illustration for 2D, while the right two figures are for 3D. In 2D, vorticity is stored on cell nodes, velocity is stored on edges, and the harmonic function is stored on centers. The curve in the left figure represents a solid object. The light purple indicates the surface mask \(\surfmask{}\) and the dark purple indicates the interior mask \(\inmask{}\). Blue arrows (both in left and right top figures) represent penalization velocity. The yellow circles (2D) or curved arrows (3D) represent its induced penalization vorticity. The right bottom figure is a 3D voxel face, for illustrating how we approximate the fluid area fraction. The purple area indicates the solid, and the dark purple circles indicate that these points are tested to be inside the solid, therefore the fluid fraction for this face is approximately \(\frac{7}{16}\). Similarly in 2D, there are two blue circles and two dark purple circles on an edge of a cell, indicating that the fluid fraction on this edge is approximately \(\frac{1}{2}\).}
    \label{fig:fluid_frac}
\end{figure}

% The discretized Eq.~\eqref{eq:incompre_finite_vol} is:
% \begin{equation}
% \label{eq:discre_imcompre_finite_vol_compact}
% \sum_{d=1}^{3} \sum_{s \in \{-1,1\}} 
% \Bigl[
%   \alpha(I_d)\,u_{d}(I_d)
%   \;-\;
%   \bigl(1 - \alpha(I_d)\bigr)\,u_{\text{solid},d}(I_d)
% \Bigr]
% \;=\; 0,
% \end{equation}
% where \(I_d = (i,j,k) + \tfrac{s}{2}\,\bm{e}_{d}\) and \(\bm{e}_{d}\) is the unit vector for dimension \(d\). \(\alpha\) represents the fluid area fraction on the corresponding voxel face.
% Then, the descretized cut cell system (expanded out in the supplementary material) is:

% \begin{equation}
% \label{eq:discre_phi_finite_vol_compact}
% \begin{aligned}
% &\sum_{d=1}^{3} \sum_{s \in \{-1,1\}}
% \Bigl[
%   -\,\alpha\bigl(I_d\bigr)\,\Phi\bigl(J_d\bigr)
% \Bigr]
% \;+\;
% \Bigl[
%   \sum_{d=1}^{3} \sum_{s \in \{-1,1\}}
%   \alpha\bigl(I_d\bigr)
% \Bigr]
% \,\Phi(i,j,k)
% \\[6pt]
% &=\;
% \Delta x
% \sum_{d=1}^{3} \sum_{s \in \{-1,1\}} \bigl(-s\bigr)
% \Bigl[
%   \alpha\bigl(I_d\bigr)\,u_{\omega, d}\bigl(I_d\bigr)
%   \;+\;
%   \bigl(1-\alpha\bigl(I_d\bigr)\bigr)\,u_{\text{solid}, d}\bigl(I_d\bigr)
% \Bigr],
% \end{aligned}
% \end{equation}
% where \(I_d = (i,j,k) + \tfrac{s}{2}\,\bm{e}_{d}\) and \(J_d = (i,j,k) + s\,\bm{e}_{d}\). 
This \rev{resulting} linear system (\rev{expanded out in the supplementary material}) can be represented by a sparse, symmetric, and diagonally dominant matrix with non-negative diagonal entries, implying that the system is SPSD. The cut cell boundary condition for the harmonic function \(\Phi\) is identical to that for the pressure \(p\) \rev{in a standard pressure-projection method}; however, the former is governed by the \textbf{Laplace equation}, while the latter follows the \textbf{Poisson equation}. This highlights the critical role of the cut cell boundary condition in determining the harmonic component in the context of vortex methods, as solutions to the Laplace equation are entirely dictated by the boundary conditions.
% while the system for \(p\) is:
% \begin{equation}
%     \label{eq:p_solve_cutcell}
%     \begin{dcases}
%     \nabla \cdot \nabla p = \textcolor{red}{\nabla \cdot \bm u_{\omega}}, & \text{in } \Omega, \\
%     \bm u = \bm u_{\omega} - \nabla p,  & \text{in } \Omega,\\
%     \oiint_{\partial C} \bm{u} \cdot \bm{n} \, dS = 0,& \text{in } \Omega \cap S, 
%     \end{dcases}
% \end{equation}

\begin{figure*}[h]
    \centering
    \includegraphics[width=1.0\textwidth]{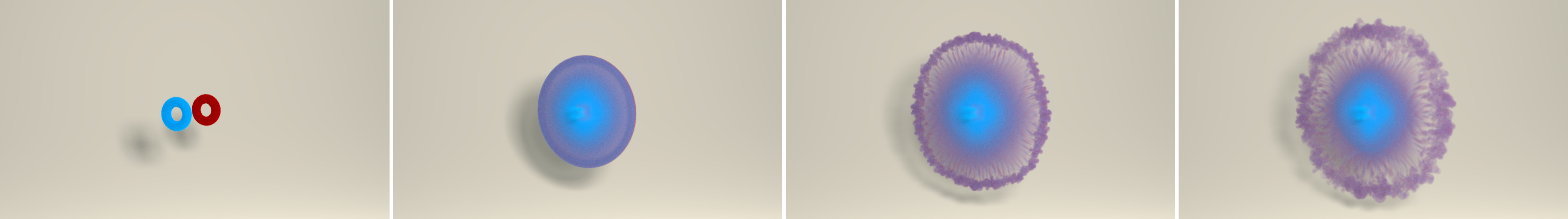}
    \caption{Head-on vortex collision. As two vortex rings collide, secondary vortices are generated, demonstrating that even with a relatively low flow map length (\(n^L = 20\)), our method effectively captures these secondary vortices. \rev{Smoke density is initialized along the rings and passively advected for visualization.} }
    \label{fig:headon}
\end{figure*}

% \begin{figure}
%     \centering
%     \includegraphics[width=0.3\textwidth]{illustration/adaptiveFlowMap_grid_2.pdf}
%     \caption{Adaptive flow map illustration. The complete trajectory (left purple curve), with a total length denoted as \(n^L\), is divided into two connected segments, separated by the purple circle at the midpoint. The length of the upper-right segment is represented as \(n^S\).}
%     \label{fig:adaptive_flowmap}
% \end{figure}

% \subsection{Fluid Area Fraction Approximation} Fluid fractions are approximated by a Monte-Carlo method based on Fast Winding Numbers \cite{barill2018fast}. The key steps are:
% \begin{enumerate}[leftmargin=*, labelindent=0pt]
%     \item \textit{Voxelization (\rev{see supplementary}):} \\
%     We voxelize a 3D mesh, accelerated by a BVH tree, and obtain a surface mask \(\surfmask{}\) representing the solid surface, an entire solid mask \(\chi_S\) and an interior mask \(\inmask{}\).
%     \item \textit{Fluid Area Fraction Computation (\rev{see supplementary})} \\
%     \rev{As illustrated in right bottom corner of Figure~\ref{fig:fluid_frac},} we compute fluid fractions on voxel faces that belong to those voxels where \(\surfmask{}\) is 1, by uniformly sample points on faces and compute the Fast Winding Numbers.
% \end{enumerate}

\subsection{Fluid Area Fraction Approximation}

\rev{Fluid fractions are approximated using a Monte Carlo method based on Fast Winding Numbers~\cite{barill2018fast}. We first voxelize a 3D mesh, accelerated by a BVH tree, to obtain a surface mask \(\surfmask{}\) representing the solid surface, as well as a full solid mask \(\chi_S\) and an interior mask \(\inmask{}\). Then, as illustrated in the bottom right corner of Figure~\ref{fig:fluid_frac}, we compute fluid fractions on voxel faces that belong to those voxels where \(\surfmask{} = 1\) by uniformly sampling points on the face and evaluating the Fast Winding Numbers. Additional implementation details are provided in the supplementary material.}

Finally, our SPSD system for harmonic solving is:
\begin{equation}
    \label{eq:harmonic_solve_cutcell_ours}
    \begin{dcases}
    \rev{\Delta} \, \Phi = 0, & \text{where } \inmask{} = 0, \\
    \bm u = \bm u_{\omega} - \nabla \Phi,  & \text{where } \alpha > 0.1,\\
    \oiint_{\partial C} \bm{u} \cdot \bm{n} \, dS = 0, & \text{where } \surfmask{} = 1. 
    \end{dcases}
\end{equation}
The same AMGPCG solver as vector potential solving (Eq.~\eqref{eq:solve_psi_numerical}) is used to solve the system.

\section{NO-SLIP BOUNDARY CONDITION}
\label{sec:noslip}
Vortex shedding, a critical phenomenon in computer graphics induced by solid objects and viscosity, is governed by no-slip solid boundary conditions. We propose a simplified, easy-to-implement, and cost-effective method for approximating no-slip conditions, based on Brinkmann penalization, that achieves visually comparable results to those produced by CFD methods. The effectiveness of our method is validated in Section \ref{subsec:validate_brinkmann}.

\paragraph{Penalized NS and its Vorticity Form}
Let \(D\) represents the entire computational domain, we start with the penalized Navier-Stokes equation \cite{mimeau2015vortex}:

\begin{equation}
\label{eq:vel_ns_pen}
%\begin{dcases}
 \frac{D \bm u }{D t} = - \frac{1}{\rho} \nabla p + \nu \Delta \bm{u} + \underbrace{\lambda \chi_S (\bm{u}_s - \bm{u})}_{\text{penalization term}}, \text{in } D.%\\[6pt]
%\nabla \cdot \bm{u} = 0 & \text{in } D.
%\end{dcases}
\end{equation}
Then we extend it into its vorticity formulation  by differentiation of Eq.\eqref{eq:vel_ns_pen}:
\begin{equation}
\label{eq:vort_ns_pen}
%\begin{dcases}
\frac{D\bm \omega}{D t} =  (\bm{\omega}\cdot\nabla)\bm{u} + \nu \Delta \bm{\omega} + \underbrace{\nabla \times [\lambda \chi_S (\bm{u}_s - \bm{u})]}_{\text{penalization term}}, \text{in } D,%\\[6pt]
%\nabla \cdot \bm{u} = 0 & \text{in } D,\\[6pt]
%\nabla \times \bm{u} = \bm{\omega} & \text{in } D.
%\end{dcases}
\end{equation}
% \begin{equation}
% \chi_S =
%     \begin{dcases}
%         1, & \text{ in solid},\\
%         0, & \text{ in fluid},
%     \end{dcases}
% \end{equation}
where \(\lambda\) is the penalization parameter which corresponds to the \rev{porosity} of the body and has the units of reciprocal time (\([1/t]\)). In our method, \(\lambda\), which could possibly represents the roughness of the solid objects, can be treated as a tunable parameter used to control the extent of vortex shedding.
Eq.\eqref{eq:vort_ns_pen} can then be solved in a time-splitting manner by separately solving the equation: % \cite{mimeau2015vortex}
\begin{equation}
\label{eq:vort_ns_pen_timesplit_1}
% \begin{dcases}
\frac{\partial \bm{\omega}}{\partial t}  = \nabla \times [\lambda \chi_S (\bm{u}_s - \bm{u})], \text{ in } D,
% \end{dcases}
\end{equation}
and the advection-stretching-diffusion equation
% \begin{dcases}
\(\frac{D \bm{\omega}}{D t} = (\bm{\omega}\cdot\nabla)\bm{u} + \frac{1}{\mathrm{Re}}\Delta \bm{\omega}, \text{ in } D,
% \end{dcases}
\) by our VPFM framework.
% We can write the solution of Eq.~\eqref{eq:vort_ns_pen_timesplit_1} using an implicit Euler scheme for the sake of numerical stability as the following \cite{coquerelle2008vortex}:
% \begin{equation}
% \label{eq:implicit_penvort}
%     \Tilde{\omega}^{n+1} = \nabla \times \Big[ \frac{\bm u^n + \lambda \Delta t \chi_S u^n_s}{1 + \lambda \Delta t \chi_S} \Big]
% \end{equation}
% A semi-implicit scheme can be achieved by setting \(\lambda = \frac{1}{\Delta t}\) \cite{rasmussen2011multiresolution}. The advection–stretching–diffusion Equation \eqref{eq:vort_ns_pen_timesplit_2} can be solved in various ways, and as described before, we solve it using the flow map. And for more details about the traditional vortex penalization methods, we refer readers to \cite{rasmussen2011multiresolution, mimeau2015vortex, hejlesen2015iterative}.
% Instead of enforcing the no-through and the no-slip conditions simultaneously, we first enforce the cut cell no-through conditions as described in Section \ref{subsec:cutcell_nothrough}. Then, we explicitly shed vortices induced by the Brinkmann penalization scheme into the flow. The details of the algorithm is given below.
\begin{figure*}[t]
\includegraphics[width=1.0\textwidth]{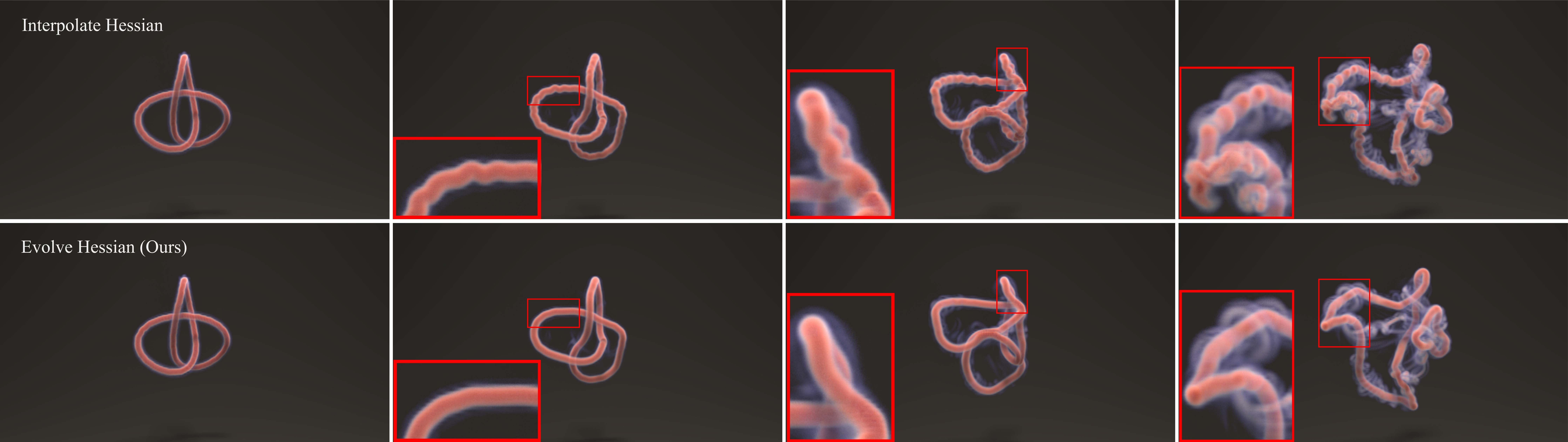}
\caption{Side view of trefoil knot. Under a relatively long flow map (\(n^L=40\)), a comparison between using Hessian interpolation \cite{zhou2024eulerian}, and our Hessian evolution, is shown here to illustrate that our evolved Hessian gives smoother vortex rings and tubes, especially for long flow maps.}
\label{fig:dTdx}
% \vspace{-0.2in}
\end{figure*}

\paragraph{Simplified Brinkmann Penalization in VPFM} We propose a simplified model of the full Brinkmann penalization scheme. Our key idea is to enforce the tangential velocity around half a grid cell from the solid boundary, rather than applying the penalization to the entire solid domain as in the traditional Brinkmann scheme \cite{mimeau2015vortex, rasmussen2011multiresolution, hejlesen2015iterative, spietz2017iterative}, especially given that the no-through boundary condition has been enforced in our previous solve.
%This is motivated by two considerations: first, we have already perfectly solved the no-through condition; second, we aim to facilitate vortex shedding, rather than strictly adhering to the exact no-slip boundary condition. 
%For achieving accurate no-slip conditions, we recommend referring to the iterative Brinkmann penalization method \cite{hejlesen2015iterative}, the Immersed Interface Method (IIM) \cite{marichal2016immersed}, or vortex panel methods \cite{willis2006unsteady}.

Without loss of generality, we focus on the case for \(x\)-direction and define \(P_{\bm{u}, x}\) as the set of all faces with normals in the \(x\)-direction that lie within the fluid domain and are directly adjacent to a solid cell. 2D and 3D illustration for \(P_{\bm{u}, x}\) are given in left and right top of \figref{fig:fluid_frac}, where their centers are marked as blue arrows. For each \(x\)-direction face located at \((i+\tfrac{1}{2}, j, k)\), we consider the \(x\)-component of velocity \(\bm u_x(i+\frac{1}{2}, j, k)\). We then define the number of adjacent solid cells surrounding this face as:
\begin{equation}
N_x (i+\tfrac{1}{2}, j, k) = \sum_{r \in \{0, 1\}} \sum_{s \in \{-1, 1\}} 
\bigl( \chi_{S}(i+r,j+s,k) + \chi_{S}(i+r,j,k+s) \bigr).
\end{equation}
The quantity \(N_x\) counts how many of the eight possible neighboring cells (around the \(x\)-direction face at \((i+\tfrac{1}{2}, j, k)\)) are solid.
Using this measure, we define the set \(P_{\bm{u}, x}\) (for \(x\)-direction) in 3D as follows:
\begin{equation}
P_{\bm{u}, x} := \Big\{ (i+\tfrac{1}{2}, j, k) \;\Big|
\sum_{r \in \{0, 1\}} \chi_S(i+r,j,k) = 0 \text{ and } N_x(i+\tfrac{1}{2}, j, k) > 0
\Big\}.
\end{equation}
In words, \(P_{\bm{u}, x}\) consists of all interior \(x\)-direction faces that are not themselves inside a solid column (i.e., the two adjacent cells along \(x\) are fluid) yet have at least one solid cell in the immediate vicinity. Thus, these faces are fluid-solid interfaces in the \(x\)-direction sense.
%A similar definition can be made for \(P_{\bm{u}, x}\) in the \(y\)- and \(z\)-directions, simply by shifting the indexing to the appropriate coordinate direction and counting neighboring solid cells in that direction accordingly.
%A 2D illustration of \(P_v\) is given in Figure~\ref{fig:fluid_frac}.

We now introduce a penalization indicator function to incorporate the effect of nearby solid cells on the fluid velocity. Since we have defined the set \(P_{\bm{u},x}\) for the \(x\)-direction, it is natural to define a corresponding penalization mask \(\chi_{\bm{u},x}\) as the characteristic function of \(P_{\bm{u},x}\). In other words, 
\begin{equation}
\chi_{\bm{u},x}(i+\tfrac{1}{2}, j, k) =
\begin{cases} 
1, & \text{if } (i+\tfrac{1}{2}, j, k) \in P_{\bm{u},x}, \\
0, & \text{otherwise}.
\end{cases}
\end{equation}

Focusing again on the \(x\)-direction for illustration, we define the penalization velocity field \(u_{pen, x}\) at the \(x\)-faces as the slip velocity as follows:
\begin{equation}
    \label{eq:upen_def}
    u_{pen, x}(I_x) = \chi_{\bm{u},x}\sum_{N \in \text{neighbors}}\frac{N_x}{8} \chi_S(N) (\bm u_{St,x}(N) - u(I_x)),
\end{equation}
where \(I_x = (i+\frac{1}{2},j,k)\), and neighbors refer to the eight neighboring voxels directly adjacent to the face \(I_x\). Two of them are marked as purple cubes at top right of Figure~\ref{fig:fluid_frac}. \(\bm u_{St}\) is the known tangential solid velocity, and \(u\) is the fluid velocity. The factor \(\frac{N_x}{8}\) represents the normalized count of adjacent solid cells influencing this face. Multiplying by \(\chi_{\bm{u},x}\) ensures that this penalization is only applied where there is at least one neighboring solid cell, i.e., on faces in \(P_{\bm{u},x}\). A similar definition can be applied in the \(y\)- and \(z\)-directions, defining \(P_{\bm{u},y}\), \(P_{\bm{u},z}\), their corresponding masks \(\chi_{\bm{u},y}\), \(\chi_{\bm{u},z}\), and the penalization velocity field \(u_{pen, y}\), \(u_{pen, z}\).

After obtaining the penalization velocity field \(\bm{u}_{pen}\), we can compute the corresponding penalization vorticity field \(\bm{\omega}_{pen}\) as follows:
\begin{equation}
    \label{eq:pen_omega_def}
    \bm{\omega}_{pen} = \lambda (\nabla \times \bm{u}_{pen}).
\end{equation}
% where \(\lambda\) is usually set as \(\frac{1}{\delta t}\) following \cite{rasmussen2011multiresolution}. 
The computed penalization vorticity field can then be treated as the same as an external force field, and interpolated back to the initial particles, allowing them to carry these newly generated (shedded) vortices.
% Note that instead of the implict scheme Eq. \eqref{eq:implicit_penvort}, we use an explicit scheme since our penalization vorticity exists only on a small portion of the computational domain (near the surface of the solid objects), therefore it would not influence the simulation stability too much. Additionally, we set \(\beta = 1\) for all 2D cases and 3D static objects, without observing any instability. In 3D, we also set \(\beta = 1\) for all static solid objects, and reduce \(\beta\) for moving solid objects, where instability may rise from moving solid boundaries.
The path integral is:
\begin{equation}
\label{eq:gamma_penomega}
\bm{\Gamma}_{pen, t} \;=\; \mathcal{F}_{t}\,\int_{0}^{t} 
  \mathcal{T}_{\tau} \left(\bm \omega_{pen} \right) \Big(\phi_{\tau}\bigl(\psi_{t}(\bm{x})\bigr),\,\tau\Big)\,d\tau.
\end{equation}
This simplified process effectively captures the vortex shedding phenomenon and its interaction with the surrounding fluid. In Section \ref{subsec:validate_brinkmann}, two 2D experiments are conducted to validate the effectiveness of our method, comparing with an original Brinkmann penalization VIC method \cite{rasmussen2011multiresolution}, and an iterative Brinkmann penalization VIC method \cite{hejlesen2015iterative}. Additionally, a 3D ablation study, where a ball encounters an inflow is performed to evaluate the efficacy of the proposed scheme in Section \ref{subsec:validate_brinkmann}.

% It is important to note that this scheme is not designed to provide high accuracy; rather, it serves as a simplified approach to mimic the behavior of no-slip boundary conditions. For achieving accurate no-slip conditions, we recommend referring to the iterative Brinkmann penalization method \cite{hejlesen2015iterative}, the Immersed Interface Method (IIM) \cite{marichal2016immersed}, or vortex panel methods \cite{willis2006unsteady}.

%% file: sec6_visc_extforce.tex
\section{VISCOSITY AND EXTERNAL FORCES}
\label{sec:visc_ext_force}
%\bo{We need to put a few sentences here to introduce the context and explain the equations. The current writing is too brief.}

In computer graphics, the viscosity term is often neglected due to the inherent numerical dissipation present in simulation methods and the community's focus on minimizing viscosity to achieve visually appealing results. As a result, the equations being solved are frequently simplified from the Navier–Stokes equations to the Euler equations. However, viscosity is a crucial factor in real-world fluid behavior. For example, \citet{li2024particle} proposed a viscosity-handling scheme in particle flow maps, specifically designed for simulating particle-laden flows, such as ink dynamics. Similarly, external forces, such as buoyancy and gravity, play a significant role in many natural phenomena. In this section, we introduce how we handle the viscosity and external forces within the VPFM framework. Our hybrid particle-grid framework naturally handles these terms by computing finite differences on \rev{the grid}, interpolating them onto particles, and then integrating them along the trajectories of particles. The accuracy of our viscosity handling is validated in Section~\ref{subsec:validate_visc} via a comprehensive analysis on the well-known benchmark, lid-driven cavity flow. The details are given as follows.

\paragraph{Basic Vorticity Formula} We start with the basic vorticity formula \cite{truesdell2018kinematics}:
\begin{equation}
\label{eq:basic_vort}
    \bm \omega (\bm x, t) = \mathcal{F}_t (\bm x) \Big[\bm \omega (\bm \psi(\bm x), 0) + \int_0^t \mathcal{F}_{\tau}^{-1}\left(\bm x_{\tau}\right) (\nabla \times \ddot{\bm{x}} _{\tau}) d\tau \Big],
 \end{equation}
where \(\ddot{\bm{x}} _{\tau}\) represents the acceleration at position \(\bm x\) and time \(\tau\).

\subsection{Viscosity}
Based on the basic vorticity formula, the viscosity can be written as a \textit{path integral} multiplied by the forward map Jacobian \(\mathcal{F}\):
\begin{equation}
\label{eq:gamma_visc}
\bm{\Gamma}_{\nu, t} \;=\; \mathcal{F}_{t}\,\int_{0}^{t} 
  \mathcal{T}_{\tau} \nu \Delta \bm{\omega} \Big(\phi_{\tau}\bigl(\psi_{t}(\bm{x})\bigr),\,\tau\Big)\,d\tau,
\end{equation}
where we actually integrate along the characteristic trajectory \(\gamma(\tau) = \phi_{\tau}\bigl(\psi_{t}(\bm{x})\bigr)\) of a particle whose location at time \(t\) is \(\bm x\). By integrating from \(0\) to \(t\), we accumulate the viscous effect along the particle path.
\subsection{External Forces}
Similarly, for any other external forces \(\bm f\) such as the gravity, they can be written as:
\begin{equation}
\label{eq:gamma_f}
\bm{\Gamma}_{\bm f, t} \;=\; \mathcal{F}_{t}\,\int_{0}^{t} 
  \mathcal{T}_{\tau} \left(\nabla \times \bm f\right) \Big(\phi_{\tau}\bigl(\psi_{t}(\bm{x})\bigr),\,\tau\Big)\,d\tau.
\end{equation}
In terms of implementation, we compute the viscosity and external forces for each particle at the current time \(\tau\), multiply it with \(\mathcal{T}_{\tau}\), and add it back to the particle's initial vorticity:
\begin{equation}
\label{eq:accumulate_init_vort}
    \bm \omega (\bm \psi(\bm x), 0) \;=\; \bm \omega (\bm \psi(\bm x), 0) + \int_{0}^{t} \mathcal{T}_{\tau} (d \Gamma) \Big(\phi_{\tau}\bigl(\psi_{t}(\bm{x})\bigr),\,\tau\Big),
\end{equation}
where \(d \Gamma\) is an infinitesimal element which stands for:
\begin{equation}
\label{eq:dgamma}
    d \Gamma \;=\; d \tau \cdot (\omegavisc{} + \nabla \times f + \bm \omega_{pen}).
\end{equation}
%The penalization term \(\bm \omega_{pen}\) will be explained in Section \ref{subsec:noslip}.

%% file: sec7_time_int.tex
\section{TIME INTEGRATION}
The main algorithm is summarized here and outlined in Alg.~\ref{alg:main}.
\begin{enumerate}[leftmargin=*]
    \item \textbf{Reinitialize Long-range Map (Step 4-7)}. \\ 
    Every $n^L$ steps, we redistribute particles uniformly in the whole computational domain. Then, we reinitialize each particle's initial vorticity \(\initvortp{}\) by a G2P process from current grid vorticity \(\currvortg{}\) according to Eq.~\eqref{eq:g2p_vort}. \rev{Finally,} all flow map Jacobians $\mathcal{T}^p_{[a,b]}$, $\mathcal{T}^p_{[b,c]}$ , $\mathcal{F}^p_{[a,b]}$, $\Fbc{}$ are reset to identity.
    % \begin{equation}
    % \label{eq:reinit_imp_long}
    %     \begin{dcases}
    %         \bm m_{q} \gets \sum_i w_{iq} \bm u_i, \\
    %         \mathcal{T}_{[q, r]} \gets \bm{I}.
    %     \end{dcases}
    % \end{equation}
    
    \item \textbf{Reinitialize Short-range Map (Step 10-12)}. \\ Every $n^S$ steps, each particle's middle vorticity \(\midvortp{}\) and its gradeint \(\nabla \midvortp{}\) is reinitialized by a G2P process, as depicted in Eq.~\eqref{eq:g2p_vort} and \eqref{eq:g2p_vort_grad}. Additionally, the former segments of the Jacobians \(\Tab{}, \Fab{}\) are set to \(\Tac{}, \Fac{}\), respectively. The latter segments of the Jacobians \(\Tbc{}, \Fbc{}\) are set to identity matrix, and the flow map Hessian \(\gradFbc{}\) is set to zero.

    \item \textbf{CFL Condition and Mesh Operations (Step 14-16)}. \\ We first compute $\delta t$ with velocity field \(\currvelg{}\) and the CFL number. Then, according to the current time, we linearly interpolate the current solid object's pose from the given pose sequences. Moreover, we compute the velocity for every vertex of the mesh by considering the current and previous vertices' positions.

    \item \textbf{Voxelization (Step 17)}. \\ 
    We voxelize the solid surface to obtain the surface mask \(\surfmask{}\). By aggregating vertex velocities of intersecting triangles, we compute normal and tangential velocities \(\normv{}\) and \(\tanv{}\). The full solid mask \(\solidmask{}\) is then flood-filled, and subtracting \(\surfmask{}\) yields the interior mask \(\inmask{}\).

    \item  \textbf{Fluid Area Fractions (Step 18)}. \\ 
    We compute non-trivial fluid fractions only on voxels marked by \(\surfmask{}\). Using a Monte Carlo approach, we sample each face to estimate fluid fractions based on the winding number.

    \item \textbf{Midpoint Method (Step 19)}. \\ 
    The midpoint method (detailed in the supplementary materials), as recommended in \cite{nabizadeh2022covector}, is used for estimating the midpoint velocity and marching the flow map quantities.
    
    \item \textbf{Particle Advection and Flow Map Quantities Evolution (Step 20)}. \\
    An RK4 scheme  (see supplementary material) is used for marching the particle positions \(\xp{}\), the latter segments of the Jacobians \(\Tbc{}\),  \(\Fbc{}\), and the Hessian \(\gradFbc{}\).
    %outlined in Alg.~\ref{alg:march}.

    \item \textbf{Vorticity Advection and P2G (Step 21)}. \\
    As outlined in Alg.~\ref{alg:advect_p2g}, we first obtain \(\Fac{}\), and \(\Tac{}\) by Jacobian connection (Eq.~\eqref{eq:connect_Jacobian}). We then advect \(\currvortp{}\) via \(\Fac{}\) (Eq.~\eqref{eq:evolve_omega}) and evolve \(\nabla \currvortp{}\) by Eq.~\eqref{eq:evolve_grad_omega}. Finally, the updated vorticity is transferred to the grid using P2G (Eq.~\eqref{eq:vort_P2G}).

    \item \textbf{\revv{Apply Forces and Diffusion (Step 22)}}. \\
    \revv{External forces and the viscosity term are computed as \(\nabla \times f\) and \(\nu \Delta \currvortg{}\), and applied to the current vorticity on the grid \(\currvortg{}\).}

    \item \textbf{Cut-cell No-through Velocity Reconstruction (Step 23-25).} \\
    Based on the current vorticity on \rev{the grid} \(\currvortg{}\), we reconstruct the cut cell no-through velocity \(\currvelg{}\) by solving the vector potential and the harmonic function. This process is explained in details in Section \ref{subsec:cutcell_nothrough}.

    \item \textbf{Accumulate Path Integral (Step 26).} \\
    As outlined in Alg.~\ref{alg:acc_path}, the path integral of the penalty vorticity, viscosity effect and all external forces stored on the initial particle vorticity \(\initvortp{}\) is updated by the current temporal increment \(\currdeltagammag{}\), multiplied by the backward Jacobian \(\Tac{}\). Details are in Section \ref{sec:noslip} and \ref{sec:visc_ext_force}.
\end{enumerate}

\begin{algorithm}
\caption{Time Integration}
\label{alg:main}
\begin{flushleft}
    \textbf{Initialize:} Velocity field $\bm{u}$, Vorticity field $\bm{\omega}$, Particle positions $\bm{x}_p$, Simulation parameters.
\end{flushleft}
\begin{algorithmic}[1]
\For{$m$ in total steps}
    \State $j \gets m \Mod {n^L}$;
    \State $l \gets m \Mod {n^S}$;
    \If{$j$ = 0}
    \State Uniformly distribute particles
    \State Reinitialize $\bm \omega_a^p$  with \(\bm \omega^g_c\)  by G2P
    \State Reset $\mathcal{T}^p_{[a,b]}$, $\mathcal{T}^p_{[b,c]}$ , $\mathcal{F}^p_{[a,b]}$ and $\mathcal{F}^p_{[b,c]}$ to identity matrix
    \EndIf
    \If{$l$ = 0}
    %\State \(\bm \omega_c^g \gets \nabla \times \bm u^g_c\)
    \State Reinitialize $\bm \omega_b^p$, \(\nabla \bm \omega_b^p\) with \(\bm \omega_c^g\) by G2P

    \State Update $\mathcal{T}_{[a, b]}^p$ and $\mathcal{F}_{[a, b]}^p$
    \State Reset $\mathcal{T}_{[b, c]}^p$, $\mathcal{F}_{[b, c]}^p$ to Identity, \(\nabla \mathcal{F}^p_{[b, c]}\) to zero 
    \EndIf
    \State Compute current time step $\delta t$ with $\bm{u}^g_c$ and CFL number

    \State Interpolate current mesh \(\mathcal{M_{\text{curr}}} \gets (\mathcal{V_\text{curr}}, Faces) \)
    \State Compute mesh velocity \(\bm u_{\mathcal{M}} \gets (\mathcal{V_{\text{curr}}} - \mathcal{V_{\text{prev}}}) / \delta t \)

    %\State $\textit{BVH} \gets \textbf{UpdateBVH}(\mathcal{M_\text{curr}})$
    \State $\chi_S, \chi_S^{\text{surf}}, \chi_S^\text{in}, \bm u_{Sn}^g, \bm u_{St}^g \gets \textbf{Voxelize}(\mathcal{M_\text{curr}}, \bm u_{\mathcal{M}})$ 
    \Statex \Comment{see supp.}
    
    \State Fluid fractions $\bm \alpha^g \gets \textbf{ComputeFractions}(\mathcal{M_\text{curr}}, \chi_S^\text{surf})$
    \Statex \Comment{\rev{see supp.}}
    
    \State $\bm{u}_\text{mid}^g \gets$ \textbf{MidPoint}$(\bm u^g_c, \bm{\omega}^g_c, \delta t)$ \Comment{see supp.}
    
    \State $\bm x^p, \mathcal{T}_{[b, c]}^p, \mathcal{F}_{[b, c]}^p, \nabla \mathcal{F}_{[b, c]}^p  \gets$ \textbf{March}$(\bm{u}_\text{mid}^g)$ \Comment{see supp.}

    \State \(\currvortg{} \gets \) \textbf{AdvectAndP2G} (\(\textit{Jacobian and Hessian}, \initvortp{}\)) \Comment{Alg. \ref{alg:advect_p2g}}

    \State \revv{\(\currvortg{}, d \Gamma_f, d \Gamma_{\nu} \gets \) \textbf{ApplyForcesAndDiffusion} (\(\currvortg{}\))}
    
    %\State Reconstruct \(\bm u_c^g \gets\) \textbf{CutcellVelRecon}(\(\bm \omega_c^g, \bm \alpha^g, \bm u_{Sn}^g, \chi_S^\text{in}\))
    \State Solve solenoidal vortical component \( \bm{u}^g_{\omega, c} \gets \nabla \times \Delta^{-1} (-\bm \omega^g_c) \)
    \Statex \Comment{Eq. \eqref{eq:solve_psi_numerical}}

    \State Solve harmonic function \(\Phi^g_c\) with \(\bm \alpha^g, \bm u_{Sn}^g, \chi_S^\text{in}\) \Comment{Eq. \eqref{eq:harmonic_solve_cutcell_ours}}

    \State Obtain final velocity on \rev{the grid} \(\bm u^g_c \gets \bm{u}^g_{\omega, c} - \nabla \Phi^g_c\)
    %\Statex \Comment{Alg. \ref{alg:cutcell_vel_recon}}
    % \State \(\bm \omega = \nabla \times \bm u\)
    \State \(\bm \omega^g_c, \bm \omega_a^p \gets \) \textbf{AccPathInt} (\( \bm u_c^g, \bm u_{St}^g, \revv{d \Gamma_f, d \Gamma_{\nu}, \initvortp{}}, \bm x^p\)) \Comment{Alg. \ref{alg:acc_path}}

\EndFor
\end{algorithmic}
\end{algorithm}

\begin{algorithm}
\caption{Advection and P2G}
\label{alg:advect_p2g}
\begin{flushleft}
    \textbf{Input:} \(\text{Flow map Jacobians } \Tab{}, \Tbc{}, \Fab{}, \Fbc{}\), \\ \( \text{Flow map Hessian } \gradFbc{}, \text{ Initial particle vorticity } \initvortp{} \) \\
    \textbf{Output:} Current grid vorticity \(\currvortg{}\)
    
\end{flushleft}
\begin{algorithmic}[1]
    \State Compute $\mathcal{F}_{[a, c]}^p \gets \mathcal{F}_{[b, c]}^p \mathcal{F}_{[a, b]}^p$
    \State Compute $\mathcal{T}_{[a, c]}^p \gets \mathcal{T}_{[a, b]}^p \mathcal{T}_{[b, c]}^p$ 

    \State Compute \(\bm \omega_c^p \gets \mathcal{F}_{[a, c]}^p \, \bm \omega_a^p\) \Comment{Eq. \eqref{eq:omega_evolve_adaptive}}
    \State Compute \(\nabla \bm \omega_c^p\) with \(\mathcal{F}_{[b, c]}^p\), \(\mathcal{T}_{[b, c]}^p\), \(\nabla \mathcal{F}_{[b, c]}^p\), \(\bm \omega_b^p\) and \(\nabla \bm \omega_b^p\) 
    \Statex \Comment{Eq. \eqref{eq:gradomega_evolve_adaptive}}

    \State Compute \(\bm \omega_c^g\) with \(\bm \omega_c^p\) and \(\nabla \bm \omega_c^p\) by P2G \Comment{Eq. \eqref{eq:vort_P2G}}
\end{algorithmic}
\end{algorithm}

% \begin{algorithm}
% \caption{Cut-cell No-through Velocity Reconstruction}
% \label{alg:cutcell_vel_recon}
% \begin{flushleft}
%     \textbf{Input:} Current vorticity \(\currvortg{}\), Fluid fraction \(\bm \alpha^g\), Normal solid velocity \(\bm u_{Sn}^g\) and Interior mask \(\chi_S^\text{in}\) \\
%     \textbf{Output:} Final velocity \(\bm u_c^g\)
    
% \end{flushleft}
% \begin{algorithmic}[1]
%     \State Solve solenoidal vortical component \( \bm{u}^g_{\omega, c} \gets \nabla \times \Delta^{-1} (-\bm \omega^g_c) \)
%     \Statex \Comment{Eq. \eqref{eq:solve_psi_numerical}}

%     \State Solve harmonic function \(\Phi^g_c\) with \(\bm \alpha^g, \bm u_{Sn}^g, \chi_S^\text{in}\) \Comment{Eq. \eqref{eq:harmonic_solve_cutcell_ours}}

%     \State Obtain final grid velocity \(\bm u^g_c \gets \bm \bm{u}^g_{\omega, c} - \nabla \Phi^g_c\)
%     %\Comment{Eq. \eqref{eq:sub_grad_Phi}}
% \end{algorithmic}
% \end{algorithm}

\begin{algorithm}
\caption{Accumulate Path Integral}
\label{alg:acc_path}
\begin{flushleft}
    \textbf{Input:} Cut-cell velocity \(\bm u^g_c\), Tangential solid velocity \(\bm u_{St}^g\),  \\ 
\revv{Force increment \(d \Gamma_f\), Viscosity increment \(d \Gamma_{\nu}\)}, Initial particle vorticity \(\initvortp{}\), Particle current positions \(\bm x_p\), \(\delta t\) 
    \\ \textbf{Output:} Current vorticity \(\bm \omega_c^g\), Initial particle vorticity \(\bm \omega_a^p\)
    
\end{flushleft}
\begin{algorithmic}[1]
    \State \(\bm \omega_c^g \gets \nabla \times \bm u^g_c\)

    \State Compute penalty velocity \(\bm u_{pen}^g\) with \(\bm u^g_c\) and \(\bm u_{St}^g\) \Comment{Eq. \eqref{eq:upen_def}}
    \State Compute penalty vorticity \(\bm \omega^g_{pen}\) \Comment{Eq. \eqref{eq:pen_omega_def}}

    %\State Compute viscosity \(\omegavisc{}_c^g\) and external forces \(\bm f^g\)

    \State Obtain increment \revv{\(\delta \Gamma^g \gets \delta t \cdot \bm{\omega}_{pen}^g + d \Gamma_f + d \Gamma_{\nu}\)}
    \Statex \Comment{Eq. \eqref{eq:dgamma}}

    \For{each particle \textit{p} \textbf{in parallel}}
        \State \(\delta \Gamma (p) \gets \textbf{Interpolate}\bigr(\delta \Gamma^g, \bm x^p(\textit{p})\bigr)\)
        \State \(\initvortp{}(\textit{p}) \gets \initvortp{}(\textit{p}) + \Tac{} \delta \Gamma (p)\) \Comment{Eq. \eqref{eq:accumulate_init_vort}}
    \EndFor

    % \For{each grid position \textit{i} \textbf{in parallel}}
    %     \State \(\delta \Gamma (i) \gets \textbf{Interpolate}\bigr(\delta \Gamma^g, \textbf{gridPos}(\textit{i}\,)\bigr)\)
    %     \State \(\currvortg{}(\textit{i}\,) \gets \currvortg{}(\textit{i}\,) + \delta \Gamma (i)\) 
    % \EndFor
    %\State \(\currvortg{}(\textit{i}\,) \gets \currvortg{}(\textit{i}\,) + \delta \Gamma^g\) 

\end{algorithmic}
\end{algorithm}

%% file: flowmap_compare_failurepoint.tex
\subsection{Vorticity Preservation with Long-term Flow Maps}
\label{subsec:vpfm_validate}
% \begin{figure*}
% \setlength{\abovecaptionskip}{0.05in}
%     \centering
%     \includegraphics[width=1.03\textwidth]{exp0_dir/combined_plot_final_energy_n_leq_60.pdf}
%     \caption{\sinan{Ours Hessian Comparison}}
%     \label{fig:exp0_w}
% \vspace{-0.05in}
% \end{figure*}
% We will validate that VPFM can achieve a robust, long-term flow map, therefore enhance its vorticity preservation ability by (1) comparing methods with the same challenging long flow map, (2) comparing methods with their optimal flow map length, (3)  a comprehensive 3D analysis for varying flow map lengths for each method.
\rev{We validate that VPFM can achieve a robust, long-term flow map, which in turn enhances its vorticity preservation ability. This is demonstrated through: (1) a comparison under the same challenging long flow map, (2) a comparison using each method’s optimal flow map length, and (3) a comprehensive 3D comparison across varying flow map lengths for different methods.}

\subsubsection{Comparison with the same Challenging Long Flow Map}
\label{subsubsec:compare_same_long_flow_map}
\begin{figure}
    \centering
    \includegraphics[width=0.5\textwidth]{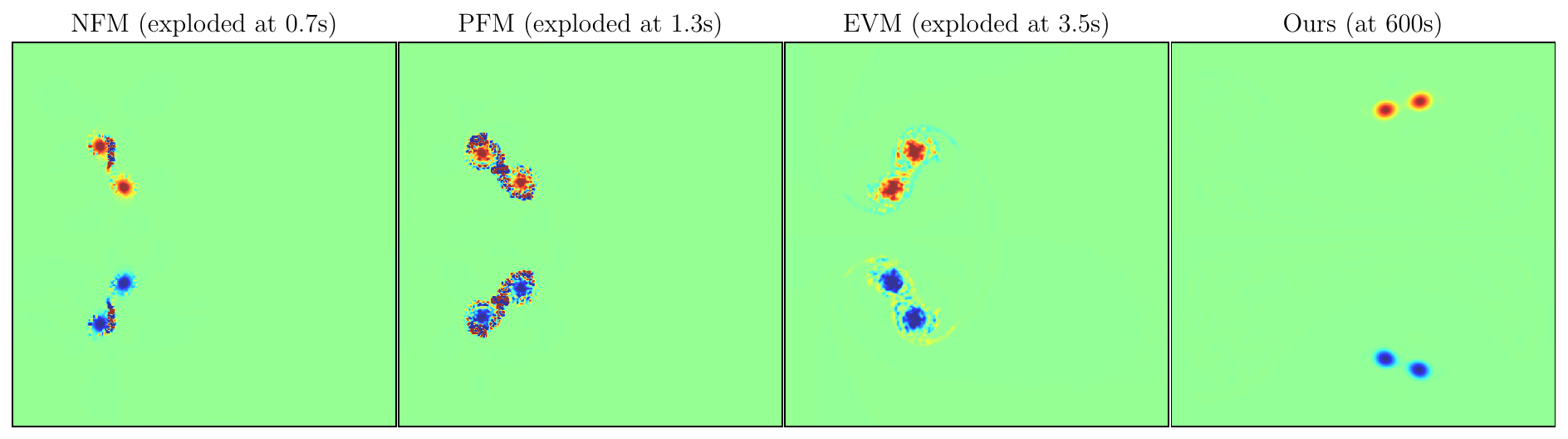}
    % \vspace{-0.7cm}
    \caption{2D leapfrog under a challenging flow map length \(n^L = 240\).}
    \label{fig:2Dleapfrog_long}
\end{figure}

\begin{figure}
    \centering
    \includegraphics[width=0.51\textwidth]{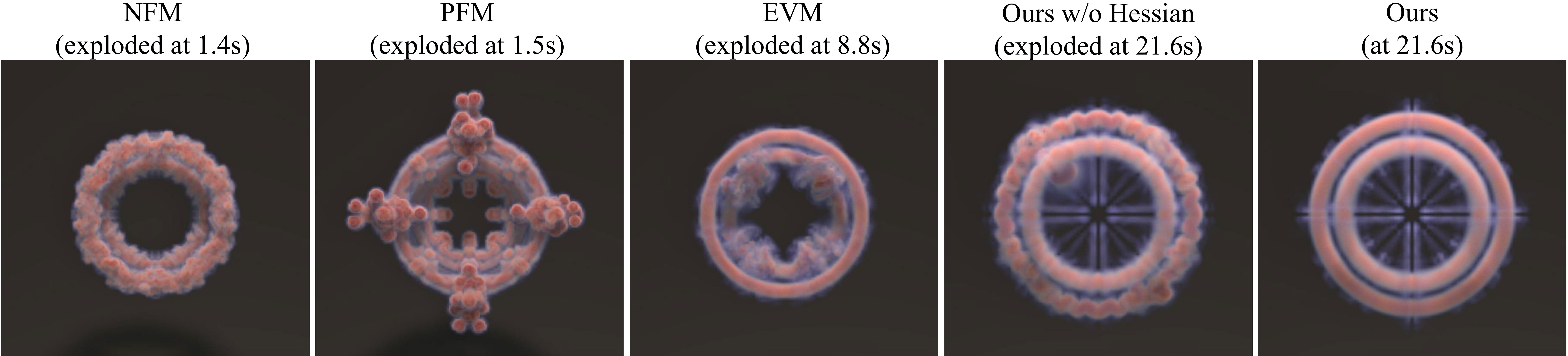}
    % \vspace{-0.6cm}
    \caption{3D leapfrog under a challenging flow map length \(n^L = 100\).}
    \label{fig:3Dleapfrog_long}
\end{figure}

In this part, we evaluate methods on a challenging long flow map. Our method remains stable for an extended time frames, whereas others either explode, or cannot preserve the vortex structure.
\paragraph{2D and 3D Leapfrog under a Challenging Long Flow Map}
As shown in Figure~\ref{fig:2Dleapfrog_long} (2D leapfrog) and \ref{fig:3Dleapfrog_long} (3D leapfrog), with results summarized in Table~\hyperlink{tab:2D3Dleapfrog}{3}, under the same challenging long flow map (240 for 2D and 100 for 3D, compared with a maximum of 20 for all previous flow-map based methods), \rev{our} method demonstrates significantly enhanced robustness, enabling substantially longer stable simulation times while effectively preserving vortex structures. In the 2D leapfrog setting, our method run \textbf{indefinitely} without exploding, whereas NFM \cite{deng2023neural}, PFM \cite{zhou2024eulerian} and EVM \cite{wang2024eulerian} exploded at 0.7s, 1.3s, and 3.5s respectively. For the 3D leapfrog setting, we achieve around \textbf{30.1$\times$}, \textbf{28.1$\times$}, and \textbf{4.8$\times$} longer stable simulation than NFM, PFM, and EVM, respectively, and our evolved Hessian nearly \textbf{doubles} the stable simulation time. \rev{The detailed initial configurations for the leapfrog experiments are described later in Section \ref{subsubsec:compare_optimal_flowmap}.}
% \begin{figure*}
%     \centering
%     \includegraphics[width=0.8\textwidth]{3D_experiments/2D3Dleapfrog_longflowmap_compare_evmold.pdf}
%     \caption{2D and 3D leapfrog with the same challenging long flow map length (240 for 2D and 100 for 3D).}
%     \label{fig:2D3D_leapfrog}
% \end{figure*}

% \begin{table}[h]
% \centering\small
% \caption{Summary of 2D and 3D leapfrog explosion time under a challenging flow map length (240 for 2D and 100 for 3D). Note that the Hessian term vanishes in 2D (discussed in Section~\ref{para:vpfm_2D}).}
% \begin{tabular}{l@{\hspace{0mm}}c@{\hspace{2mm}}c@{\hspace{2mm}}c}
% \hlineB{3}
% Method & 2D Explosion Time (s) &  3D Explosion Time (s) \\
% \hlineB{2}
% NFM & 0.7 & 1.4 \\
% \hlineB{2}
% PFM & 1.3 & 1.5 \\
% \hlineB{2}
% EVM & 3.5 & 8.8 \\
% \hlineB{2}
% Ours w/o Hessian & - & 21.6 \\
% \hlineB{2}
% Ours & $\bm \infty$ & \textbf{42.2} \\
% \hlineB{3}
% \end{tabular}
% \label{tab: 2D3Dleapfrog}
% \end{table}

\label{tab:2D3Dleapfrog}
\begin{figure}[h]
\centering
\begin{minipage}[t]{0.6\linewidth}
\centering\small
\begin{tabular}{l@{\hspace{1mm}}c@{\hspace{2mm}}c}
\toprule
Method & 2D Expl. Time & 3D Expl. Time \\
\midrule
NFM & 0.7 & 1.4 \\
PFM & 1.3 & 1.5 \\
EVM & 3.5 & 8.8 \\
Ours w/o Hess. & -- & 21.6 \\
Ours & $\infty$ & \textbf{42.2} \\
\bottomrule
\end{tabular}
\end{minipage}%
\hfill
\begin{minipage}[t]{0.3\linewidth}
\vspace{-1.45cm}
\hypertarget{tab:2D3Dleapfrog}{}
\captionof{table}{Summary of 2D and 3D leapfrog explosion times under a challenging flow map length (240 for 2D and 100 for 3D). The Hessian term vanishes in 2D (see Section~\ref{para:vpfm_2D}).}
\end{minipage}
\end{figure}

% \vspace{-0.8cm}

\begin{figure}
    \centering
    \includegraphics[width=0.5\textwidth]{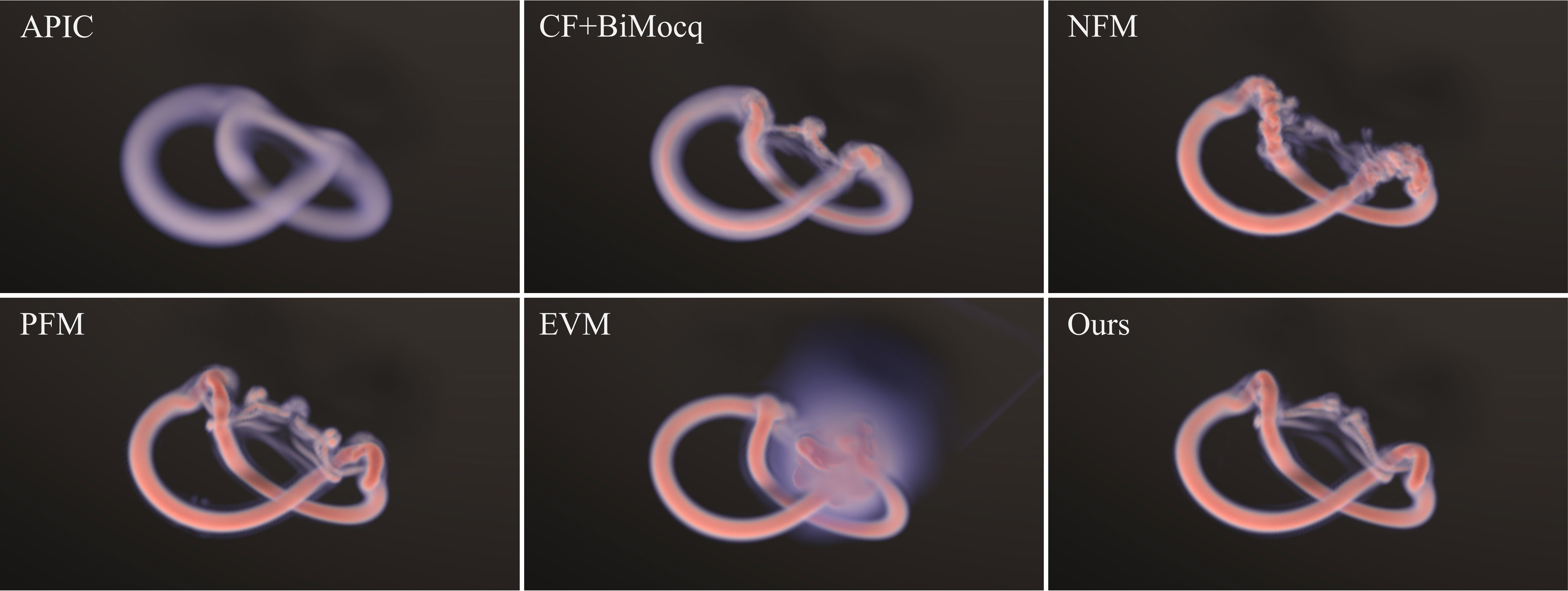}
    \caption{Comparison of Hopf link. At a relatively long flow map length (\(n^L = 30\)), our method effectively preserves the vortex ring structure and maintains smoothness at the twists, surpassing the performance of other methods. PFM produces comparable results. EVM explodes and becomes unstable early in the simulation. Although NFM preserves the vortex ring structure, it fails to maintain smoothness at the twists. CF+BiMocq \cite{nabizadeh2022covector, qu2019efficient}, and APIC \cite{jiang2015affine} are unable to effectively preserve the vortex ring structure.}
    \label{fig:hopflink_compare}
\end{figure}

\begin{figure}
    \centering
    \includegraphics[width=0.5\textwidth]{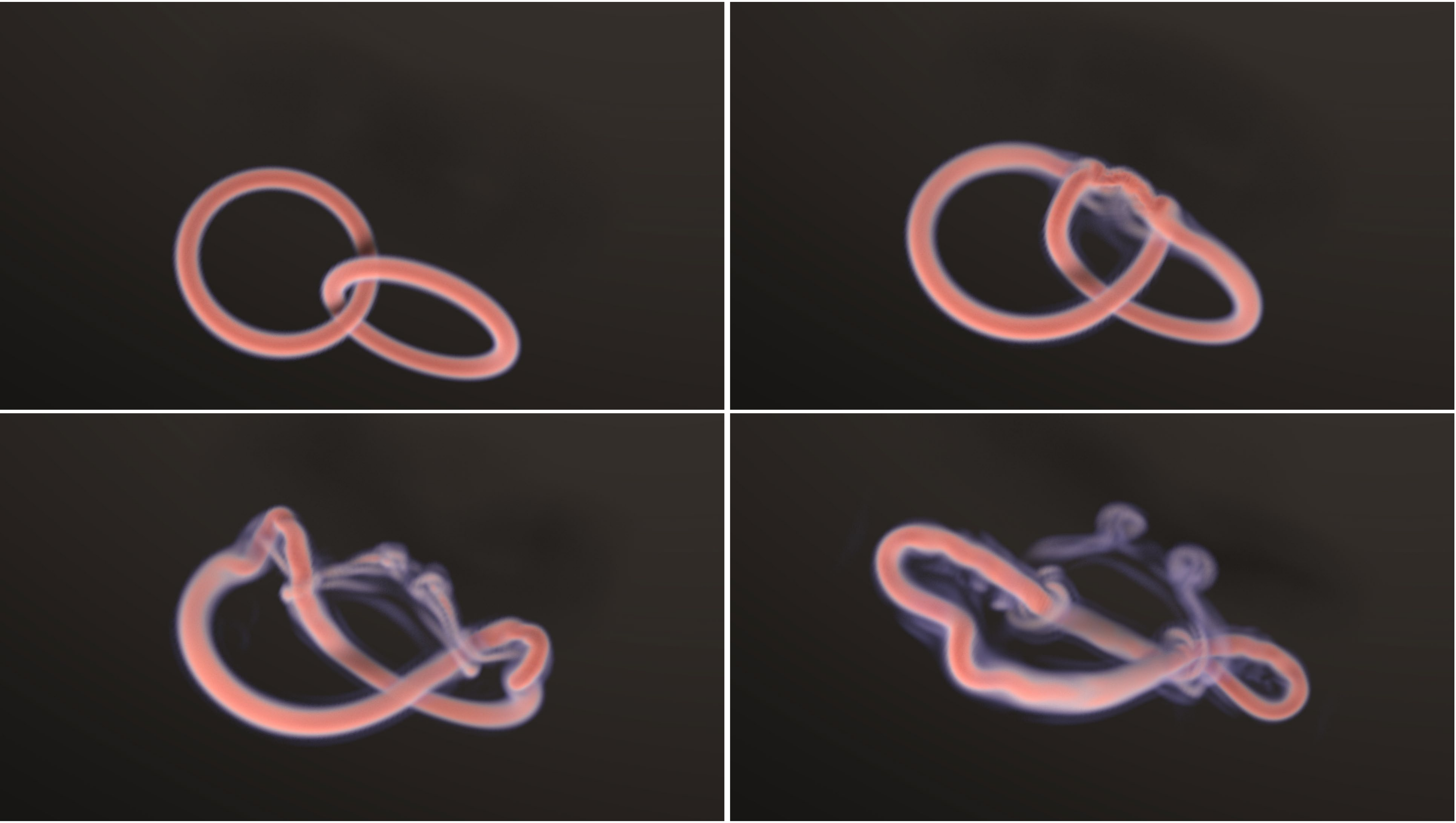}
    \caption{The famous Hopf link. Two vortex rings initially linked together naturally evolve to unknit themselves. Our results align closely with those reported in \cite{villois2020irreversible}.}
    \label{fig:hopflink}
\end{figure}

\paragraph{Hopf Link (3D)}
As shown in Figure~\ref{fig:hopflink}, initially, two vortex rings are linked together in a configuration known as the famous Hopf link. \rev{The first ring is centered at $(0.35, 0.5, 0.5)$ and lies in the $x$–$y$ plane, while the second is centered at $(0.65, 0.5, 0.5)$ and lies in the $x$–$z$ plane. Both rings have a major radius of $0.18$ and a minor radius (mollification support) of $0.0168$, and are initialized with opposite vorticity strengths of \(\pm 2 \times 10^{-2}\).} Over time, they untangle and reconnect to form a single vortex ring. In Figure~\ref{fig:hopflink_compare}, we compare different methods using the same, relatively long flow map length of 30\rev{. Our} method generates a smoother untangled vortex ring with fewer artifacts while effectively preserving the vorticity. Furthermore, our results are consistent with those reported in \cite{villois2020irreversible}.

\subsubsection{Comparison with each method's optimal flow map length}
\label{subsubsec:compare_optimal_flowmap}
Three comparisons are conducted, each using the optimal flow map length for the respective method, as shown in Table~\ref{tab:optimal_length}. Compared to the state-of-the-art, our method achieves \textbf{12} times longer flow map in 2D leapfrog, \textbf{2}-\textbf{3} times longer in 3D leapfrog, and \textbf{3.3} times longer in the 3D trefoil knot.

\begin{table}
\centering\small
\caption{Optimal flow map lengths (\(n^L\)) used in Section \ref{subsubsec:compare_optimal_flowmap}. Most values are adopted from the respective papers. For a fair comparison, we implemented an NFM version without the neural buffer and observed that \(n^L = 20\) caused the 2D leapfrog simulation to explode. Consequently, \(n^L = 10\) was used for the 2D leapfrog case and \(n^L = 12\) for the trefoil knot. For 3D leapfrog with EVM, we experimentally found that using \(n^L = 30\) is better than 20.}
\begin{tabular}{l@{\hspace{2mm}}c@{\hspace{2mm}}c@{\hspace{2mm}}c@{\hspace{2mm}}c}
\hlineB{3}
Method & 2D Leapfrog &  3D Leapfrog & Trefoil knot \\
\hlineB{2}
BFECC & 1 & - & -\\
\hlineB{2}
VIC & 5 & - & -\\
\hlineB{2}
APIC & 1 & 1 & 1\\
\hlineB{2}
CF+BiMocq & 5 & 5 & 5 \\
\hlineB{2}
NFM & 10 & 20 & 12 \\
\hlineB{2}
PFM & 20 & 20 & 12 \\
\hlineB{2}
EVM & 20 & 30 & 12 \\
\hlineB{2}
Ours & 240 & 60 & 40 \\
\hlineB{3}
\end{tabular}
\label{tab:optimal_length}
\end{table}

\begin{figure}
% \hspace{-0.28\textwidth}
\centering
\makebox[0pt][l]{\hspace{-0.25\textwidth}\includegraphics[width=0.52\textwidth]{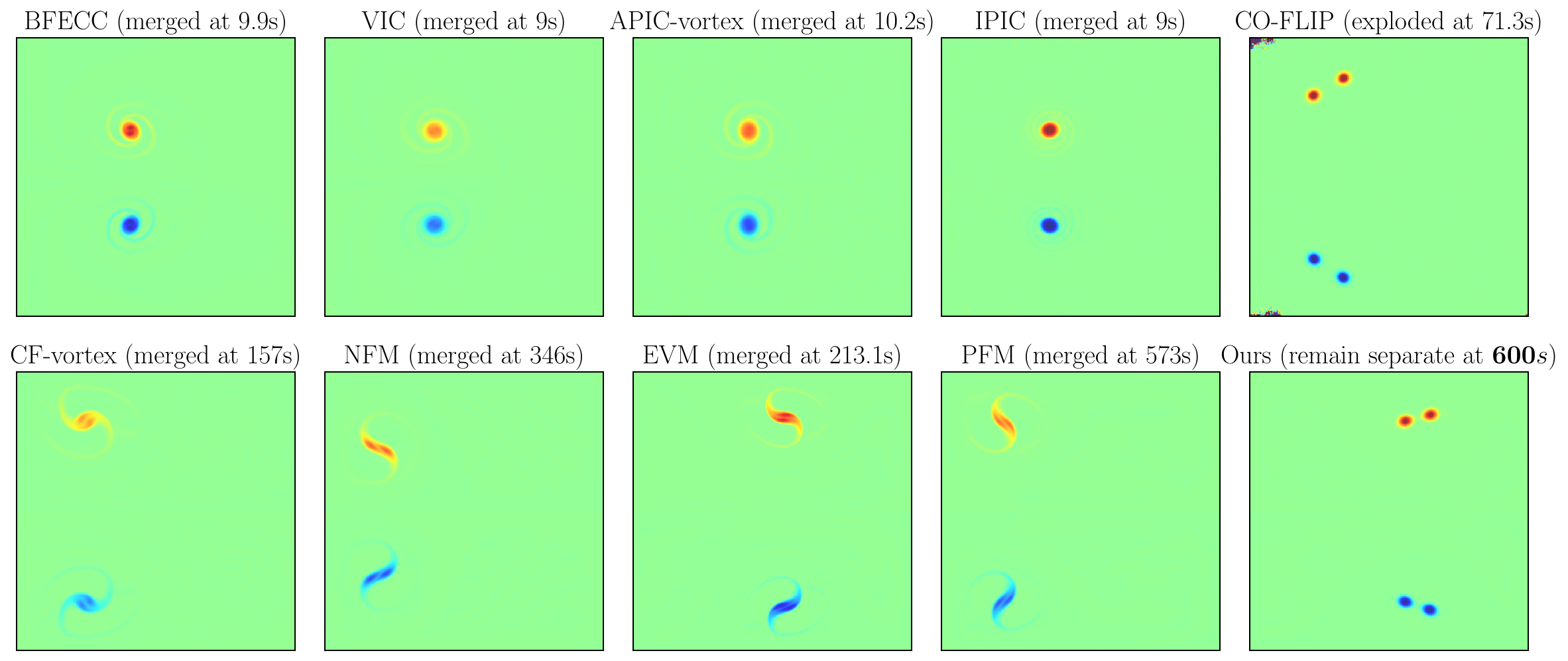}}
\caption{Comparison of 2D leapfrogging vortices. Time is indicated in simulation world time (frames × timestep). Our method successfully maintains the separation of two pairs of vortices at 600s, while PFM, EVM, NFM, CF-vortex and \rev{CO-FLIP} show noticeably weaker performance. APIC-vortex, VIC, and BFECC, \rev{IPIC} are unable to maintain vortex separation over an extended period. \rev{CF-vortex and APIC-vortex represent Covector Fluids and APIC carrying vorticity instead of impulse and velocity, respectively.}}
\label{fig:2d_leapfrog}
% \vspace{-0.2in}
\end{figure}

\begin{figure}[h]
    \centering
    \hspace*{-0.5cm}
    \includegraphics[width=0.52\textwidth]{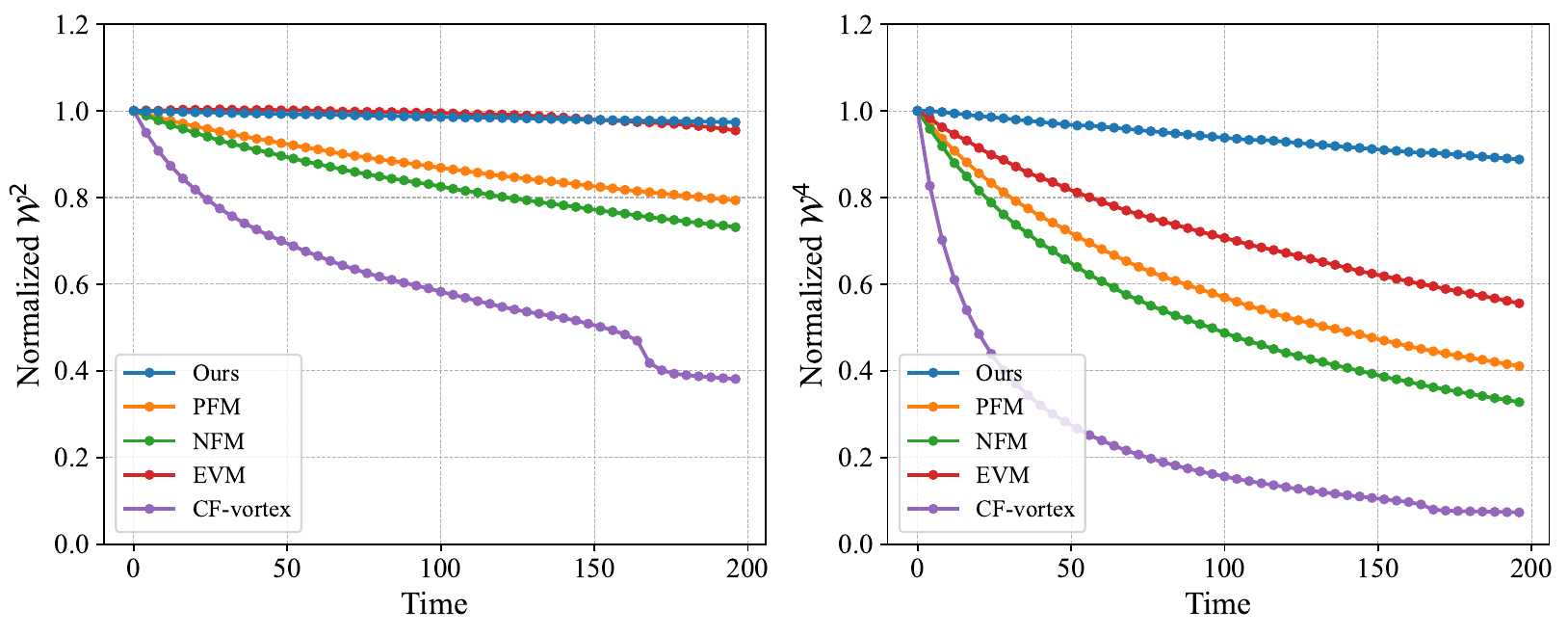}
    \caption{\rev{Casimir invariants in 2D leapfrog. We present the entropy (left), i.e., second moment of vorticity and fourth moment of vorticity (right).}}
    \label{fig:casimir_2D}
\end{figure}

\paragraph{Leapfrog (2D)}
\rev{Adopting the same initial setting as NFM \cite{deng2023fluid},} two pairs of vortices with opposite vorticity are released from the left side of the domain, engaging in a leapfrog-like motion. Ideally, this motion should continue indefinitely without dissipation. In our simulation, the vortex pairs remain separate for \textbf{613} seconds before merging, significantly surpassing the performance of other methods. The corresponding vortex structure maintaining times are as follows: PFM (573s), EVM (213.1s), NFM  (346s), the vortex version of Covector Fluids  (157s), \rev{CO-FLIP \cite{nabizadeh2024coflip} (71.3s)}, the vortex version of APIC (10.2s), \rev{IPIC \cite{sancho2024impulse} (9s)}, Vortex-In-Cell \cite{rossinelli2008vortex} (9s), and BFECC advection \cite{kim2005flowfixer} (9.9s), as illustrated in \figref{fig:2d_leapfrog}. \rev{To eliminate the influence of initial conditions, we also run VPFM using the initial velocity field given by CO-FLIP's implementation. Under this setup, VPFM is able to preserve vortex structures for over 600 seconds, compared to the 500 seconds reported in the CO-FLIP paper. Under our initial setting, we also evaluate the metrics introduced to graphics by CO-FLIP—namely, the \textit{Casimir invariants}, which are theoretically conserved throughout the simulation. Specifically, we report the normalized entropy (second moment of vorticity) and fourth moment of vorticity in Figure~\ref{fig:casimir_2D}. Our method is able to maintain the entropy almost perfectly and exhibits only a slight decrease in the fourth moment of vorticity, whereas several baseline methods struggle to preserve these conserved quantities.}

\paragraph{Leapfrog (3D)}

\begin{figure}[t]
    \centering
    \includegraphics[width=0.5\textwidth]{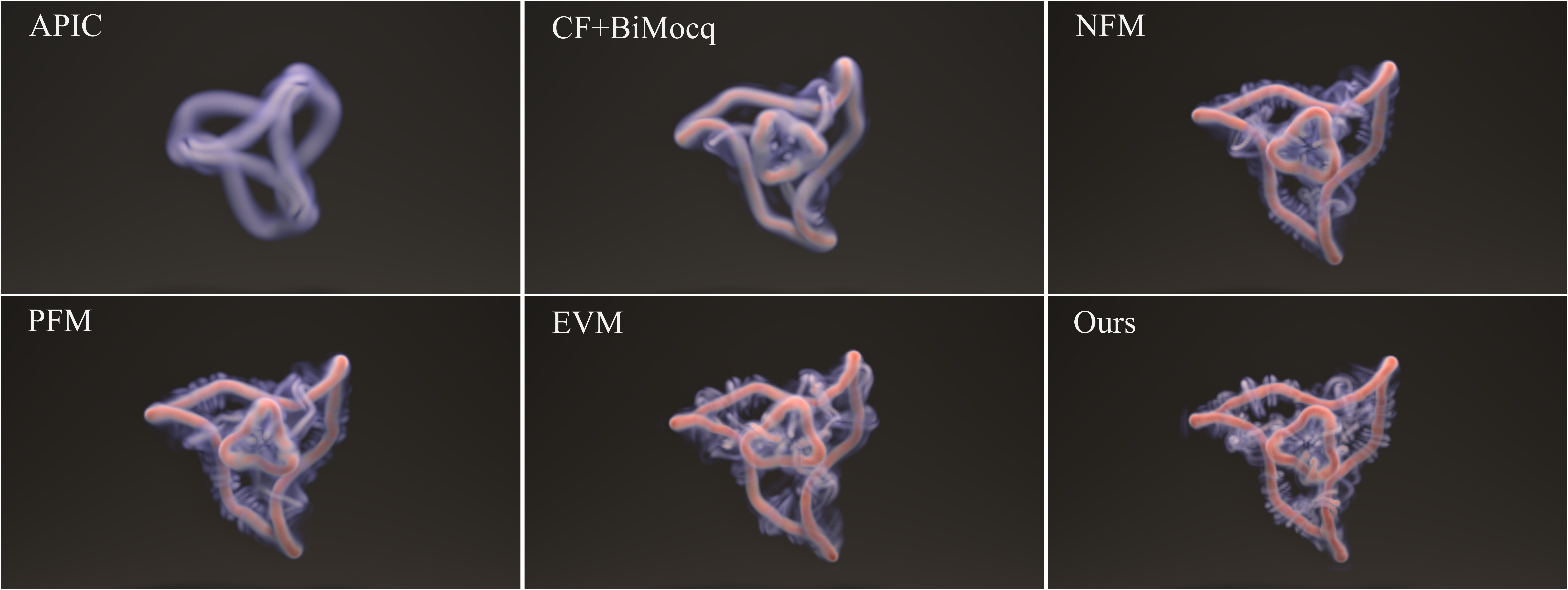}
    \caption{Front view of the trefoil knot experiment, demonstrating the vorticity preservation capability of our method, highlighted by the deeper color of the vortex tubes.}
    \label{fig:trefoil}
\end{figure}

% \vspace{-0.8cm}

\begin{figure*}
    \centering
    \includegraphics[width=1.01\textwidth]{3D_experiments/3D_leapfrogs_compare_7_large_small_final.pdf}
    \caption{Comparison of 3D leapfrog simulations across different methods in a shorter domain. Our method maintains the separation of the vortex rings even when they reach the right boundary, outperforming other methods, which merge the rings after at most five leaps.}
    \label{fig:3d_leapfrog_compare}
\end{figure*}

\begin{figure*}[h]
    \centering
    \includegraphics[width=1.01\textwidth]{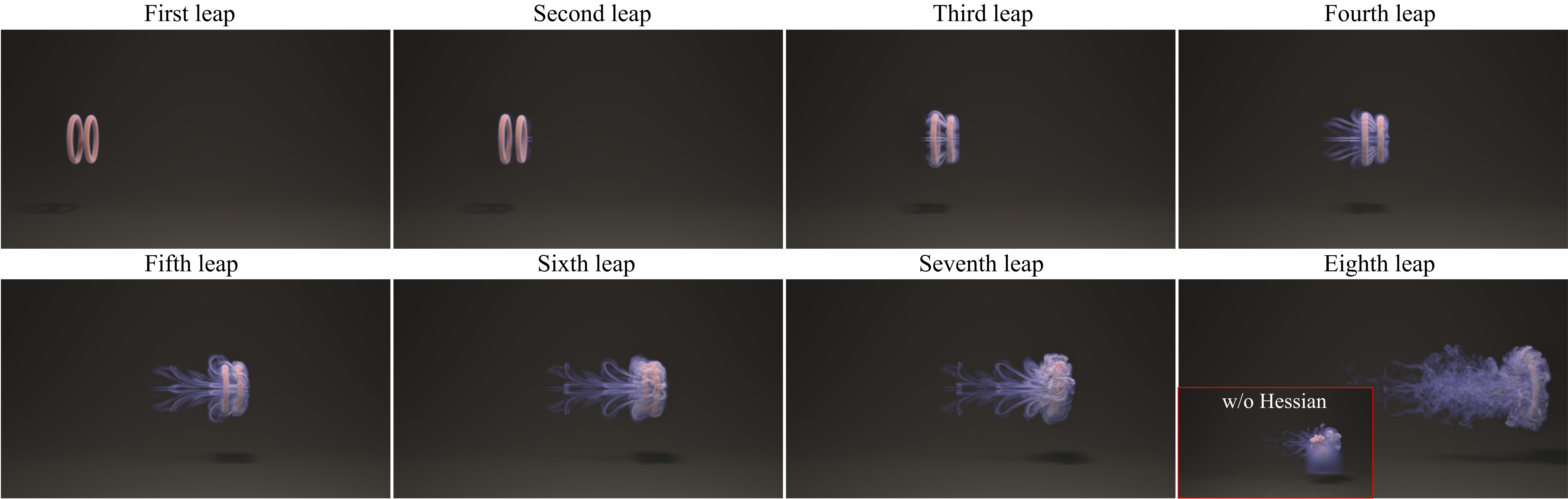}
    \caption{3D leapfrog simulation in a longer domain, using the same settings as the shorter domain comparison in Figure~\ref{fig:3d_leapfrog_compare}. The extended domain allows the vortex rings to complete \textbf{seven leaps} without hitting the right boundary.}
    \label{fig:3d_leapfrog}
\end{figure*}

 Analogous to the 2D leapfrog, in 3D, two vortex rings are released from the left side and undergo a leapfrogging-like motion that, ideally, continues indefinitely. \rev{The initial setup follows that of NFM, except that we shift the two rings slightly to the left, placing them at \(x = 0.1\) and \(x = 0.23125\), respectively, to allow for a longer simulation before the rings reach the right boundary.} Our method maintains the separation of the two rings through the \textbf{seventh leap} (shown in \figref{fig:3d_leapfrog}, surpassing other methods that can only do so for up to five leaps (see \figref{fig:3d_leapfrog_compare}). We note that the comparison is conducted in a shorter simulation domain to reduce experimental cost, our rings remain separated upon reaching the domain’s right boundary. To further validate our approach, we simulate in a longer domain with the same settings and \(\delta x\) (see \figref{fig:3d_leapfrog}), where the two vortex rings complete seven leaps without merging. Notably, our evolved Hessian method enhances the performance of PFM \cite{zhou2024eulerian} as well and one more leap is observed. Further details on this enhancement are provided in Section~\ref{para:enhance_pfm_with_hessian}. Optimal \(n^L\) used here are listed in Table~\ref{tab:optimal_length}.
 %The optimal flow map lengths for each method, as reported in their papers, are as follows: PFM \cite{zhou2024eulerian} (\(n^L = 20\)), EVM \cite{wang2024eulerian} (\(n^L = 30\)), NFM \cite{deng2023neural} (\(n^L = 20\)), CF+BiMocq \cite{nabizadeh2022covector, qu2019efficient} (\(n^L = 5\)), APIC \cite{jiang2015affine} (10.2s, \(n^L = 1\)). For EVM, we experimentally found that using \(n^L = 30\) is better than 20.

% \begin{wrapfigure}{r}{0.3\textwidth}
% \centering
% \includegraphics[width=0.3\textwidth]{3D_experiments/normalized_energy_curves.pdf}
%     \caption{Trefoil energy curve}
%     \label{fig:trefoil_energy}
% %\vspace{-0.1in}
% \end{wrapfigure}

% \begin{figure}
%     \centering
%     \includegraphics[width=0.4\textwidth]{3D_experiments/compare_energy_only_time_redfont_250.pdf}
%     \caption{Comparison of normalized energy curves in the trefoil knot across different methods, demonstrating that our method achieves better energy preservation over time through  long-term flow mapping.}
%     \label{fig:trefoil_energy}
% \end{figure}

\begin{figure}[h]
\centering
\begin{minipage}{0.6\linewidth}
    \includegraphics[width=\linewidth]{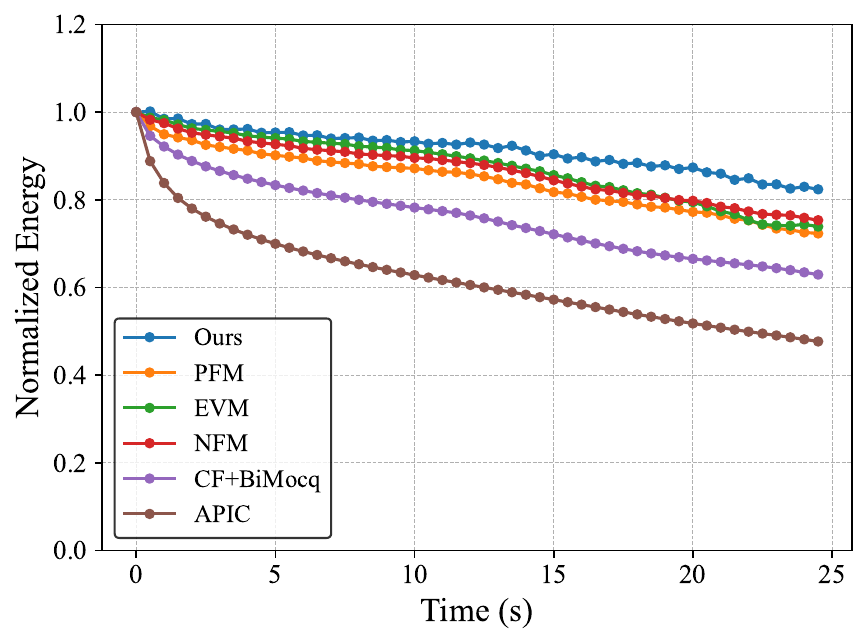}
\end{minipage}%
\hfill
\begin{minipage}{0.38\linewidth}
    \vspace{-2.4em}
    \captionof{figure}{Comparison of normalized energy curves in the trefoil knot across different methods, demonstrating that our method achieves better energy preservation over time through long-term flow mapping.}
    \label{fig:trefoil_energy}
\end{minipage}
\end{figure}

\paragraph{Trefoil Knot (3D)}
\label{para:trefoil}
The trefoil knot comparison is presented here, using the same setting as \citet{nabizadeh2022covector}. We compare qualitatively by \figref{fig:trefoil}, and quantitatively by the energy curve in \figref{fig:trefoil_energy}. It can be seen that via the long-term flow map (\(n^L=40\)), our method preserves vorticity and energy better than other methods, as evidenced by the deeper vorticity colors in the figure and the higher energy curve. A side view of this experiment, comparing the Hessian evolution (ours) and the Hessian interpolation \cite{zhou2024eulerian} under a same long flow map length (\(n^L = 40\)), is given in \figref{fig:dTdx}. Smoother vortex tubes and rings, and stabler simulation are achieved by our evolved Hessian. Optimal \(n^L\) used here are listed in Table~\ref{tab:optimal_length}.

\subsubsection{A Comprehensive 3D Analysis of Flow Map Lengths \(n^L\)}
\label{subsubsec:comprehensive_3D}
\begin{figure}[!t]
    \centering
    \setlength{\abovecaptionskip}{0.05in}

    \includegraphics[width=0.46\textwidth]{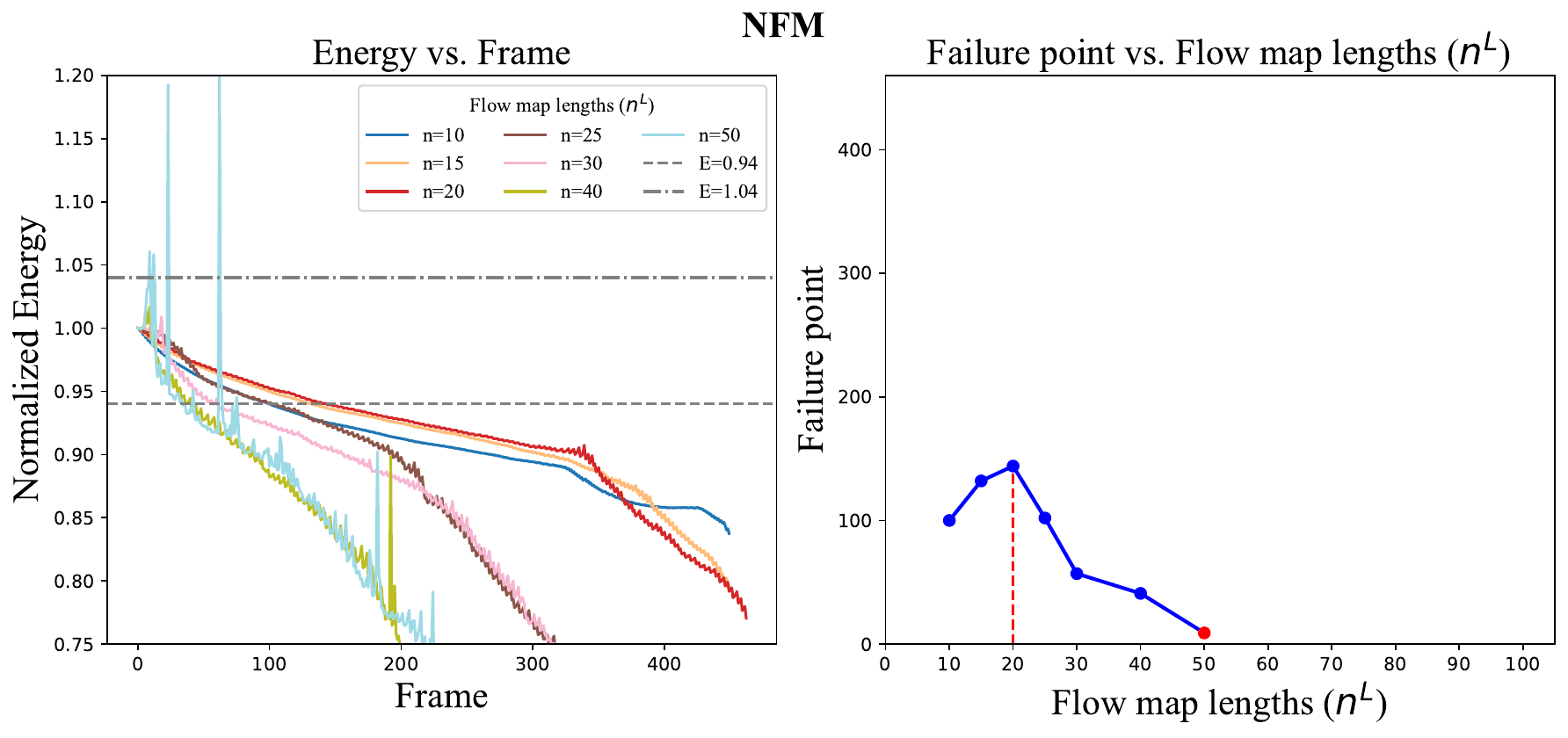}
    %\vspace{-0.05in}
    \label{fig:exp0_w_nfm}

    \includegraphics[width=0.46\textwidth]{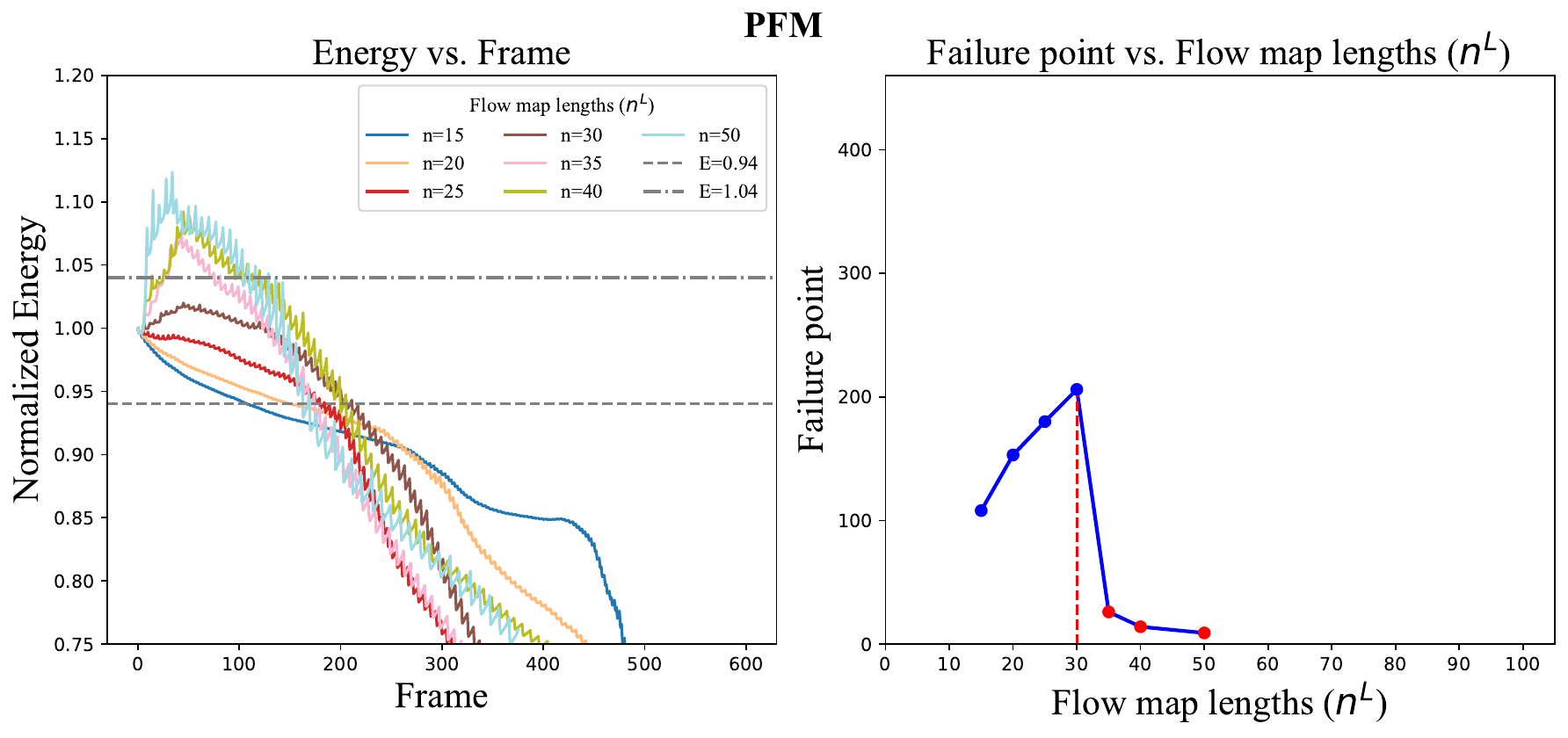}
    %\vspace{-0.05in}
    \label{fig:exp0_w_pfm}

    \includegraphics[width=0.46\textwidth]{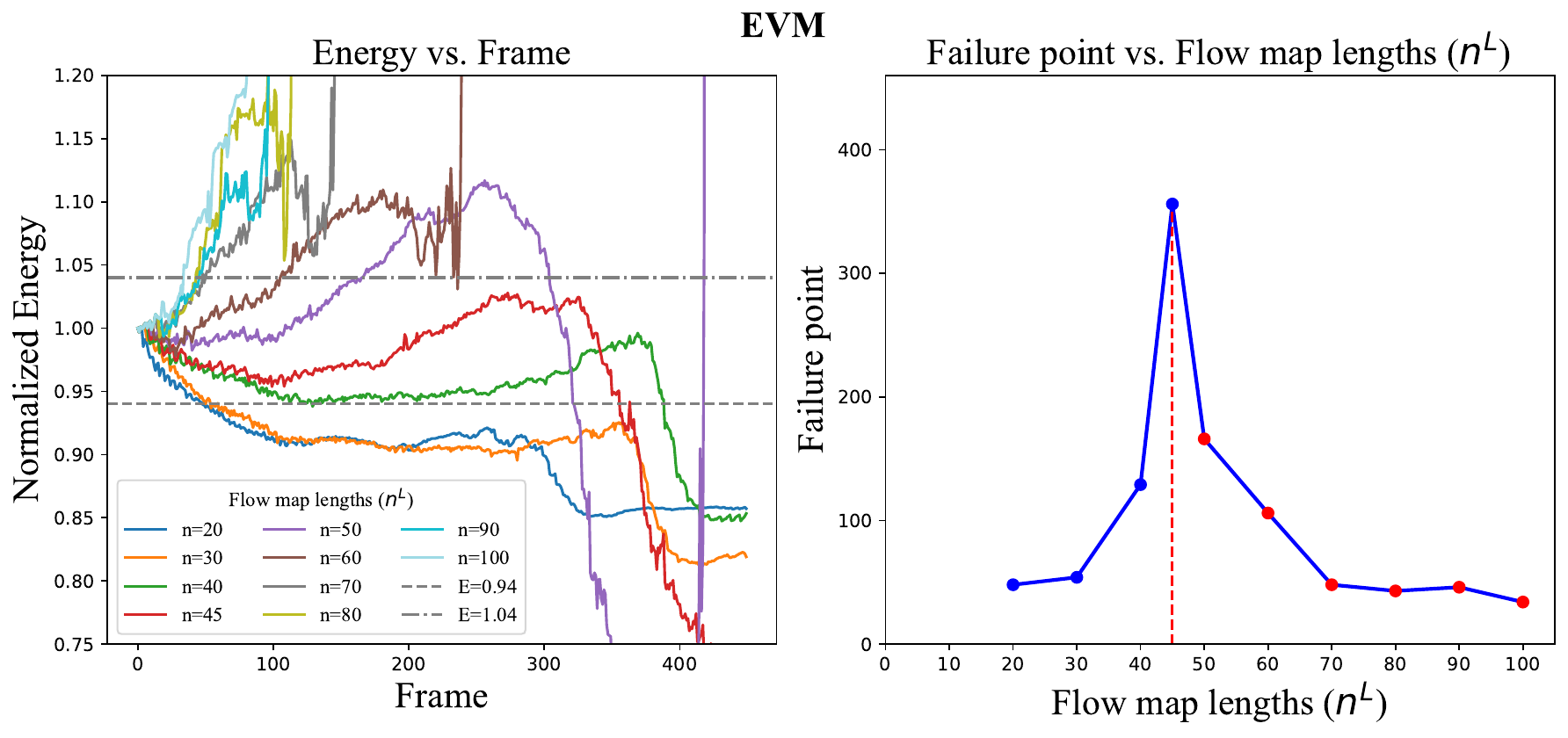}
    %\vspace{-0.05in}
    \label{fig:exp0_w_evm}

    \includegraphics[width=0.46\textwidth]{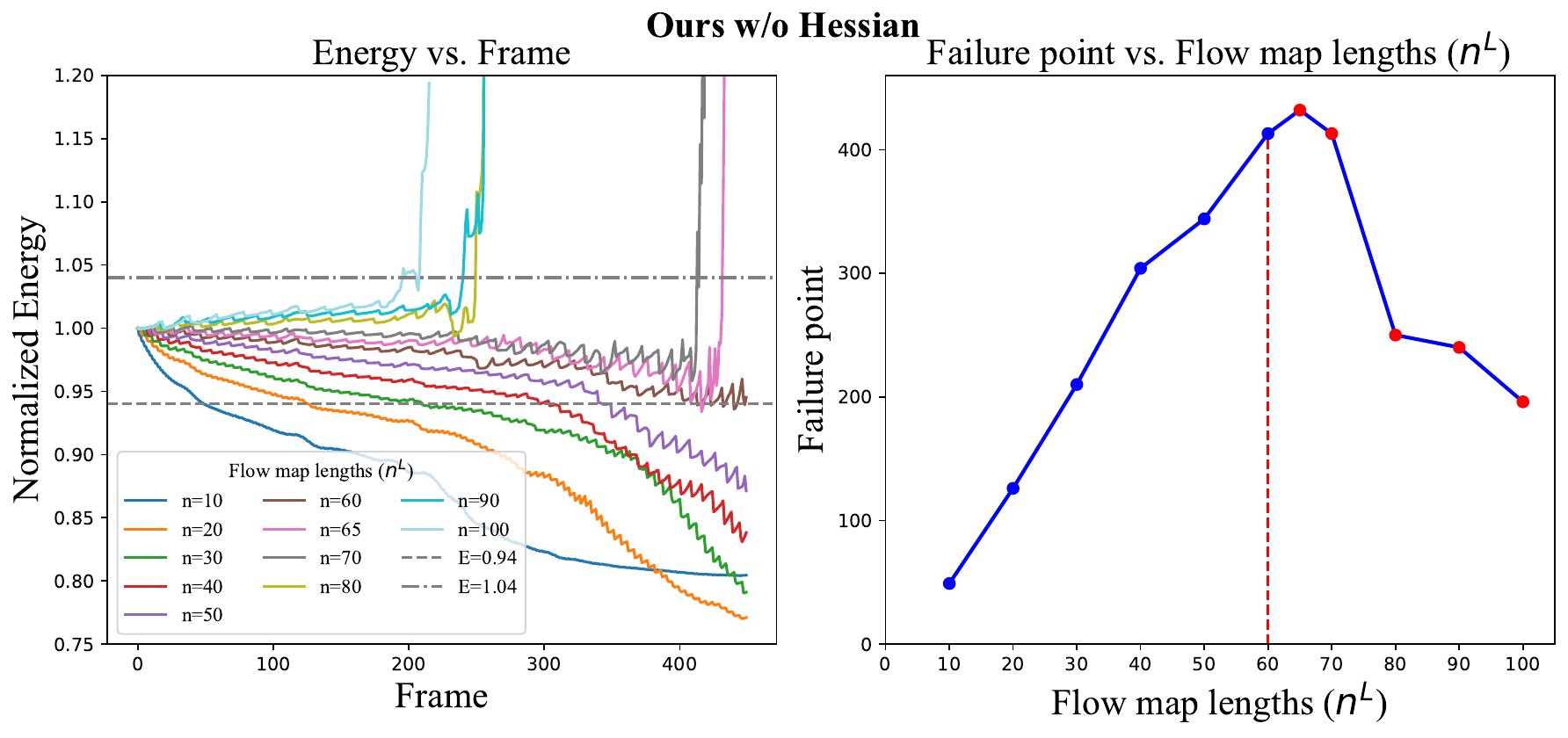}
    %\vspace{-0.05in}
    \label{fig:exp0_w_ours_noHessian}

    \includegraphics[width=0.46\textwidth]{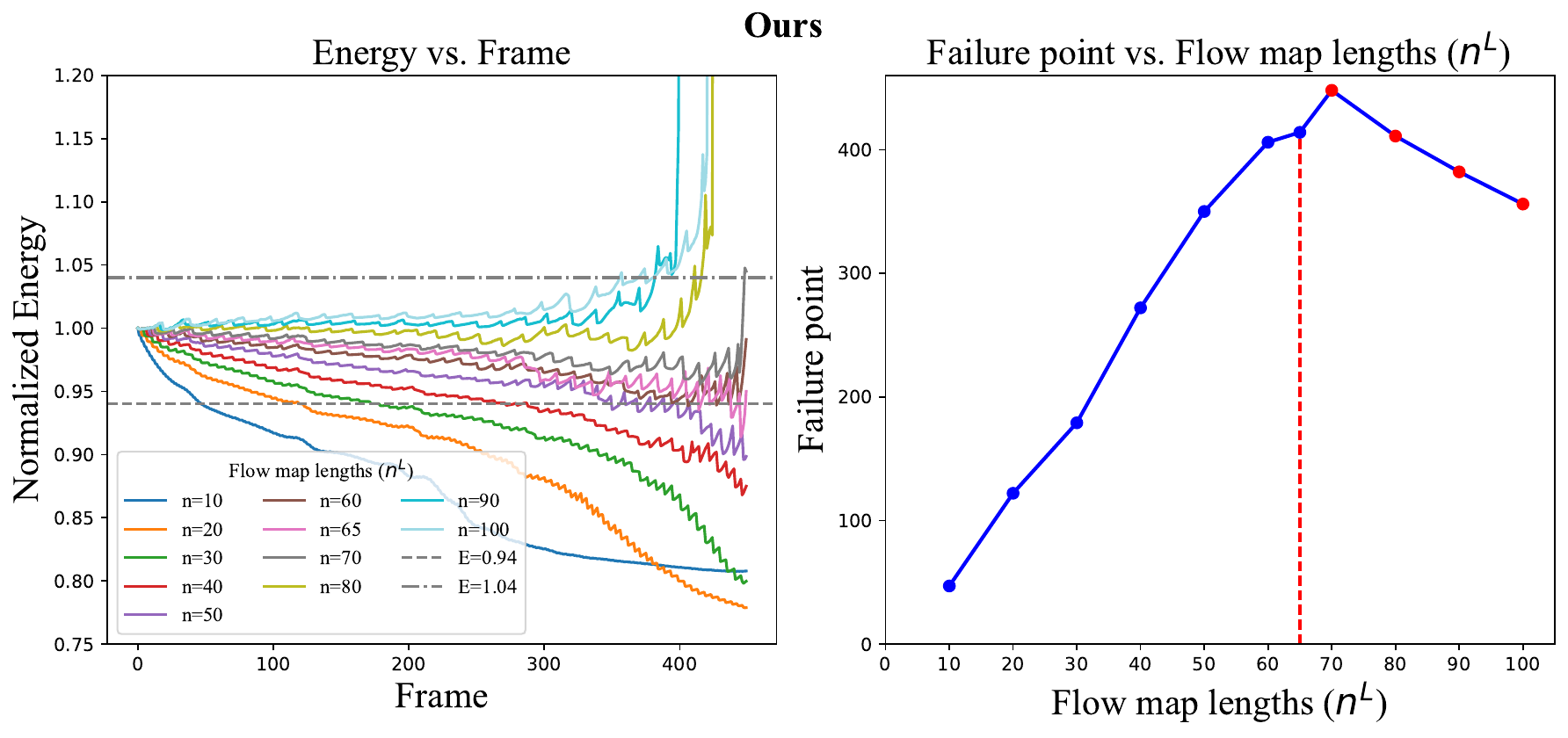}
    %\vspace{-0.05in}
    \label{fig:exp0_w_ours}

    \caption{A comprehensive 3D analysis on the long flow map length \(n^L\). In the right column, the right shift of the red dashed line indicates that our method can significantly extend the flow map length. The curve peaks' upward shift indicates that our extended flow map largely reduces numerical dissipation.}
    \label{fig:exp0_all_in_one}
\end{figure}

\begin{figure}
    \centering
    \hspace{-0.5cm}
    \includegraphics[width=0.5\textwidth]{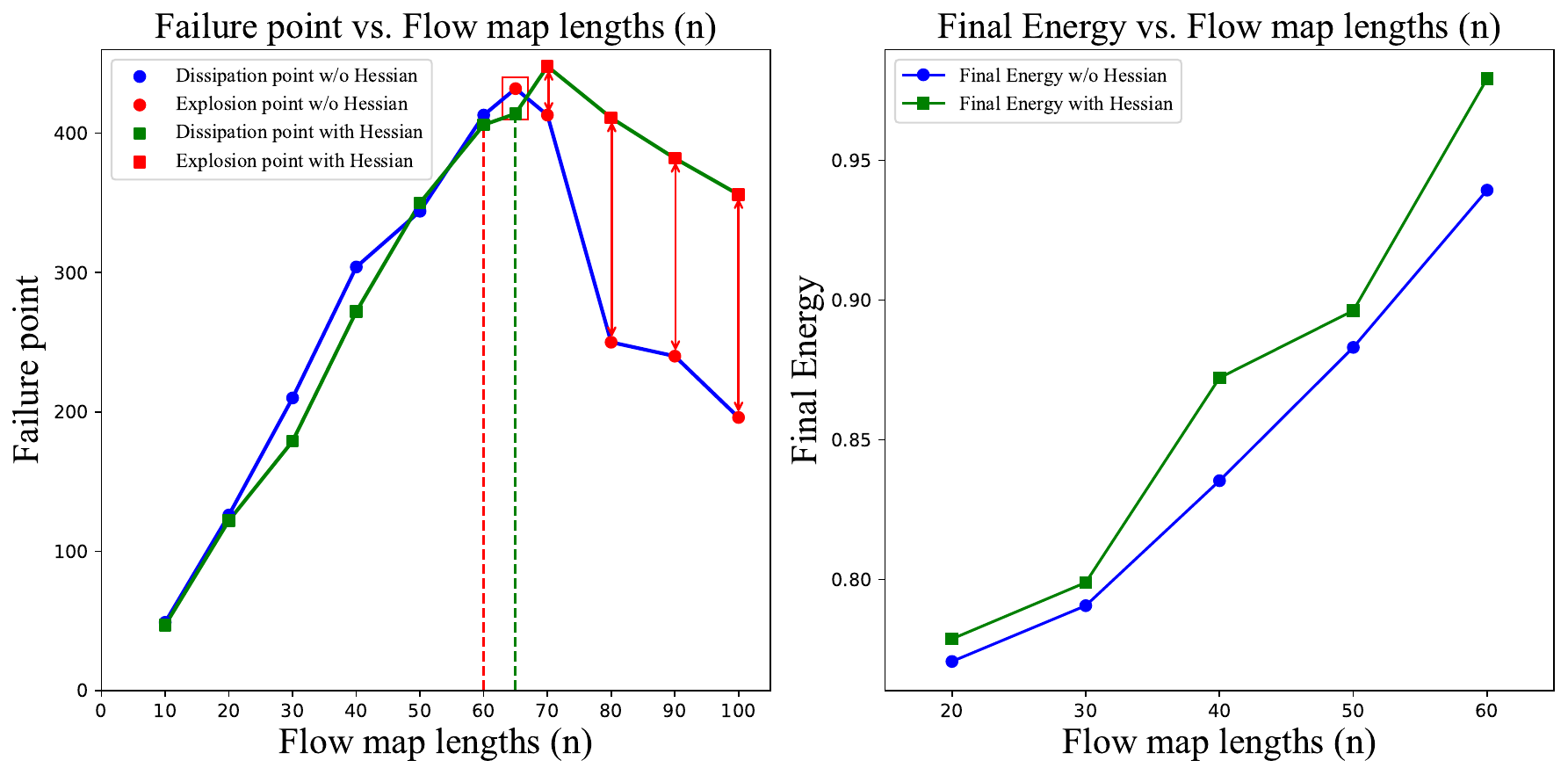}
    % \vspace{-0.8cm}
    \caption{An ablation study on the Hessian term. We compare the curves of the failure points in the left, with the red rectangle indicating that our method fails at \(n^L = 65\) without the Hessian term. The red arrows highlight that incorporating the Hessian term significantly extends the stable simulation time. The right plot shows the final energy for moderate \(n^L\), demonstrating that the Hessian term helps preserve energy during the simulation.}
    \label{fig:hessian_subplot23}
\end{figure}

\begin{figure}
\hspace{-0.03\textwidth}
\includegraphics[width=0.49\textwidth]{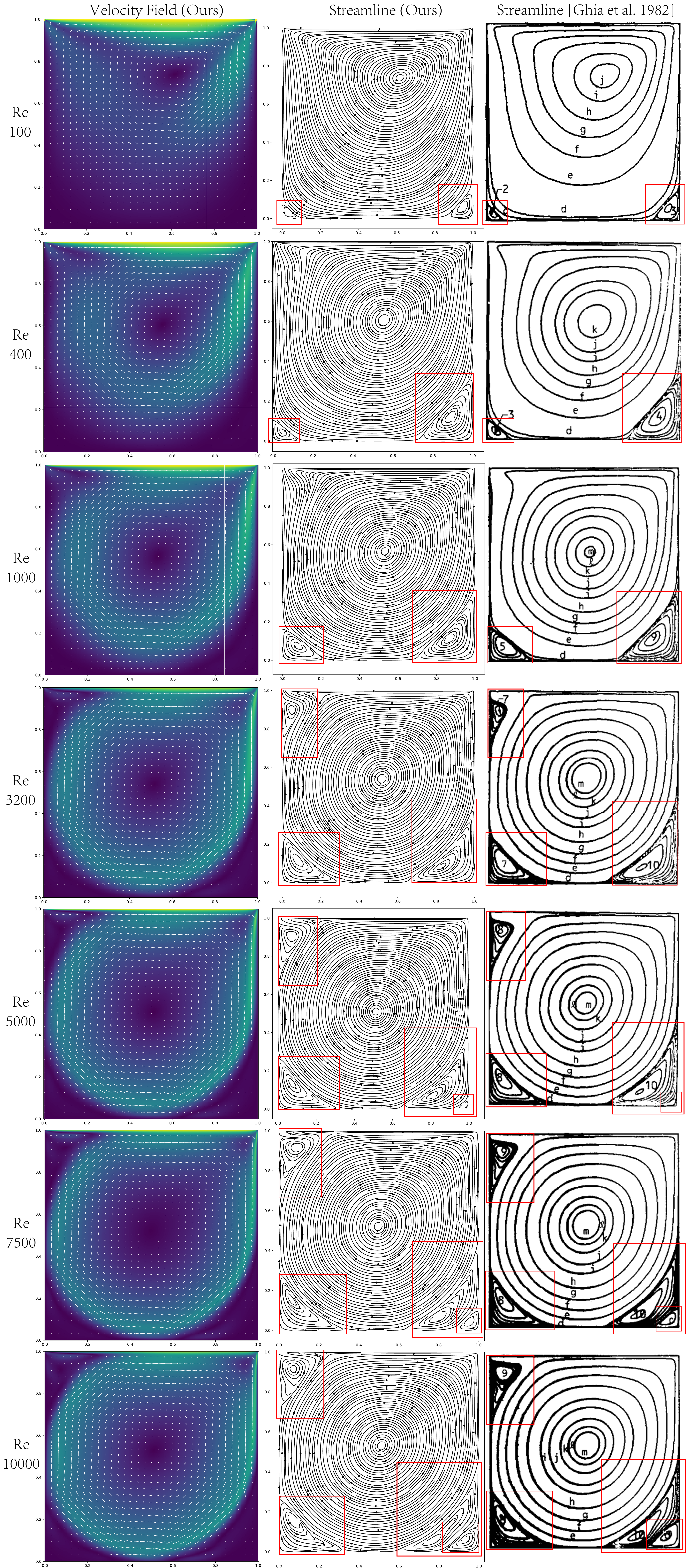}
\caption{Lid-driven cavity flow. The left shows the velocity fields, while the middle and right compare the streamlines of our results with \cite{ghia1982high} under different Re. Red rectangles represents similar secondary vortices observed. More secondary vortices are generated as Re increases.}
\label{fig:lid_driven}
\end{figure}

 % In this study, we demonstrate that our approach extends the flow map length by up to \textbf{2x} compared to previous impulse-based methods and \textbf{1.5x} compared to prior vorticity-based approaches. Moreover, the energy conservation ability are improved \textbf{1.2x} compared with EVM. 
 In Figure~\ref{fig:exp0_all_in_one}, we present a comprehensive 3D Leapfrog experiment where we vary the flow map lengths (i.e., \( n^L \)) and compute each time frame's normalized energy for each of the following method: NFM, PFM, EVM, Ours w/o Hessian and Ours with Hessian.
 %This experiment demonstrates that our method can extend \( n^L \) to be \textbf{two times} longer than previous flow map methods—including NFM~\cite{deng2023neural}, PFM~\cite{zhou2024eulerian}, and EVM~\cite{wang2024eulerian}. Furthermore, the inclusion of the evolved  Hessian term enhances the stability of the simulation when \( n^L \) is large and reduces numerical dissipation at moderate \( n^L \) values.
For different flow map lengths, the left column presents the energy curves, while the right column highlights the failure points. Ideally, the energy should conserve and be a straight line (\(energy = 1\)). Therefore, we define failure points as:

\begin{itemize}[leftmargin=*, labelindent=0pt]
    \item \textbf{Dissipation Failure (Blue Circles)}:  
    A dissipation failure point occurs when the energy curve first intersects the horizontal line at \( \text{energy} = 0.94 \), indicating that the energy has decreased excessively. This signifies a loss of energy due to numerical dissipation.

    \item \textbf{Explosion Failure (Red Circles)}:  
    An explosion failure point occurs when the energy curve first intersects the horizontal line at \( \text{energy} = 1.04 \). This indicates that the simulation may have become unstable with excessive energy growth.
\end{itemize}
 Explosion failures are considered more serious; therefore, if the energy curve intersects both horizontal lines, only the explosion failure is plotted.
 Several observation can be made by Figure~\ref{fig:exp0_all_in_one}:
 \begin{enumerate}[leftmargin=*, labelindent=0pt]
     \item Typically, as we expected, when \(n^L\) increases, the numerical dissipation failure points increase, i.e., we experience less numerical dissipation.
     \item However, with less numerical dissipation, explosion failure is more likely to occur when \(n^L\) is large, i.e., the simulation is more likely to become unstable.
     \item There is an optimal \(n^L\), that can keep this experiment stable, while experience the least numerical dissipation. This is indicated by a red vertical dashed line in the right column of Figure~\ref{fig:exp0_all_in_one}.
     \item This optimal value of \(n^L\) (the x-coordinate of the red dashed line) experiences a considerable \textbf{right shift} from previous methods to ours (top to bottom), implying our method can handle longer flow map.
     \item The numerical dissipation failure point of this optimal value of \(n^L\) (the y-coordinate of the blue circle corresponding to the red dashed line) experiences a considerable \textbf{upward shift}, implying our method effectively reduces the numerical dissipation by extending the flow map length.
 \end{enumerate}

Now we justify the Hessian term. In summary, the Hessian term enhances the simulation stability at larger \( n^L \) values and reduces numerical dissipation at moderate \( n^L \) values, as reflected by the following perspectives:
\begin{enumerate}[leftmargin=*, labelindent=0pt]
    \item At \( n^L = 65 \), the simulation without the Hessian term explodes, while the simulation with the Hessian term remains stable, indicated by the red rectangle in the left of Figure~\ref{fig:hessian_subplot23}.
    \item For all \( n^L > 65 \), the simulation with the Hessian term takes significantly longer (as shown by the gaps indicated by the red arrows in the left of Figure~\ref{fig:hessian_subplot23}) to explode compared to that without it.
    \item For all \( 20 \leq n^L \leq 60 \), the final energy with the Hessian term is higher than that without it, as shown in the right of Figure~\ref{fig:hessian_subplot23}.
    %\item The energy curves with the Hessian term (fourth subplot) are much more consistent and compact compared to those without the Hessian term (first subplot).
\end{enumerate}
Some more experimental results of this comprehensive analysis are summarized in Table~\ref{tab:all_in_one_summary}.

\begin{table*}
\label{tab:flow_map_table}
\caption{Summary of experiment given in Section \ref{subsubsec:comprehensive_3D}. The longest flow map length represents the largest \(n^L\) that keeps this experiment stable. The best flow map length is the \(n^L\) which achieves the largest number of leaps in 3D leapfrog. The maximum final energy of this experiment across different \(n^L\) and its corresponding \(n^L\) are listed in the second to right column.}
\centering\small
\begin{tabularx}{\textwidth}{Y  Y  Y  Y  Y Y Y }
\hlineB{2}
Methods &  Longest flow map length \(n^L\) & Best flow map length \(n^L\) (Qualitative) & Peak of Explosion Failure Point (frame) & Peak of Numerical Failure Point (frame) & Max Final Energy w/o Explosion and \(n^L\) & Number of leap in 3D Leapfrog (Qualitative)\\
\hlineB{1.5}
\textbf{Ours}  & \textbf{65} & \textbf{60} &  \textbf{449} &  \textbf{415} & \textbf{0.950, 65} & \textbf{7}\\
\hlineB{1}
\textbf{Ours w/o Hessian}  & 60 & \textbf{60} &  433 & 414 & 0.945, 60 & 6 (exploded)\\ 
\hlineB{1}
\textbf{EVM}  & 40 & 30 &  177 & 350 & 0.853, 40 & 5\\
\hlineB{1}
\textbf{PFM}  & 30 & 20 &  20 & 208 & 0.620, 30 & 5\\
\hlineB{1}
\textbf{NFM}  & 20 & 20 &  10 & 141 & 0.792, 20 & 5\\
\hlineB{2}
\end{tabularx}
\label{tab:all_in_one_summary}
\end{table*}

% \begin{table}
% \label{tab:dTdx_time}
% \centering\small
% \begin{tabularx}{0.48\textwidth}{Y  Y  }
% \hlineB{2}
% Methods for Computing Hessian &  Wall Clock Time per Frame (s/fr)  \\
% \hlineB{1.5}
% \textbf{Interpolation by neighbors, \cite{morris1997modeling}}  & 0.105 \\
% \hlineB{1}
% \textbf{Ours (Evolve Hessian by Flow Maps)}  & 0.142 \\
% \hlineB{2}
% \end{tabularx}
% \vspace{5pt}
% \caption{\sinan{Pending} Time comparison for computing Hessian}
% \vspace{-0.2in}
% \end{table}

% \begin{figure}
%     \centering
%     \includegraphics[width=0.5\textwidth]{3D_experiments/leapfrog50compare_14.pdf}
%     \caption{3D leapfrog reinit = 50 compare at 400fr}
%     \label{fig:leapfrog50compare}
% \end{figure}

% \begin{figure}
%     \centering
%     \includegraphics[width=0.47\textwidth]{3D_experiments/100compare_leapfrog_hessian_12.pdf}
%     \caption{(Hessian) 3D leapfrog reinit = 100 compare at 211fr}
%     \label{fig:100compare_leapfrog}
% \end{figure}